%% file: main.tex
\documentclass[a4paper,11pt]{article}
\pdfoutput=1 

\usepackage{jcappub} 

\usepackage[T1]{fontenc} 
\usepackage{xcolor}
\include{mycommands}

\title{Constraining Ultra-light Axions with Galaxy Cluster Number Counts}

\author[a,b]{J. Diehl}
\author[b,c]{and J.Weller}

\affiliation[a]{Max-Planck-Institute for Physics, 80805 Munich, Germany}
\affiliation[b]{Universitäts-Sternwarte, Fakultät für Physik, Ludwig-Maximilians Universität München, Scheinerstr. 1, 81679 München, Germany}
\affiliation[c]{Max Planck Institute for Extraterrestrial Physics, Giessenbachstrasse, 85748 Garching, Germany}

\emailAdd{diehl@mpp.mpg.de}
\emailAdd{jochen.weller@lmu.de}

\abstract{In this paper we investigate the potential of current and upcoming cosmological surveys to constrain the mass and abundance of ultra-light axion (ULA) cosmologies with galaxy cluster number counts. ULAs, sometimes also referred to as Fuzzy Dark Matter, are well-motivated in many theories beyond the Standard Model and could potentially solve the $\Lambda$CDM small-scale crisis. Galaxy cluster counts provide a robust probe of the formation of structures in the Universe. Their distribution in mass and redshift is strongly sensitive to the underlying linear matter perturbations. In this forecast paper we explore two scenarios, firstly an exclusion limit on axion mass given a no-axion model and secondly constraints on an axion model. With this we obtain lower limits on the ULA mass on the order of $\ma \gtrsim 10^{-24}$\,eV. However, this result depends heavily on the mass of the smallest reliably observable clusters for a given survey. Cluster counts, like many other cosmological probes, display an approximate degeneracy in the ULA mass vs. abundance parameter space, which is dependent on the characteristics of the probe. These degeneracies are different for other cosmological probes. Hence galaxy cluster counts might provide a complementary window on the properties of ultra-light axions.}


\begin{document}
\maketitle
\flushbottom

\section{Introduction}

In the current standard model of cosmology, consisting of a cosmological constant, $\Lambda$, and a cold dark matter component, CDM, 26\% \cite{Planck2018} of the total mass-energy content of the universe consists of cold dark matter. Most commonly it is considered to be a yet undiscovered elementary particle, which is cold in the sense that its velocity dispersion is non-relativistic. This constituent is referred to as dark matter, because so far only its gravitational imprint has been detected. While being remarkably successful on larger scales, $\Lambda$CDM simulations show significant discrepancies to observations at galactic scales, known as missing satellite problem (e.g. \cite{Bigdudes2, Bigdudes101, klypin1999missing}), too-big-to-fail problem (e.g. \cite{1103.0007, toobigtofail2, toobigtofail3, toobigtofail4}) and cusp-core problem (e.g. \cite{NFW, Emi9, cuspcore3, cuspcore4, cuspcore5}). Even though these problems might be solved invoking baryonic physics (e.g. \cite{baryons1, baryons2, baryons3, baryons4, baryons5, baryons6, baryons7, baryons8, baryons9, baryons10, baryons11, baryons12, baryons13, baryons14, baryons15, baryons16, baryons17, baryons18}), this tension of observations with $\Lambda$CDM predictions on small scales makes modifications on these scales appealing.

One possible modification is substituting (part of) the CDM with so-called fuzzy dark matter (FDM) \cite{FDM}. This dark matter variant consists of very light, scalar bosonic fields. These fields have similar properties as axions in QCD and are therefore often called axion-like particles (ALPs) or ultra-light axions (ULAs). Many theories beyond the standard model --- most notably many string theories --- can naturally produce cosmologically relevant mass densities, i.e. $\Oma \sim \Omega_{\rm dm}$, of these particles over a wide range of particle masses \cite{BigDudes}. For string theories in particular, a single theory often predicts non-thermal production of many different ULA particles ("axiverse" \cite{axiverse}) in the interesting axion mass range $\ma \sim 10^{-22}$\;eV, their number depending on the existence of anti-symmetric tensors in the theory and the topology of the compactified extra dimensions \cite{string1, string2, string3}. Since we only use mass and energy density as ULA parameters, our approach is theory independent. For referring to the particles, we will use the term ULA throughout, $\ma$ for the mass of the particle and $\Oma$ for the energy density in axions or $\F = \Oma / \Omega_{\rm dm}$ for the fractional axion energy density with respect to the overall dark matter density $\Omega_{\rm dm}$. 

For masses around $\ma \sim 10^{-22}$\;eV the deBroglie wavelength $\lambda_{\rm dB}$ of the particles becomes of order of $2\;$kpc \cite{BigDudes}, i.e. macroscopic, thus preventing formation of density cusps below $r < \lambda_{\rm dB}$ at the centre of galaxies and effectively suppressing structures on scales smaller than $\lambda_{\rm dB}$ \cite{FDM, hwang2009axion, poulin2018cosmological}.  The effect is similar to, but not degenerate with the thermal free-streaming induced suppression in warm dark matter (WDM) models, like for models with massive neutrinos. Another key signature of ULAs is their distinct early-time and late-time behaviour: Being scalar fields ULAs obey the Klein-Gordon (KG) equation. Due to the tiny mass $\ma \ll H$, compared to the Hubble scale $H$ at early times, the KG equation for the ULA field is over-damped, the field has a slow-roll phase and behaves like dark energy. At late times, when $\ma \gg H$ the KG equation oscillates rapidly. Here an effective fluid approximation \cite{HGMF} is used, in order to cycle-average over the oscillations. The ULA field then behaves like a DM-like particle with a scale dependent sound speed. \citep[e.g.][]{JC-paper, HGMF, hlozek2018, hwang2009axion}

To address the inconsistencies of the $\Lambda$CDM model on scales of stellar streams and dwarf galaxies, a mass range of around $\ma \sim 10^{-22}$\;eV is required \cite{marshpop, marshsilk, schive2014, bernal2018, deng2018, broadhurst2020}. At lower masses relative heights of the acoustic peaks in the cosmic microwave background (CMB) temperature power spectrum \cite{HGMF} as well as CMB polarisation \cite{Bozek2014} provide a lower limit of $\ma \sim 10^{-23}$\;eV. A similar constraint is obtained from the UV luminosity function of high redshift galaxies \cite{UVlum1, UVlum2, UVlum3}. Significantly higher masses are ruled out by the existence of an old central globular clusters within the ultra-faint dwarf galaxy Eridanus-II. This places interesting constraints on the ULA DM mass around $\ma \sim 10^{-20}$\;eV \cite{MarshNiemeyer, Zoutendijk}.  ULAs would also impact the orbital decay of compact binaries \cite{Poddar20} and via black hole superradiance the spin of black holes \cite{BHSR}. The existence of spinning super-massive and solar-mass black holes places limits on high ULA masses at $\ma \sim 10^{-11}$\;eV $- 10^{-19}$\;eV \cite{Marsh:SMBH}, M87* specifically provides a limit around $4 \times 10^{-21}$\;eV \cite{Davoudiasl:2019nlo}. Lyman-$\alpha$ forest measurements \cite{Lymana, lymana2, lymana3, lymana4, lymana5, Rogers:2020ltq} provide very tight constraints up to $\ma \sim 10^{-20}$\;eV, potentially limiting the capability of ULAs to solve $\Lambda$CDM small scale problems, although various astrophysical effects could alleviate this bound \cite{BigDudes, garzilli2017cutoff}. Another interesting possibility to constrain axion properties is through observations of the x-ray spectrum of galaxy clusters \cite{Krippendorf1,Krippendorf2,Krippendorf3}. Here, due to the presence of a magnetic field in galaxy clusters, axions and photons interact and leave a distinct imprint on the x-ray spectrum.

One of the main motivations to introduce cold dark matter in the modelling of the Universe is to explain the observed large-scale structure. This was early on investigated with dark matter particles originating from extensions of the standard model of particle physics \cite{1977PhLB...69...85H,1982PhRvL..48..223P,1982Natur.299...37B}
leaving imprints on the formation of structures on various astrophysical scales \cite{1984AdSpR...3j.379E}. Axions have been considered for a long time with respect to their influence on the structure formation process \cite{1983PhRvL..51..935M}. Since the nature of dark matter plays an important role in structure formation, we can exploit measures of the distribution of the large-scale structure to constrain these models \cite{LSSprobesDM6, LSSprobesDM1, LSSprobesDM2, LSSprobesDM4, LSSprobesDM5, LSSprobesDM7, LSSprobesDM3}. One of the most sensitive probes of the process of structure formation in the Universe is the distribution of galaxy clusters in mass and redshift \cite{1992ApJ...398L..81B,1993ApJ...407L..49B}. In recent years clusters of galaxies have developed into a state-of-the art cosmological probe \cite{2003A&A...398..867S,2014A&A...571A..20P,Planck2015,2019ApJ...878...55B,scalrel3,2020PhRvD.102b3509A}. With the currently flying Spektr-RG space observatory and eRosita instrument \cite{2010AIPC.1248..543P} and the upcoming Euclid mission of the European Space Agency (ESA) \cite{2011arXiv1110.3193L,2010arXiv1001.0061R} thousands of clusters will be observed and can provide powerful constraints for cosmological models \cite{Euclid2016}.  

In this paper we will investigate the potential of galaxy cluster counts to constrain properties of ULAs. The paper is structured as follows: In chapter \ref{subsec:pathtoNC} we review the calculation of cluster number counts from the halo mass function and our implementation of the Monte Carlo analysis. In chapter \ref{sec:results} our results are presented, distinguishing between the two cases whether or not ULAs are detectable with a given survey. We conclude in chapter \ref{sec:conclusion}.

\section{Modelling Galaxy Cluster Counts in the Presence of Ultra-light Axions} \label{subsec:pathtoNC}

To forecast cluster number counts, or to compare observed counts to a given cosmological model, we need to know the distribution of halos for a given mass $M$ and redshift $z$. \cite{PressSchechter} obtained a first estimate by simply calculating the probability of the formation of a halo and assuming that the underlying distribution of the density fluctuations is Gaussian. This approach relies on a critical density threshold. If a linear perturbation is above this threshold the corresponding region starts to collapse and forms a massive halo. This simple estimation was later put on a more sound footing by the excursion set approach \cite{Bond:HMF,ShethTormen}. A more realistic treatment, taking into account all subtleties of structure formation, is to exploit cold dark matter N-body simulations \cite{Springel, Jenkins, Tinker:HMF}. This allows calibrating the mass function, where the functional form is still motivated by the analytical estimates. Typically the mass function is given in terms of the multiplicity function $f(\sigma)$ \cite{Tinker:HMF} as 
\begin{equation}
    \frac{dn}{d\ln M} = f(\sigma) \frac{d\ln \sigma^{-1}}{d\ln M} \frac{\rho_{m, 0}}{M},  
\end{equation}
where $\sigma(R,z)$ is the r.m.s.~density fluctuation smoothed on a scale $R$ and at a given redshift $z$. $R$ is related to the mass of the halo via the background density of the Universe, i.e. $M=4\pi/3\rho_{m, 0} R^3$. The mass variance $\sigma^2(R) = 1 / \left( 2 \pi^2 \right) \int_0^{\infty} dk P(k) k^2 W(kR)^2$ is obtained from the linear matter power-spectrum $P(k)$ applying a real-space spherical top hat window function $W(kR)$.

In \cite{Tinker:HMF} the multiplicity function is calibrated with multiple detailed simulations and parameterised in the form, $f_\mathrm{Tinker} (\sigma) = A \left( \left( \frac{\sigma}{b} \right)^{-a} + 1 \right) e^{-c/\sigma^2}$. Here $a$, $b$ and $c$ depend on redshift $z$ and the chosen value for the over-density in halos $\Delta$. We use $\Delta = 200$ throughout. The formulation in terms of the multiplicity function makes it possible to study the cosmological universality of the mass function. This provides $5\%$ accuracy over a reasonable parameter range assuming a $\Lambda$CDM framework

In order to compute the mass function to predict the number of clusters at a given mass $M$ and redshift $z$, we first need to compute matter power spectra for ULA cosmologies. The matter power spectrum is calculated from matter density fluctuations as a function of scale $k$ and redshift $z$. For this we need to solve the system of coupled differential equations for the fluctuations of the different components of the cosmological model. An early overview of this is given in \cite{MaBertschinger}. The solution of the hierarchy of Boltzmann equations is computed by Boltzmann solvers such as \cite{CAMB} or \cite{CLASS}. To include the effects of ULAs we need to extend the standard hierarchy of perturbation equations by the ULA component. 

ULAs obey the Klein-Gordon equations of motion for a scalar field $\phi$ and small perturbations $\delta \phi$. The perturbed equation in synchronous gauge and Fourier space is given by \cite{KGeq, HGMF}:
\begin{equation} \label{eq:KG}
    \ddot{\delta \phi} + 2 \frac{\dot{a}}{a}\dot{\delta \phi} + \left( \ma^2 a^2 + k^2 \right) \delta \phi =  - \frac{1}{2} \dot{\phi} \dot{\zeta},
\end{equation}
where $\zeta$ is the scalar potential of the synchronous gauge \cite{MaBertschinger}.\footnote{Usually this parameter is called $h$. We changed it to avoid confusion with the reduced Hubble parameter.} Here $a$ refers to the scale factor and overdots are derivatives w.r.t.\ conformal time. Eq. (\ref{eq:KG}) can be expressed in terms of the over-density $\delta_a = \delta \rho_a / \bar{\rho}_a$ and the velocity $u_a = (1+ w_a) v_a$, which are given in terms of the ULA background density $\bar{\rho}_a$, the velocity perturbation $v_a$ and equation of state parameter $w_a$ \cite{HGMF}:
\begin{align}
    \dot{\delta}_a &= -k u_a - (1+w_a) \dot{\zeta}/2 - 3 \dot{a} / a (1-w_a) \delta_a - 9 \dot{a} / a (1-c_{\rm ad}^2) u_a / k \\
    \dot{u}_a &= 2 \dot{a} / a u_a + k \delta_a + 3 \dot{a} / a (w_a - c_{\rm ad}^2) u_a,
\end{align}
with adiabatic sound speed $c_{\rm ad}^2 = w_a - \frac{\dot{w}_a}{3 \dot{a} / a (1 + w_a)}$.

At late times these equations oscillate heavily due to $\ma \gg H$, making an exact solution impossible. Therefore \cite{HGMF} employ a WKB approximation to average over the fast time scale. This effective fluid approximation leads to averaged equations \cite{HGMF}
\begin{align}
    \dot{\delta}_a &= - k u_a - \dot{\zeta}/2 - 3 \dot{a} / a c_a^2 \delta_a - 9 \dot{a} / a c_a^2 u_a / k \\
    \dot{u}_a &= - \dot{a} / a u_a + c_a^2 k \delta_a + 3 \dot{a} / a c_a^2 u_a, 
\end{align}
with time-averaged ULA sound speed in perturbations $c_a^2 = \frac{k^2/4\ma^2 a^2}{1 + k^2/4\ma^2 a^2}$ \cite{hwang2009axion}. Once we have $\delta_a$, the matter power-spectrum can be calculated via $P(k) \equiv \langle |\delta_a(k)|^2 \rangle$. Due to the scale dependent sound speed of ULAs, structure growth does not evolve equally with redshift on all scales, so the standard, scale-independent growth factor description is no longer valid \cite{growthfactor}.

There are modifications for the two well-known Boltzmann solvers CAMB \citep{CAMB} and CLASS \citep{CLASS} accounting for light scalar fields as a dark matter candidate: AxionCAMB\footnote{Publicly available at \url{https://github.com/dgrin1/axionCAMB}.} \citep{HGMF} and CLASS.FreeSF\footnote{Publicly available at \url{https://github.com/lurena-lopez/class.FreeSF}.} \citep{SFDM-CLASS} (for a comparison see \cite{SFAxcompare, JC-paper}). In this work we use AxionCAMB.

In AxionCAMB the transition to the effective fluid approximation is made at $\ma~=~3 H$ which has been shown to reproduce the exact solution reasonably well \cite{JC-paper}. For more implementation details, see \cite{HGMF}.

\begin{figure}
	\includegraphics[width=\columnwidth]{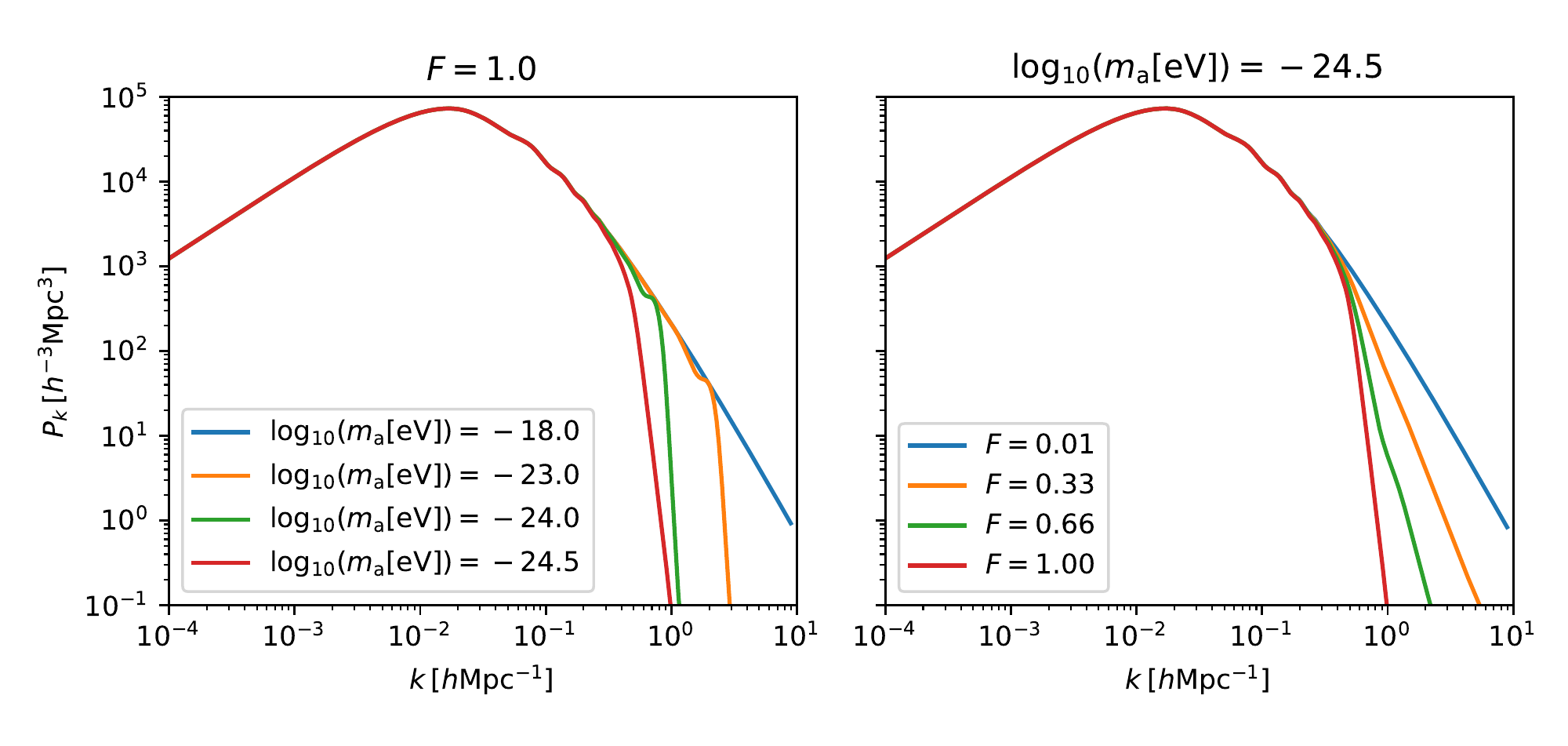}
    \caption{Influence of ULA mass $\ma$ (left) and fractional axion density in ULAs $\F = \Oma / \Omega_{\rm dm}$ (right) on the total matter power spectrum for redshift $z = 0$.}
    \label{fig:MPS_paramtest}
\end{figure}

In figure \ref{fig:MPS_paramtest} we show the total matter power spectra for different ULA cosmologies at redshift $z=0$. We choose the cosmological parameters to be from the Planck cosmology \cite{Planck2018} with fractional energy density in baryons $\Omega_b=0.0495$ and in dark matter $\Omega_{\rm dm} = 0.267 $, scalar spectral index $n_s=0.966$, Hubble parameter $H_0=100\,h\, \rm{km\, s^{-1} Mpc^{-1}}$ with $h=0.672$, scalar amplitude $\ln \left( A_s \times 10^{10} \right) = 3.04$ and optical depth of reionisation $\tau_{\rm re} = 0.0515$. On the left side we see a very distinct cutoff at a scale $k_{\rm J}\propto \ma^{1/2}F^{1/4}$ depending on the ULA mass and fraction. This scale is governed by the de Broglie wavelength of the ULA particles, suppressing structure formation at larger scales with decreasing axion mass. On the left hand side in figure \ref{fig:MPS_paramtest} the increase of the cut-off scale (decreasing k) with increasing axion mass $\ma$ is very noticeable. However, if we consider the changes with increasing axion fraction $F$, on the right hand side in figure \ref{fig:MPS_paramtest}, we have to keep in mind that we are using the \textit{total} matter power spectrum, with $P\propto (1-F)^2P_{\rm dm}+F^2P_{\rm a}$. Hence the cut-off scale becomes more visible for higher axion fraction. Since the scale itself depends only mildly on the fraction, $k_{\rm J} \propto F^{1/4}$, the position of the cut-off barely changes with $F$. The models with $\ma=10^{-18} {\rm eV}$ (left) and $F = 0.01$ (right) are indistinguishable from $\Lambda$CDM in the given parameter range.

With the matter power-spectrum we can now calculate the halo mass function. In order to calculate the halo mass function we adapt an approach previously exploited in \cite{Hagstotz2018}. We should note that simulations for ULA cosmologies using a pseudo-spectral method with - for the first time - subsequent determination of the halo mass function have been presented in a recent paper \cite{May}. The authors only find significant quantitative differences in the power spectrum and halo distribution, compared to $\Lambda$CDM, at much lower halo masses as the ones we consider here. Although they consider larger axion masses than our analysis, we do not expect significant changes on the relevant scales for cluster cosmology beyond linear effects. The effects of ULAs on linear scales are well modelled in AxionCAMB. Implementing ULAs firstly requires modelling of the transfer function and growth of structures \cite{FDM,marshsilk} and secondly the implementation of an adjusted critical linear over-density in order to account for the fact that no structures are formed below the Jeans mass of the ULA cosmology. Note that the Jeans mass $M_J$ for ULA cosmologies  is proportional to $\ma^{-3/2}$ \cite{FDM, Jeans1, Jeans2, Jeans3, Jeans4, Jeans5, Jeans6}. Different approaches to achieve this exist \cite{marshsilk, du2016, 2020arXiv201102116K}. They all have in common that there are no significant effects on the mass function for halo masses above $10^{11}h^{-1}M_\odot$. So for the purpose of this paper it is sufficient to model correctly the growth and transfer function for the ULA cosmologies and these are obtained by the application of the Boltzmann solver including axions \cite{HGMF}. We can then apply this appropriately calculated matter power spectrum to the $\Lambda$CDM mass function \cite{Tinker:HMF}.

\begin{figure}
	\includegraphics[width=\columnwidth]{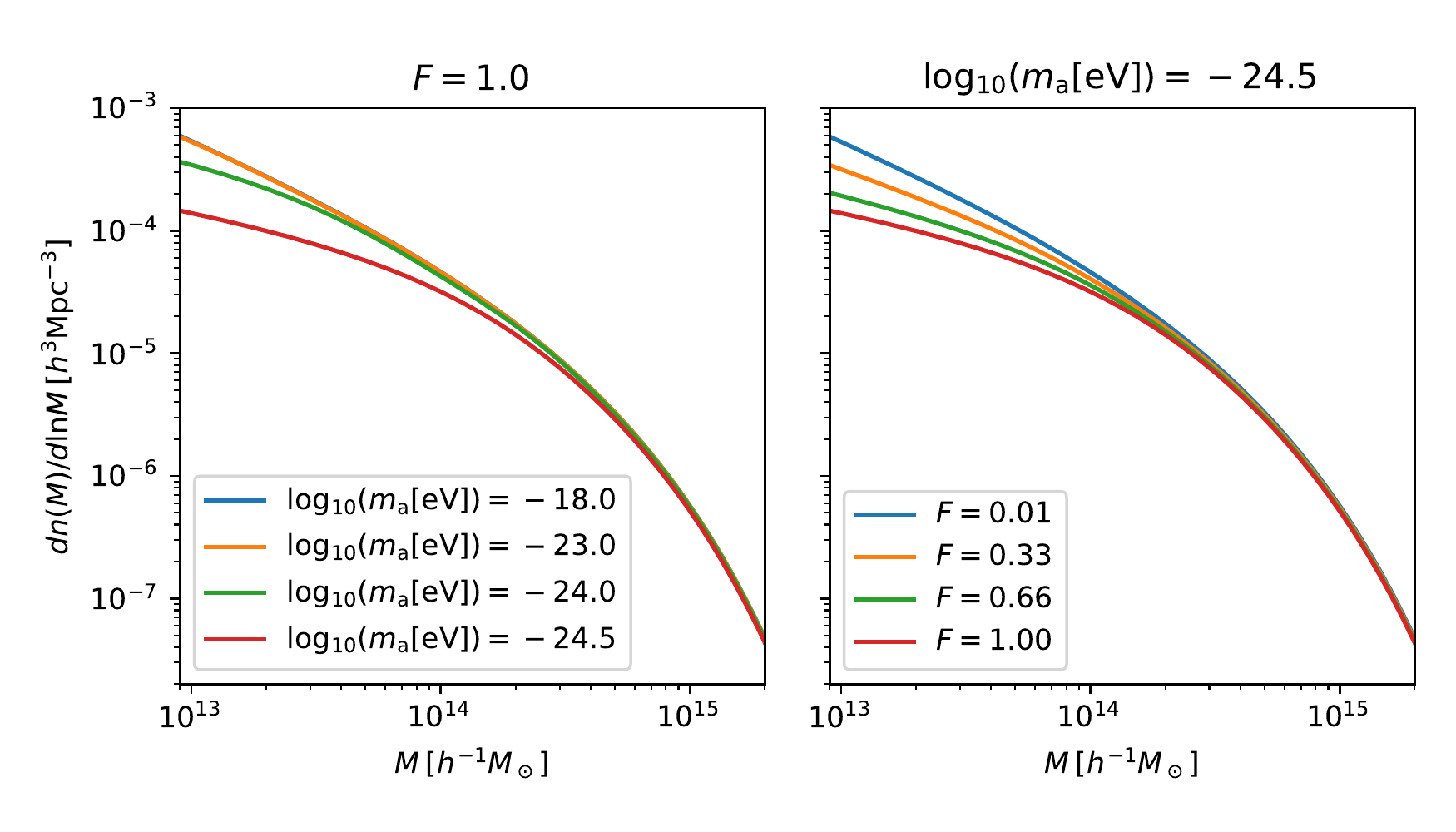}
    \caption{Influence of ULA mass $\ma$ (left) and fractional dark matter density in ULAs $\F = \Oma / \Omega_{\rm dm}$ (right) on the halo mass function for $z = 0$.}
    \label{fig:HMF_paramtest}
\end{figure}

Figure \ref{fig:HMF_paramtest} shows the influence of ULA parameters on the halo mass function in the mass range relevant for galaxy clusters. As expected, the influence of ULAs is larger for smaller halo masses with the ULA mass $m_{\rm a}$ determining the mass range of halos where significant suppression relative to $\Lambda$CDM occurs. The ULA abundance, parametrised by $F = \Omega_{\rm a} / \Omega_{\rm dm}$, determines how strong the suppression is. Since the models $\ma = 10^{-18} {\rm eV}$ and $\ma=10^{-23} {\rm eV}$ (figure \ref{fig:HMF_paramtest}, left) are almost indistinguishable from each other, one can already see that surveys sensitive to cluster masses $> 10^{13} h^{-1} M_\odot$ will possibly not be  able to distinguish these ULA masses from $\Lambda$CDM cosmology. In general the models with $\ma=10^{-18} {\rm eV}$ (left) and $F = 0.01$ (right) are indistinguishable from $\Lambda$CDM in the given mass range.

In order to predict the number of observed clusters we first need to multiply the halo mass function with the observed volume of the survey. For this we need to 
calculate the co-moving volume element $\frac{dV}{dz d\Omega} = r(z)^2/H(z)$, where $r(z)$ is the co-moving coordinate distance and $H(z)$ the Hubble parameter. 
Clusters of galaxies are typically detected and observed via either their Sunyaev-Zel'dovich signature in the cosmic microwave background \cite{1969Ap&SS...4..301Z,1972CoASP...4..173S,1970Ap&SS...7...20S,2014A&A...571A..29P,2016A&A...594A..27P,Planck2015}, their x-ray flux \cite{2000ApJS..129..435B,2004A&A...425..367B,2012MNRAS.423.1024M,2013MNRAS.430..134W} or their optical richness \cite{1958ApJS....3..211A,1989ApJS...70....1A,2007ApJ...660..239K,2016ApJS..224....1R}.
Since true masses of galaxy clusters are not directly accessible with these observations, it is required to assume a scaling relation between the observable and the true mass of the cluster. While the general form of the scaling relation can be motivated for x-ray or Sunyaev-Zel'dovic detected clusters by the underlying physics \cite{2003PhRvD..68h3506B,2014A&A...571A..20P,2007ApJ...668..772M,2010MNRAS.407.2339S}, the situation is less clear for the relation between the optical richness and the mass \cite{rozo14}. Here we choose to explore the ability of future optical clusters surveys to constrain ULA parameters, though we expect that our results hold also for other types of observations. Hence, we follow a generic approach to the mass-observable scaling relation (as e.g. in \cite{obsmass,rozo14, mana13, scalrel2, scalrel3}):
\begin{equation} \label{eq:scalrel}
    \ln \frac{M}{M_{\rm norm}} = a_M + \alpha_M \ln \frac{O}{O_{\rm norm}},
\end{equation}
where $M$ is the true underlying mass of the galaxy cluster, $O$ is the prediction for the observable given $M$, $a_M$ and $\alpha_M$ are scaling relation parameters, and $O_{\rm norm}$ and $M_{\rm norm}$ are normalisation factors. Note that in this generic forecast exercise we keep the mass-observable relation fixed with redshift. We normalise the scaling relation at $M_{\rm norm} = 10^{14} h^{-1} M_\odot$. For our analysis we choose the normalisation by adapting the value from observations of the galaxy cluster richness, with $O_{\rm norm} = N^{\rm gal}_{\rm norm} = 40$ \citep{rozo14}. We further assume that the observed value of the observable $O^{\rm obs}$ follows a log-normal distribution around the "true" theoretical prediction for the observable $O$ \cite{obsmass}:
\begin{equation}
    p(O^{\rm obs}|O) = \frac{1}{\sqrt{2 \pi \sigma^{2}_{\ln{O^{\rm obs}|M}}}} \exp{ \left[-x^2(O, O^{\rm obs}) \right]}
\end{equation}
with 
\begin{equation}
    x(O, O^{\rm obs}) = \frac{\ln{O^{\rm obs}} - \ln{O(M)}}{\sqrt{2 \sigma^{2}_{\ln{O^{\rm obs}|M}}}}.
\end{equation}
Here $\sigma_{\ln O^{\rm obs}|M}$ is the scatter in $\ln{O^{\rm obs}}$ for fixed $M$ and we have used $\sigma_{\ln O^{\rm obs}|M} = \sigma_{\ln O^{\rm obs}|O}$.

We can predict the number of galaxy clusters in bins of the observable $[O^{\rm obs}_i, O^{\rm obs}_{i+1}]$ and redshift $[z_j,z_{j+1}]$ with \cite{mana13}:
\begin{equation} \label{eq:NCs}
    N_{i j} = \Delta \Omega \int_{z_j}^{z_{j+1}} dz \frac{dV}{dz\;d\Omega} \int_{- \infty}^{\infty} d \ln O \frac{1}{\alpha_M} \frac{dn}{d\ln{O} } \frac{1}{2} \left( \mathrm{erfc}(x_i) - \mathrm{erfc}(x_{i+1}) \right),
\end{equation}
using $x_i = x(O, O^{\rm obs}_i)$, the sky coverage of the survey $\Delta\Omega$ and having expressed the halo mass function in terms of the observable via eq.~(\ref{eq:scalrel}). We would like to emphasize that the choice of parameterization of the scaling relation is quite general. As pointed out below we work with mass limits instead of observable limits in the analysis presented here. This allows the results to be transferable to other survey set-ups, like for example SZ or x-ray observations. The crucial parameter however is the scatter in the mass observable relation. This parameter defines how many clusters scatter into the mass-bin from neighbouring bins or the observational limits. Hence the uncertainty in the scattering parameter will be degenerate with cosmological parameters, as pointed out for the case of dark energy in \cite{obsmass}.

\begin{figure}
\centering
    \includegraphics[width=\columnwidth]{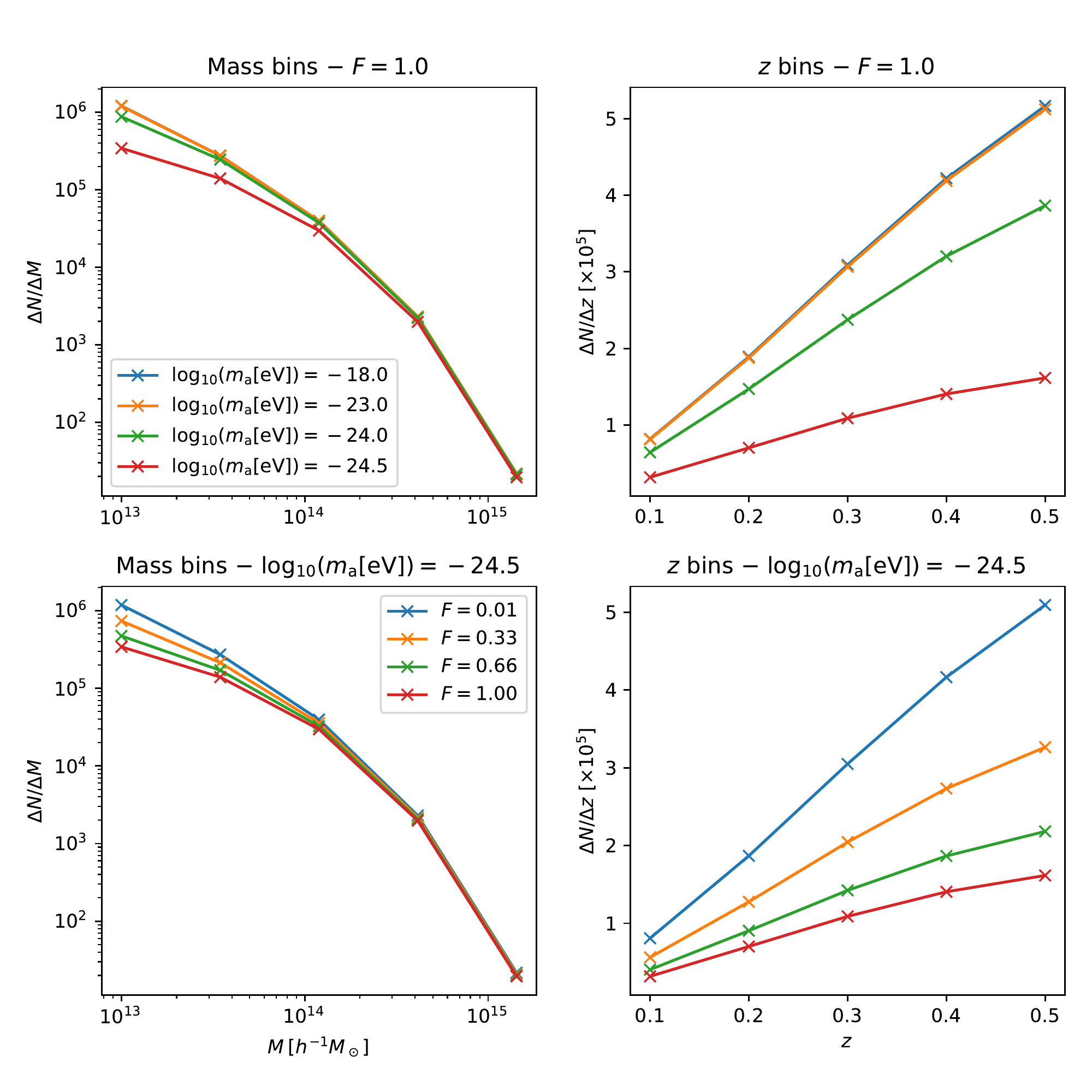}
    \caption{Influence of ULA mass $\ma$ with constant ULA abundance $\F$, and vice versa, on cluster number counts in mass (left) and redshift (right) bins.  The number of clusters per bin is plotted as an x at the lowest mass in each bin. Note that in mass bins we integrate over the full redshift range, i.e $z \in [0.1,0.6]$ and in redshift bins over the full mass range, i.e. $M \in [1 \times 10^{13},5 \times 10^{15}]\,h^{-1}M_{\odot}$. The legends each apply for both panels in a row.}
    \label{fig:NC_paramtest}
\end{figure}

Throughout our analysis we assume survey parameters comparable to next generation cosmological surveys such as the Euclid mission of the European Space Agency ESA \citep{Euclid2016}. Specifically we assume a sky coverage of $\Delta \Omega = 10^{4}\;\mathrm{deg}^2$ and a redshift range $0.1<z<0.6$ with five equally sized bins in redshift $z$ and in the mass $\log_{10}M$, resulting in a total of 25 bins of galaxy cluster counts. In order to understand the influence of the limiting mass of a survey, we explore the lower bound from values at  $10^{13}\,h^{-1}M_{\odot}$ to more realistic values of $5 \times 10^{14}\,h^{-1}M_{\odot}$. Since for any realistic cosmology there are no clusters of masses above $5 \times 10^{15}\,h^{-1}M_{\odot}$ expected, we set the upper integration bound to this value. Figure \ref{fig:NC_paramtest} shows the predicted number counts for different ULA parameters. For the standard cosmological parameters we use Planck 2018 values \cite{Planck2018}, as listed above. In order to explore the influence of the ULA parameters, in figure \ref{fig:NC_paramtest} we set the observable-mass relation parameters to the no bias, ideal scaling and no scatter values, i.e.~$a_M = 0$, $\alpha_M = 1$ and $\sigma_{\ln O^{\rm obs}|M} \approx 0$.

Since cluster number counts are a binned version of the halo mass function convolved with the purely geometrical volume element, the dependence on the ULA parameters for number counts follows the trend observed in the halo mass function as shown in figure \ref{fig:NC_paramtest}. Here we can already see that the lowest mass bins are crucial for the analysis. Firstly, because this is where ULAs produce the biggest relative differences in NCs and secondly, because the highest number of absolute counts also has the biggest impact on the value of the likelihood. In redshift we see an increasing number of clusters due to a growing co-moving volume. The suppression of the halo mass function only becomes dominant at much higher redshifts. $ \ma = 10^{-18}$\;eV in the top and $F = 0.01$ in the bottom panels are indistinguishable from a $\Lambda$CDM model.

\section{Modelling the Likelihood and Forecasting Constraints on Ultra-light Axion Parameters} \label{sec:results}

\subsection{Sampling of the Likelihood}

The dominant error of galaxy number counts follows a Poisson statistics, hence we need to compute the Poissonian log-likelihood (compare e.g. \cite{Planck2015}) 

\begin{equation}
    \sum_k \ln{P(\Nobs^k|\mathbf{\theta})} = \sum_k \Nobs^k \ln{\Nth^k} - \Nth^k + \textrm{const.},
\end{equation}

with $\mathbf{\theta}$ a vector of cosmological parameters of interest. The sum is over all observational bins, which for our analysis is a grid of 5 bins each in cluster mass and redshift. Note that here we ignore the contribution due to sample covariance \cite{Lima:2004wn}. Of course any precision cosmological analysis should include this term. However, the sample covariance is proportional to the square of the number density of observed clusters. Since we will concentrate our forecast on observational programs which cover a quarter of the sky, we do not expect this contribution to make a big difference for our analysis, even in the low mass bins.

We want to analyse the ability of future surveys to constrain the ULA parameters. For this we assume fiducial model galaxy cluster counts calculated according to section~\ref{subsec:pathtoNC}. In order to efficiently sample the posterior likelihood we employ a Monte-Carlo Markov Chain (MCMC) algorithm.

Since we do not have prior knowledge of the prior probability for ULA parameters we use uninformative flat priors as shown in table \ref{tab:MCparams}. For the standard cosmological parameters we use the Planck covariances from the 2018 data release \cite{Planck2018} as priors. For the numerically demanding probability analysis we use the dark matter content $\Omega_{\rm dm}$, the scalar primordial perturbation amplitude $A_{\rm s}$ and the spectral index $n_{\rm s}$ as free parameters.  All other cosmological parameters are fixed.  They are the parameters, which most affect the matter power spectrum and hence the halo mass function. Note that we fix the Planck covariance to the $\Lambda$CDM case. For a detailed investigation with real data, of course a joint analysis between the two probes should be performed. Including the ULA parameters would most likely widen the prior range from Planck. Given that we also fix parameters, like the baryon contents of the universe, we expect this approximation to be sufficient for the forecast analysis presented here.
Priors on observable-mass distribution parameters have been taken from the analysis by Rozo et al.~\cite{rozo14} (their table 4). An overview for all the parameters and their priors used in our analysis can be found in table \ref{tab:MCparams}.

\begin{table}[t]
    \centering
    \begin{tabular}{|c|c|c|c|c|} \hline
        Type & Symbol & Definition & Fiducial Value & Prior \\ \hline
        ULAs & $\log_{10}\left( \ma[\rm eV] \right)$ & Logarithm of ULA mass & various & various \\
         & $F$ & Fraction of DM in ULAs & various & $[0,1]$ \\ \hline
        Cosmology & $\Omega_{\rm dm}$ & DM energy density & $0.267$ & Planck \\ 
         & $\ln \left( A_s \times 10^{10} \right)$ & Amplitude of primordial perturb. & $3.04$ & Planck \\ 
          & $n_s$ & Scalar spectral index & $0.966$ & Planck \\ \hline
        obs.-mass & $\alpha_M$ & Scaling Exponent & $1.06 \pm 0.11$ & Gauss \\ 
        distrib. & $a_M$ & Scaling Offset & $0.75 \pm 0.1$ & Gauss \\ 
         & $\sigma_{\ln O^{\rm obs}|M}$ & Scaling Scatter & $0.45 \pm 0.1$ & Gauss \\ 
        \hline
    \end{tabular}
    \caption{Parameters for posterior analysis. \textit{Planck} denotes Gaussian priors obtained from Planck 2018 \cite{Planck2018} covariance matrix. Fiducial values for cosmological parameters also come from \cite{Planck2018}, the observable-mass distribution parameters from \cite{rozo14}. The prior in ULA mass is logarithmically flat but needs to be shifted for various values for $M_{\rm min}$ to not sample too much of the unconstrained parameter space.}
    \label{tab:MCparams}
\end{table}

The analysis is conducted using the affine-invariant code \textit{emcee} \cite{emcee}, convergence is tested using the indicator from \cite{Gelman1996}.

\subsection{Lower Limits on the Axion Mass}
\begin{figure}
    \centering
    \includegraphics[width=\columnwidth]{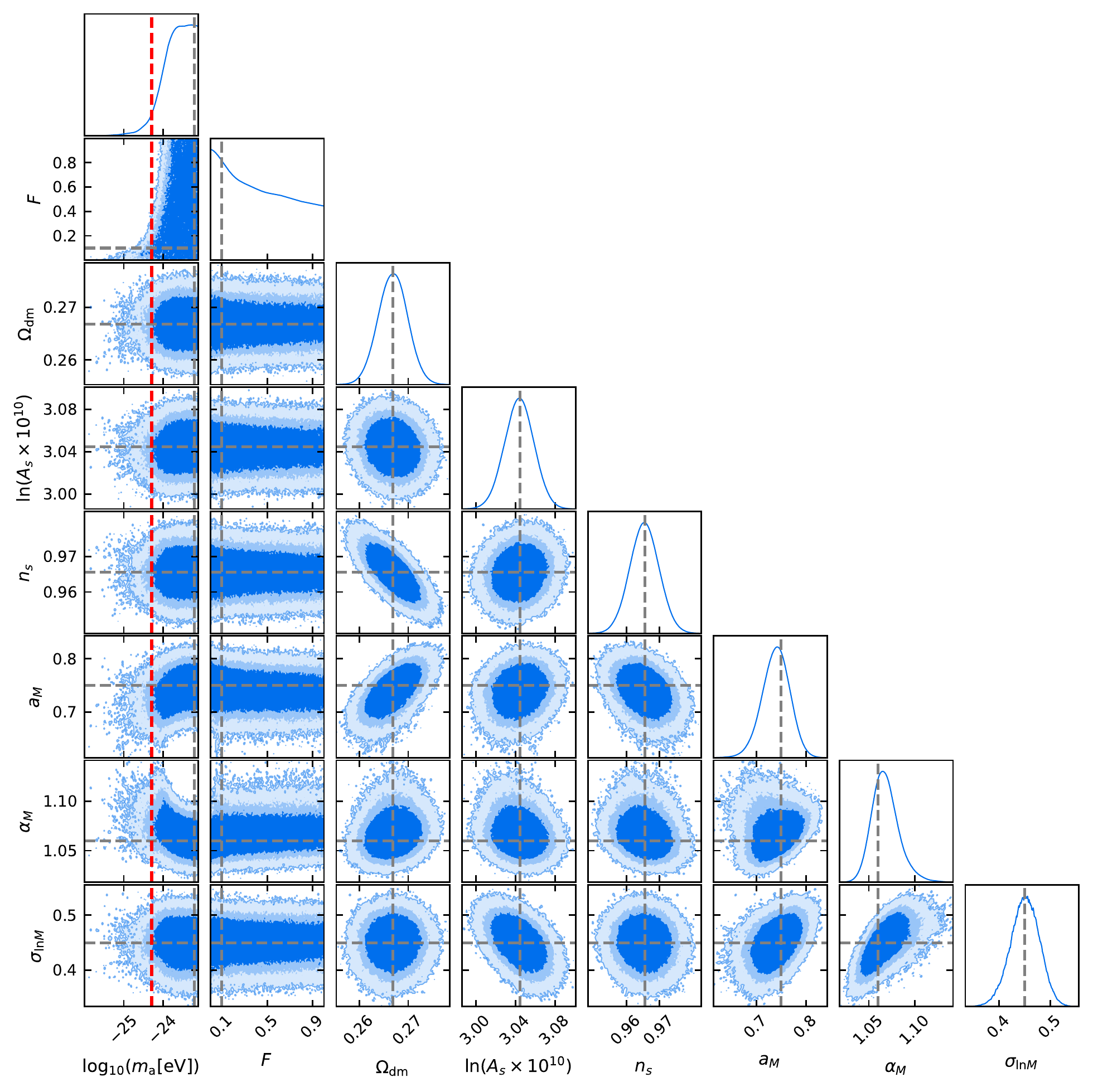}
    \caption{Posterior for ULA parameters indistinguishable from $\Lambda$CDM ($\log_{10}\left( \ma[\rm eV] \right) = -23.2$, $F = 0.1$). Dashed grey lines denote fiducial parameter values, the dashed red line is the $95\%$ exclusion limit of the marginalised 1D density for $\log_{10}\left( \ma[\rm eV] \right)$. Minimum cluster mass observable $10^{14}\, h^{-1}M_\odot$. Different shades of blue from dark to light denote 68$\,\%$, 95$\,\%$ and 99$\,\%$ posterior contours respectively.}
    \label{fig:posterior1}
\end{figure}

We would first like to address the question up to which limit galaxy cluster counts can exclude an axion cosmology given a $\Lambda$CDM fiducial cosmology. Hence we perform a likelihood analysis with fiducial values on the standard cosmological parameters from \cite{Planck2018} and an axion mass an order of magnitude above the expected exclusion limit.
For relatively large particle masses ULA cosmologies cannot be distinguished from $\Lambda$CDM anymore as figure \ref{fig:posterior1} shows. The posterior for ULA mass is not Gaussian, instead it resembles a logistic function, running into the upper prior bound of the axion mass $\ma$. This makes sense, since large axion mass cosmologies are indistinguishable from a $\Lambda$CDM model. Lower axion masses are ruled out for a fiducial $\Lambda$CDM model, due to their large impact on the halo mass function at the mass scales probed by galaxy clusters. For large abundances $F$ the differences in the posterior primarily arise due to number count differences in the lowest-mass bins of the survey. Smaller $F$ have less impact on the level of suppression, therefore smaller ULA masses are allowed for lower abundance values $F$. Above this exclusion threshold the posterior is relatively flat for the ULA mass. The preference for small ULA abundances shown in the posterior distribution of $F$ is caused by a larger unconstrained region in $\ma$ for small $F$.

The dark matter density and scalar spectral index do not portray significant degeneracy with ULA parameters. Only $A_s$ experiences a non-significant shift to slightly smaller values when the lower bound on axion mass is approached. 

Cluster counts provide additional constraining power to the scaling relation parameters $a_M$ and $\alpha_M$. The marginalised 1-$\sigma$ limits shrink from $\pm 0.1$ and $\pm 0.11$ to $\pm 0.03$ and $\pm 0.02$ respectively. A significant degeneracy between the two parameters and ULAs is visible, for small ULA masses $a_M$ is shifted to lower, $\alpha_M$ to higher values. This negatively impacts the constraining capability of the probe, making tight priors on these two parameters desirable. We performed an analysis with fixed scaling relation parameters. In this case the exclusion on the axion mass changes from $\log_{10}(m_a[\rm{eV}])=-24.29$ to $\log_{10}(m_a[\rm{eV}])=-24.23$, corresponding to a 15$\%$ increase of the mass limit. We want to stress that this is of course an unrealistic scenario but serves as a limiting case what in the best circumstances is achievable. This could be achieved by complementary observations of the galaxy cluster masses, for example with gravitational lensing observations. We also investigated the case of doubling the spread of the prior range on the scaling parameters. In this case there is barely a change on the exclusion limit.

Given a $\Lambda$CMD cosmology as fiducial model the analysis can only produce a lower limit for the axion mass and not place any constraints on its fractional abundance. For the lower limit we use the $95\%$ exclusion limit of the marginalised 1D posterior probability in $\log_{10}\left( \ma[\rm eV] \right)$. In order to investigate different survey possibilities we consider different lower mass limits and redshift ranges. Note that current optical and SZ surveys have typical mass limits between $10^{14}\,h^{-1}M_\odot - 5\times 10^{14}\,h^{-1}M_\odot$. \cite{Planck2015,Staniszewski:2008ma,2007ApJ...660..239K,2016ApJS..224....1R}. Nevertheless in the following analysis we will also include a lower mass limit of $5\times 10^{13}h^{-1}M_\odot$ in order to assess the ability of future x-ray \cite{2012MNRAS.422...44P} and optical surveys \cite{Euclid2016}. Table \ref{tab:malimits} summarises $\log_{10}\left( \ma[\rm eV] \right)$ exclusion limits for different minimal mass $M_{\rm min}$ of the observed clusters. Our analysis indicates that $M_{\rm min}$ is by far the most important survey parameter when it comes to the ability to constrain axion mass with a certain data-set. An increase in the surface angle of the survey $\Delta \Omega$ increases the total number of observed clusters and should therefore have a small positive impact on the constraints for real surveys. The same is true for reaching higher redshifts. Since fractional differences of different axion models do not have a significant redshift dependency however (compare figure \ref{fig:NC_paramtest}), this effect is subdominant compared to being able to observe clusters with lower masses. The left panels of figure \ref{fig:NC_paramtest} show that this is where relative differences between different axion cosmologies are biggest, which translates to our findings for exclusion limits of $\ma$ in table \ref{tab:malimits}.

To investigate the influence of the redshift range of the survey on the constraints, we additionally conducted an analysis with $0.1 < z < 1.0$ and $M_{\mathrm{min}} = 1 \times 10^{14}\,h^{-1}M_{\odot}$. We increased the number of redshift bins from five to nine, all other parameters remaining equal. This increases our $95\%$ exclusion limit on $\mathrm{log}_{10}\left( \ma[\mathrm{eV}] \right)$ from $-24.29$ to $-24.21$, which is a factor $1.2$ increase in sensitivity in non-logarithmic units. Two effects explain this slightly more constraining exclusion limit: Increasing the redshift range to higher redshifts increases the number of clusters. Firstly this increased number of clusters reduces the Poisson error of the likelihood and hence improves the constraint. In addition ULA effects are slightly more pronounced at higher redshifts, as can be inferred from figure \ref{fig:NC_paramtest}, leading to better constraints for a larger redshift range.

\begin{table}[t]
    \centering
    \begin{tabular}{|c|c|} \hline
        $M_{\rm min} [h^{-1}M_{\odot}]$ & $95\%$ exclusion for $\log_{10}\left( \ma[\rm eV] \right)$ \\ \hline
        $5 \times 10^{13}$ & $-24.09$ \\ \hline
        $1 \times 10^{14}$ & $-24.29$ \\ \hline
        $5 \times 10^{14}$ & $-24.98$ \\ \hline
    \end{tabular}
    \caption{$\log_{10}\left( \ma[\rm eV] \right)$ exclusion limits for different minimal mass $M_{\rm min}$ of the observed clusters.}
    \label{tab:malimits}
\end{table}

\subsection{Constraining the Axion Mass}

Instead of asking which exclusion limit we can obtain given specific survey parameters, we may also wonder how a possible detection might unfurl and be impacted by parameters of the survey. To answer this question we adopt a fiducial model of $F = 0.4$, $\log_{10}\ma[{\rm eV}] = -24.3$.
Although this specific model is already under strain from other probes it is interesting to see if such a model could be "detected" with galaxy cluster counts as a cosmological probe.

\begin{figure}
    \centering
    \includegraphics[width=0.6\textwidth]{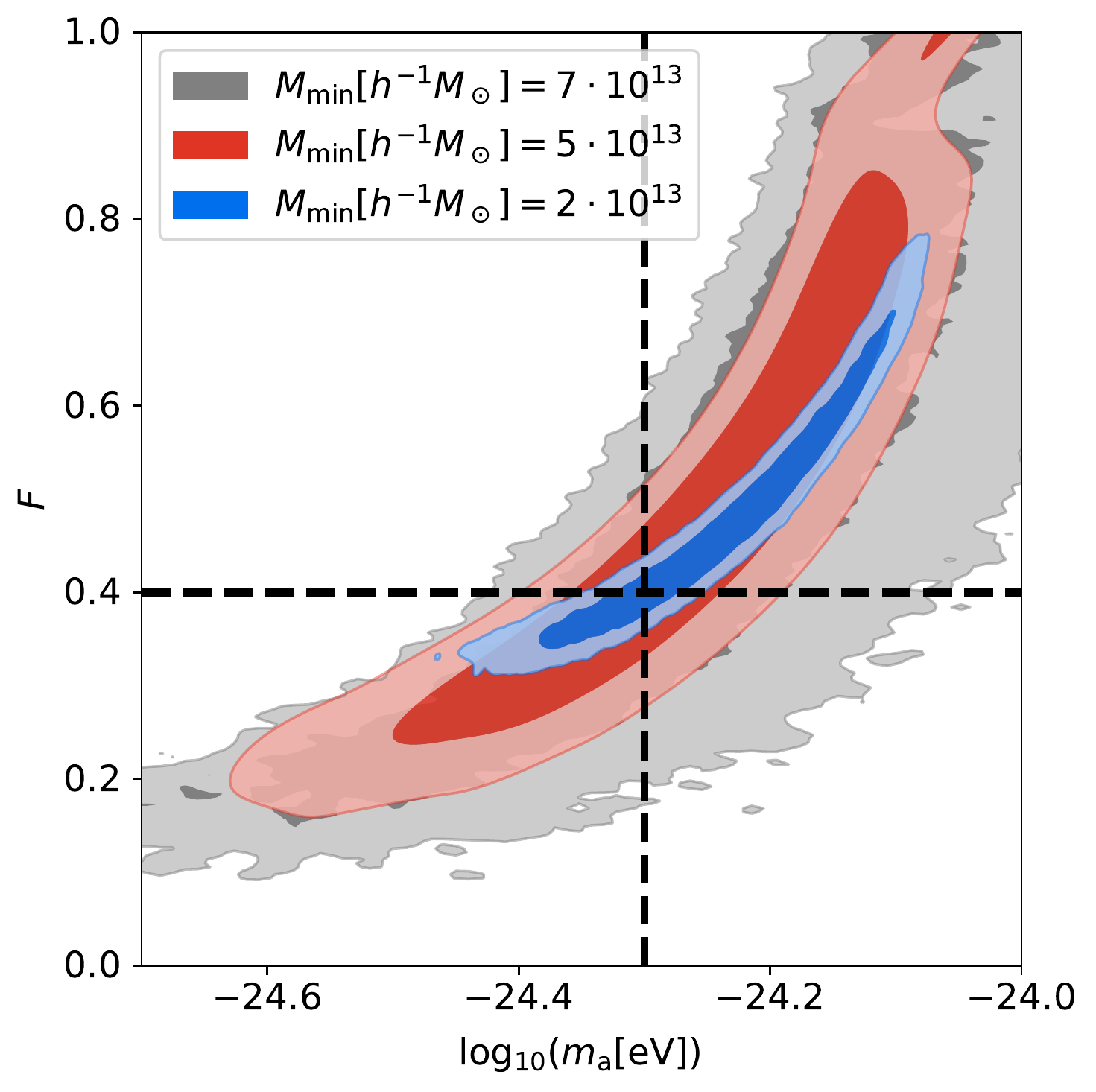}
    \caption{$68\%$ and $95\%$ joint posterior contours on the ULA abundance $F$ and ULA mass $\ma$ for Changing shape of different minimum observed galaxy cluster mass. Dashed lines denote the fiducial model.}
    \label{fig:degeneracyshape}
\end{figure}

ULAs with these parameters could possibly be distinguished from $\Lambda$CDM by cluster surveys with minimal observed clusters masses $M_{\rm min} \lesssim 1 \times 10^{14} \,h^{-1}\,M_\odot$. For $M_{\rm min}$ values slightly below this limit, it is not surprising to find a significantly degenerate posterior. In section \ref{subsec:pathtoNC} we showed that the effect from the abundance $F$ and ULA mass $\ma$ on the matter power spectrum or the halo mass function look similar. In the limit where ULAs are only barely detectable this similarity naturally leads to a degeneracy in the $\ma$ -- $F$ parameter plane as figure \ref{fig:degeneracyshape} shows. For reduced minimal observed cluster masses $M_{\rm min}$ the joint posterior in $\ma$ and $F$ has a smaller volume and is less steep. One can understand this with the effect of the ULA parameters on the halo mass function in figure \ref{fig:HMF_paramtest}. Let us consider two ULA models, which look degenerate for a survey with a limiting mass of $M_{\rm min} = 7 \times 10^{13}\,h^{-1}\,M_\odot$, e.g. a fiducial model with $F_1 = 0.4$ and $\log_{10}m_{\rm a,1}[{\rm eV}] = -24.3$ and a second model with $F_2 = 0.8$ and $\log_{10}m_{\rm a,2}[{\rm eV}] = -24.1$. For a different survey with a smaller limiting mass of $M_{\rm min} = 2 \times 10^{13}\,h^{-1}\,M_\odot$, these two models are not degenerate anymore. The second survey has an additional mass bin with $2 \times 10^{13}\,h^{-1}\,M_\odot \leq M_{\rm min} \leq 7 \times 10^{13}\,h^{-1}\,M_\odot$. Here the fiducial model is highly suppressed in comparison to the second model. Therefore lowering the abundance $F$ of the second model will result in a better match with the fiducial model and hence the posterior will be less steep in the $\log_{10}m_{\rm a}$ -- $F$ parameter space. The reason for this behaviour essentially is a result of the different effect of the axion mass $\ma$ and the axion fraction $F$ on the total matter power spectrum as discussed in section \ref{subsec:pathtoNC}. Reducing the limiting mass $M_{\rm min}$ of the survey for the same axion mass range is increasing the constraining power of the survey and therefore the posterior volume is smaller. Of course we would like to stress that it is very challenging to use "clusters" in a meaningful way down to such low masses, but we demonstrate here that this could reveal new physics.

\begin{figure}
    \centering
    \includegraphics[width=\columnwidth]{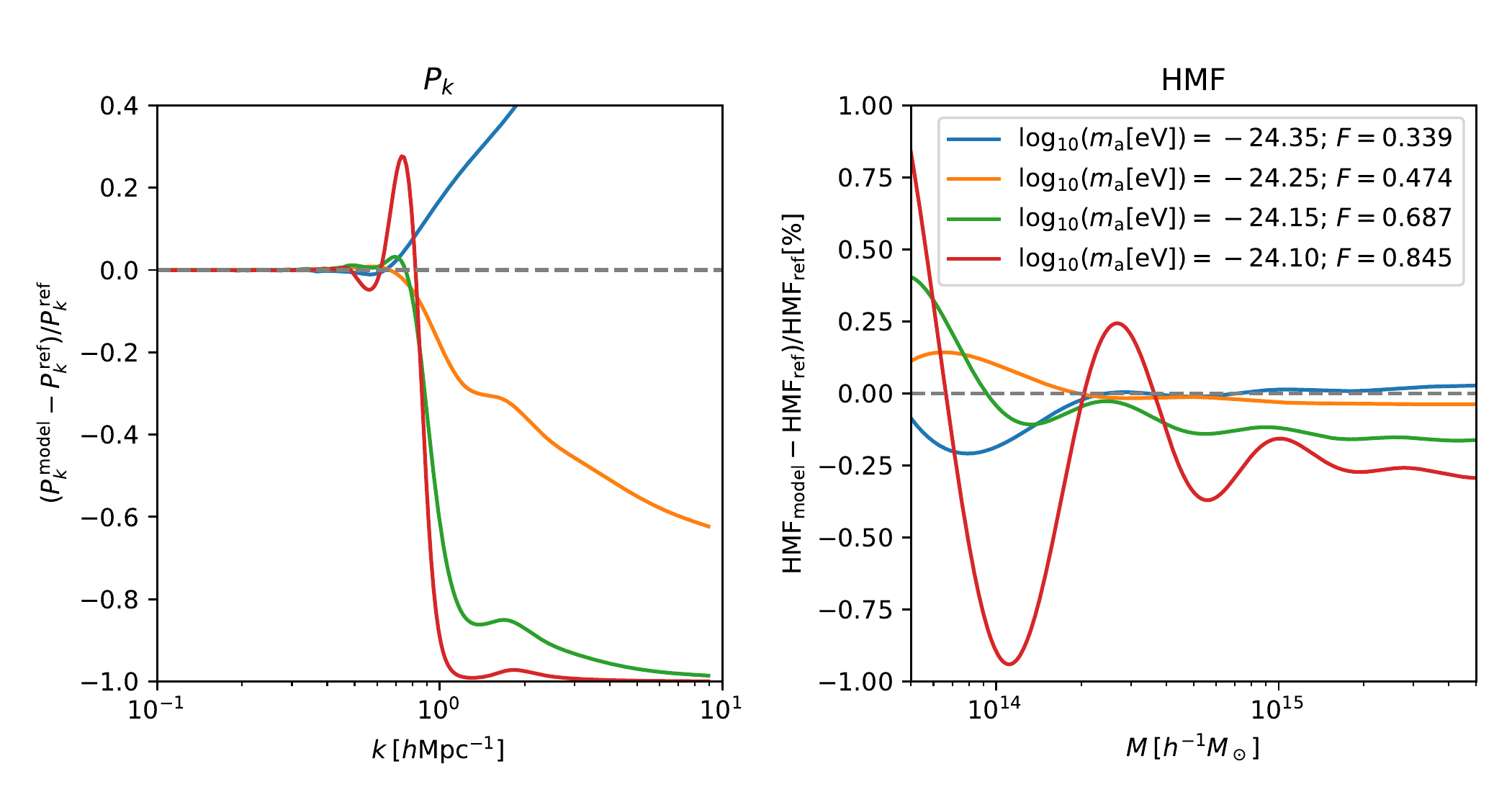}

    \caption{The relative difference in the power spectrum $P_k$ and the halo mass function, at $z=0$, for different models along the degeneracy line for  a limiting mass of $M_\mathrm{min} = 5 \times 10^{13}\,h^{-1}\,M_\odot$. The reference model is given by $\log_{10}m_{\rm a}[{\rm eV}] = -24.3$ and $F = 0.4$ . The relatively big, short oscillations in the power spectrum, $P_k$, lead to longer oscillations in the halo mass function at the $1\%$ level.}
    \label{fig:pkdust}
\end{figure}

Additionally figure \ref{fig:degeneracyshape} displays slight multi-modality for the abundance $F\sim 1$. This is not a boundary effect but arises due to small oscillations in the matter power spectrum. The amplitude and position of these oscillations in $k$ space depend on ULA parameters (compare figure \ref{fig:pkdust}). These oscillations translate to percent level differences over a wide mass range in the halo mass function, which can not be resolved by typical mass bins of cluster counts. Therefore multi-modalities in the posterior arise, which depend on both ULA as well as survey parameters. Oscillation-induced multi-modalities should be a generic problem of ULA probes from the large scale structure, if not observational effects smear out the differences in the posterior. Further investigation of this effect for different large scale structure probes would be desirable.

\section{Conclusion} \label{sec:conclusion}

In this analysis we forecast constraints from galaxy cluster number counts on ultra-light axion cosmologies for upcoming surveys. We have used AxionCAMB to calculate the matter power spectrum and the Tinker halo mass function to estimate the halo mass function including ULAs as dark matter. We have forecasted galaxy cluster number counts in mass and redshift bins using a standard approach, which involved modelling of scaling relation parameters and scatter in the observable. This way we could predict cluster number counts for different standard cosmological and ULA parameters. Our Monte Carlo analysis involves the eight-dimensional parameter space $\{\ma, \Omega_a / \Omega_{\rm dm}, \Omega_{\rm dm}, A_s, n_s, a_M, \alpha_M, \sigma_{\ln O^{\rm obs}|M}\}$. We chose various fiducial values for ULA parameters to investigate which region in the $\ma$ -- $\Omega_a / \Omega_{\rm dm}$ parameter space can be constrained given specific survey parameters. We saw a slight increase in constraining power including higher redshifts, the most important survey parameter to obtain competitive constraints however turned out to be the limiting mass of the surveys $M_{\rm min}$. Varying this parameter we obtained $95 \%$ exclusion limits for the ULA mass, typically in the region of $\ma=10^{-25}-10^{-24}\;\rm{eV}$. This of course makes sense, since the largest effects of axions on the power spectrum are on smaller scales or smaller masses.

Investigating the posterior for ULAs models which are detectable by a given survey, we find an anisotropic, curved shape. Its slope depends sensitively on the limiting mass, $M_{\rm min}$, of the survey and multimodalities from ULA induced oscillations in the matter power spectrum occur, which may or may not be smoothed out by observational effects for a specific probe.

A natural next step is to extended our analysis to more realistic cluster observables and to obtain constraints from existing observational campaigns such as the Dark Energy Survey \cite{DES} and Planck \cite{Planck2018}.

To obtain more accurate constraints for cosmological probes involving the halo mass function, calibration of the halo mass function with respect to ULA simulations might be necessary. First steps in this direction have been taken (e.g. \cite{axionsim}, \cite{May}) during the development of this work.
However, we do not expect that this is a major source of uncertainty for galaxy clusters as a cosmological probe. The dominant error, in particular for halo masses below $10^{13}h^{-1}\rm{M}_\odot$, will be the calibration of the mass observable relation and much focus will be given to this task in the future. Here the important quantity is the scatter of the mass-observable relation. The better the scatter is known the larger will be the ability to constrain cosmological parameters and also the axion quantities.

With Galaxy cluster counts alone, it is difficult to improve upon existing constraints from CMB observables or the Lyman-$\alpha$ forest. Cluster counts suffer - similarly to all other cosmological probes for ULAs - under an approximate degeneracy between ULA mass and abundancy. While this degeneracy line looks similar in all cases, its precise form does not only depend on the probe used but also on controllable survey parameters. In case of an ULA detection in the parameter range around $\ma \sim 10^{-22}$\eV, different probes could be combined to alleviate this degeneracy. Here cluster counts could provide an important contribution in particular if they can be extended to smaller halo masses.

\acknowledgments
We thank Steffen Hagstotz for allowing us to use his software developments. We also want to thank him, David Marsh and Simon May for valuable discussions during the course of this work.
Funded by the Deutsche Forschungsgemeinschaft (DFG, German Research Foundation) under Germany's Excellence Strategy -- EXC-2094 -- 390783311. \\

\include{bibliography.bib}


\bibliography{bibliography}
\bibliographystyle{JHEP}



\end{document}

%% file: mycommands.tex
\newcommand{\ma}{m_{\rm a}}

\newcommand{\Oma}{\Omega_{\rm a}}

\newcommand{\Nobs}{N_{\mathrm{obs}}}
\newcommand{\Nth}{N_{\mathrm{th}}}

\newcommand{\eV}{\ensuremath{\;\mathrm{eV}}}
\newcommand{\F}{F}

\newcommand{\prl}{Physical Review Letters}
\newcommand{\prd}{Physical Review D}
\newcommand{\nat}{Nature}
\newcommand{\apj}{Astrophysical Journal}
\newcommand{\apjl}{Astrophysical Journal Letters}
\newcommand{\aap}{Astronomy \& Astrophysics}
\newcommand{\apss}{Astrophysics and Space Science}
\newcommand{\mnras}{Monthly Notices of the Royal Astronomical Society}
\newcommand{\apjs}{Astrophysical Journal Supplement}
\newcommand{\jcap}{Journal for Cosmology and Astroparticle Physics}

%% file: bibliography.bib
@article{HGMF,
      author         = "Hlozek, Renée and Grin, Daniel and Marsh, David J. E. and Ferreira, Pedro G.",
      title          = "{A search for ultralight axions using precision cosmological data}",
      journal        = "Phys. Rev.",
      volume         = "D91",
      year           = "2015",
      number         = "10",
      pages          = "103512",
      doi            = "10.1103/PhysRevD.91.103512",
      eprint         = "1410.2896",
      archivePrefix  = "arXiv",
      primaryClass   = "astro-ph.CO",
      SLACcitation   = "
}

@ARTICLE{SFDM-CLASS,
       author = {{Ure{\~n}a-L{\'o}pez}, L. Arturo and {Gonzalez-Morales}, Alma X.},
        title = "{Towards accurate cosmological predictions for rapidly oscillating scalar fields as dark matter}",
      journal = {Journal of Cosmology and Astro-Particle Physics},
     keywords = {Astrophysics - Cosmology and Nongalactic Astrophysics, General Relativity and Quantum Cosmology},
         year = "2016",
        month = "Jul",
       volume = {2016},
          eid = {048},
        pages = {048},
          doi = {10.1088/1475-7516/2016/07/048},
        archivePrefix = {arXiv},
       eprint = {1511.08195},
        primaryClass = {astro-ph.CO},
       adsurl = {https://ui.adsabs.harvard.edu/abs/2016JCAP...07..048U},
      adsnote = {Provided by the SAO/NASA Astrophysics Data System}
}

@article{SFAxcompare,
	Author = {Ure{\~n}a-L{\'o}pez, L. Arturo},
	Doi = {10.3389/fspas.2019.00047},
	Issn = {2296-987X},
	Journal = {Frontiers in Astronomy and Space Sciences},
	Pages = {47},
	Title = {Brief Review on Scalar Field Dark Matter Models},
	Url = {https://www.frontiersin.org/article/10.3389/fspas.2019.00047},
	Volume = {6},
	Year = {2019},
	Bdsk-Url-1 = {https://www.frontiersin.org/article/10.3389/fspas.2019.00047},
	Bdsk-Url-2 = {https://doi.org/10.3389/fspas.2019.00047}
}

@article{CAMB,
      author         = "Lewis, Antony and Challinor, Anthony and Lasenby, Anthony",
      title          = "{Efficient computation of CMB anisotropies in closed FRW models}",
      journal        = "Astrophys. J.",
      volume         = "538",
      year           = "2000",
      pages          = "473-476",
      doi            = "10.1086/309179",
      eprint         = "astro-ph/9911177",
      archivePrefix  = "arXiv",
      primaryClass   = "astro-ph",
      SLACcitation   = "
}

@ARTICLE{CLASS,
       author = {{Blas}, Diego and {Lesgourgues}, Julien and {Tram}, Thomas},
        title = "{The Cosmic Linear Anisotropy Solving System (CLASS). Part II: Approximation schemes}",
      journal = {Journal of Cosmology and Astro-Particle Physics},
     keywords = {Astrophysics - Cosmology and Nongalactic Astrophysics},
         year = "2011",
        month = "Jul",
       volume = {2011},
          eid = {034},
        pages = {034},
          doi = {10.1088/1475-7516/2011/07/034},
archivePrefix = {arXiv},
       eprint = {1104.2933},
 primaryClass = {astro-ph.CO},
       adsurl = {https://ui.adsabs.harvard.edu/abs/2011JCAP...07..034B},
      adsnote = {Provided by the SAO/NASA Astrophysics Data System}
}

@article{Tinker:HMF,
      author         = "Tinker, Jeremy L. and others",
      title          = "{Toward a halo mass function for precision cosmology: The Limits of universality}",
      journal        = "Astrophys. J.",
      volume         = "688",
      year           = "2008",
      pages          = "709-728",
      doi            = "10.1086/591439",
      eprint         = "0803.2706",
      archivePrefix  = "arXiv",
      primaryClass   = "astro-ph",
      SLACcitation   = "
}

@ARTICLE{PressSchechter,
   author = {{Press}, W.~H. and {Schechter}, P.},
    title = "{Formation of Galaxies and Clusters of Galaxies by Self-Similar Gravitational Condensation}",
  journal = "Astrophys. J.",
     year = 1974,
    month = feb,
   volume = 187,
    pages = {425-438},
      doi = {10.1086/152650},
   adsurl = {http://adsabs.harvard.edu/abs/1974ApJ...187..425P},
  adsnote = {Provided by the SAO/NASA Astrophysics Data System}
}

@article{LeeShandarin,
	doi = {10.1086/305710},
	url = {https://doi.org/10.1086\%2F305710},
	year = 1998,
	month = {jun},
	publisher = {{IOP} Publishing},
	volume = {500},
	number = {1},
	pages = {14--27},
	author = {Jounghun Lee and Sergei F. Shandarin},
	title = {The Cosmological Mass Distribution Function in the Zeldovich Approximation},
	journal = {The Astrophysical Journal},
	abstract = {}
}
@ARTICLE{Bond:HMF,
   author = {{Bond}, J.~R. and {Cole}, S. and {Efstathiou}, G. and {Kaiser}, N.},
    title = "{Excursion set mass functions for hierarchical Gaussian fluctuations}",
  journal = "Astrophys. J.",
 keywords = {Computational Astrophysics, Gauss Equation, Dark Matter, Density Distribution, Many Body Problem, Mass Distribution, Monte Carlo Method},
     year = 1991,
    month = oct,
   volume = 379,
    pages = {440-460},
      doi = {10.1086/170520},
   adsurl = {http://adsabs.harvard.edu/abs/1991ApJ...379..440B},
  adsnote = {Provided by the SAO/NASA Astrophysics Data System}
}

@ARTICLE{Allen+2011,
       author = {{Allen}, Steven W. and {Evrard}, August E. and {Mantz}, Adam B.},
        title = "{Cosmological Parameters from Observations of Galaxy Clusters}",
      journal = {Annual Review of Astronomy \& Astrophysics},
     keywords = {Astrophysics - Cosmology and Extragalactic Astrophysics},
         year = "2011",
        month = "Sep",
       volume = {49},
       number = {1},
        pages = {409-470},
          doi = {10.1146/annurev-astro-081710-102514},
archivePrefix = {arXiv},
       eprint = {1103.4829},
 primaryClass = {astro-ph.CO},
       adsurl = {https://ui.adsabs.harvard.edu/abs/2011ARA&A..49..409A},
      adsnote = {Provided by the SAO/NASA Astrophysics Data System}
}

@article{Euclid2016,
      author         = "Sartoris, B. and others",
      title          = "{Next Generation Cosmology: Constraints from the Euclid Galaxy Cluster Survey}",
      journal        = "Mon. Not. Roy. Astron. Soc.",
      volume         = "459",
      year           = "2016",
      number         = "2",
      pages          = "1764-1780",
      doi            = "10.1093/mnras/stw630",
      eprint         = "1505.02165",
      archivePrefix  = "arXiv",
      primaryClass   = "astro-ph.CO",
      SLACcitation   = "
}

@article{Hagstotz2018,
      author         = "Hagstotz, Steffen and Costanzi, Matteo and Baldi, Marco and Weller, Jochen",
      title          = "{Joint halo-mass function for modified gravity and massive neutrinos – I. Simulations and cosmological forecasts}",
      journal        = "Mon. Not. Roy. Astron. Soc.",
      volume         = "486",
      year           = "2019",
      number         = "3",
      pages          = "3927-3941",
      doi            = "10.1093/mnras/stz1051",
      eprint         = "1806.07400",
      archivePrefix  = "arXiv",
      primaryClass   = "astro-ph.CO",
      SLACcitation   = "
}

@ARTICLE{emcee,
       author = {{Foreman-Mackey}, Daniel and {Hogg}, David W. and {Lang}, Dustin and {Goodman}, Jonathan},
        title = "{emcee: The MCMC Hammer}",
      journal = {Pub. of the Astro. Soc. of the Pacific},
     keywords = {Astrophysics - Instrumentation and Methods for Astrophysics, Physics - Computational Physics, Statistics - Computation},
         year = "2013",
        month = "Mar",
       volume = {125},
       number = {925},
        pages = {306},
          doi = {10.1086/670067},
archivePrefix = {arXiv},
       eprint = {1202.3665},
 primaryClass = {astro-ph.IM},
       adsurl = {https://ui.adsabs.harvard.edu/abs/2013PASP..125..306F},
      adsnote = {Provided by the SAO/NASA Astrophysics Data System}
}

@article{Planck2018,
      author         = "Aghanim, N. and others",
      title          = "{Planck 2018 results. VI. Cosmological parameters}",
      collaboration  = "Planck",
      year           = "2018",
      note         = "astro-ph/1807.06209",
      archivePrefix  = "arXiv",
      eid = {arXiv:1807.06209},
      journal = {arXiv e-prints},
      primaryClass   = "astro-ph.CO",
      SLACcitation   = "
}

@article{Gelman1996,
 ISSN = {10170405, 19968507},
 URL = {http://www.jstor.org/stable/24306036},
 author = {Andrew Gelman and Xiao-Li Meng and Hal Stern},
 journal = {Statistica Sinica},
 number = {4},
 pages = {733--760},
 publisher = {Institute of Statistical Science, Academia Sinica},
 title = {POSTERIOR PREDICTIVE ASSESSMENT OF MODEL FITNESS VIA REALIZED DISCREPANCIES},
 volume = {6},
 year = {1996}
}

@article{Bigdudes101,
      author         = "Moore, B. and Ghigna, S. and Governato, F. and Lake, G. and Quinn, Thomas R. and Stadel, J. and Tozzi, P.",
      title          = "{Dark matter substructure within galactic halos}",
      journal        = "Astrophys. J.",
      volume         = "524",
      year           = "1999",
      pages          = "L19-L22",
      doi            = "10.1086/312287",
      eprint         = "astro-ph/9907411",
      archivePrefix  = "arXiv",
      primaryClass   = "astro-ph",
      SLACcitation   = "
}

@article{Bigdudes2,
      author         = "Dubinski, John and Carlberg, R. G.",
      title          = "{The Structure of cold dark matter halos}",
      journal        = "Astrophys. J.",
      volume         = "378",
      year           = "1991",
      pages          = "496",
      doi            = "10.1086/170451",
      SLACcitation   = "
}

@ARTICLE{1103.0007,
       author = {{Boylan-Kolchin}, Michael and {Bullock}, James S. and
         {Kaplinghat}, Manoj},
        title = "{Too big to fail? The puzzling darkness of massive Milky Way subhaloes}",
      journal = {Mon. Not. Roy. Astron. Soc.},
     keywords = {Galaxy: halo, galaxies: abundances, cosmology: theory, dark matter, Astrophysics - Cosmology and Extragalactic Astrophysics, Astrophysics - Galaxy Astrophysics},
         year = "2011",
        month = "Jul",
       volume = {415},
       number = {1},
        pages = {L40-L44},
          doi = {10.1111/j.1745-3933.2011.01074.x},
archivePrefix = {arXiv},
       eprint = {1103.0007},
 primaryClass = {astro-ph.CO},
       adsurl = {https://ui.adsabs.harvard.edu/abs/2011MNRAS.415L..40B},
      adsnote = {Provided by the SAO/NASA Astrophysics Data System}
}

@article{NFW,
      author         = "Navarro, Julio F. and Frenk, Carlos S. and White, Simon
                        D. M.",
      title          = "{A Universal density profile from hierarchical
                        clustering}",
      journal        = "Astrophys. J.",
      volume         = "490",
      year           = "1997",
      pages          = "493-508",
      doi            = "10.1086/304888",
      eprint         = "astro-ph/9611107",
      archivePrefix  = "arXiv",
      primaryClass   = "astro-ph",
      SLACcitation   = "
}

@article{Emi9,
	Author = {Moore, Ben},
	Da = {1994/08/01},
	Date-Added = {2019-08-15 08:41:36 +0000},
	Date-Modified = {2019-08-15 08:41:36 +0000},
	Doi = {10.1038/370629a0},
	Id = {Moore1994},
	Isbn = {1476-4687},
	Journal = {Nature},
	Number = {6491},
	Pages = {629--631},
	Title = {Evidence against dissipation-less dark matter from observations of galaxy haloes},
	Ty = {JOUR},
	Url = {https://doi.org/10.1038/370629a0},
	Volume = {370},
	Year = {1994},
	Bdsk-Url-1 = {https://doi.org/10.1038/370629a0}
}

@article{wiggleZ,
    author = {Blake, Chris and others},
    title = "{The WiggleZ Dark Energy Survey: the selection function and z=0.6 galaxy power spectrum}",
    journal = {Mon. Not. Roy. Astron. Soc.},
    volume = {406},
    number = {2},
    pages = {803-821},
    year = {2010},
    month = {07},
    issn = {0035-8711},
    doi = {10.1111/j.1365-2966.2010.16747.x},
    url = {https://doi.org/10.1111/j.1365-2966.2010.16747.x},
    eprint = {http://oup.prod.sis.lan/mnras/article-pdf/406/2/803/18720013/mnr0406-0803.pdf},
}

@article{Bozek2014,
      author         = "Bozek, Brandon and Marsh, David J. E. and Silk, Joseph
                        and Wyse, Rosemary F. G.",
      title          = "{Galaxy UV-luminosity function and reionization
                        constraints on axion dark matter}",
      journal        = "Mon. Not. Roy. Astron. Soc.",
      volume         = "450",
      year           = "2015",
      number         = "1",
      pages          = "209-222",
      doi            = "10.1093/mnras/stv624",
      eprint         = "1409.3544",
      archivePrefix  = "arXiv",
      primaryClass   = "astro-ph.CO",
      SLACcitation   = "
}

@article{Lymana,
      author         = "Kobayashi, T. and Murgia, R. and De Simone,
                        A. and Iršič, V. and Viel, M.",
      title          = "{Lyman-$\alpha$ constraints on ultralight scalar dark
                        matter: Implications for the early and late universe}",
      journal        = "Phys. Rev.",
      volume         = "D96",
      year           = "2017",
      number         = "12",
      pages          = "123514",
      doi            = "10.1103/PhysRevD.96.123514",
      eprint         = "1708.00015",
      archivePrefix  = "arXiv",
      primaryClass   = "astro-ph.CO",
      reportNumber   = "SISSA-33-2017-FISI",
      SLACcitation   = "
}

@article{BigDudes,
      author         = "Hui, Lam and Ostriker, Jeremiah P. and Tremaine, Scott
                        and Witten, Edward",
      title          = "{Ultralight scalars as cosmological dark matter}",
      journal        = "Phys. Rev.",
      volume         = "D95",
      year           = "2017",
      number         = "4",
      pages          = "043541",
      doi            = "10.1103/PhysRevD.95.043541",
      eprint         = "1610.08297",
      archivePrefix  = "arXiv",
      primaryClass   = "astro-ph.CO",
      SLACcitation   = "
}

@article{WHITE,
      author         = "Grin, Daniel and others",
      title          = "{Gravitational probes of ultra-light axions}",
      year           = "2019",
      journal = {arXiv e-prints},
      pages = {arXiv:1904.09003},
      eprint         = "1904.09003",
      archivePrefix  = "arXiv",
      primaryClass   = "astro-ph.CO"
}

@ARTICLE{JeansAnalysis,
       author = {{Hayashi}, Kohei and {Obata}, Ippei},
        title = "{Non-sphericity of ultra-light axion dark matter halos in the Galactic dwarf spheroidal galaxies}",
      journal = {arXiv e-prints},
     keywords = {Astrophysics - Cosmology and Nongalactic Astrophysics, Astrophysics - Astrophysics of Galaxies, High Energy Physics - Phenomenology},
         year = "2019",
        month = "Feb",
          eid = {arXiv:1902.03054},
        pages = {arXiv:1902.03054},
archivePrefix = {arXiv},
       eprint = {1902.03054},
 primaryClass = {astro-ph.CO},
       adsurl = {https://ui.adsabs.harvard.edu/abs/2019arXiv190203054H},
      adsnote = {Provided by the SAO/NASA Astrophysics Data System}
}

@ARTICLE{MarshNiemeyer,
       author = {{Marsh}, David J.~E. and {Niemeyer}, Jens C.},
        title = "{Strong Constraints on Fuzzy Dark Matter from Ultrafaint Dwarf Galaxy Eridanus II}",
      journal = {arXiv e-prints},
     keywords = {Astrophysics - Cosmology and Nongalactic Astrophysics, Astrophysics - Astrophysics of Galaxies, High Energy Physics - Phenomenology},
         year = "2018",
        month = "Oct",
          eid = {arXiv:1810.08543},
        pages = {arXiv:1810.08543},
archivePrefix = {arXiv},
       eprint = {1810.08543},
 primaryClass = {astro-ph.CO},
       adsurl = {https://ui.adsabs.harvard.edu/abs/2018arXiv181008543M},
      adsnote = {Provided by the SAO/NASA Astrophysics Data System}
}

@article{SKA,
	doi = {10.1088/0264-9381/30/22/224011},
	url = {https://doi.org/10.1088\%2F0264-9381\%2F30\%2F22\%2F224011},
	year = 2013,
	month = {nov},
	publisher = {{IOP} Publishing},
	volume = {30},
	number = {22},
	pages = {224011},
	author = {T J W Lazio},
	title = {The Square Kilometre Array pulsar timing array},
	journal = {Classical and Quantum Gravity},
}

@article{Marsh:SMBH,
      author         = "Stott, Matthew J. and Marsh, David J. E.",
      title          = "{Black hole spin constraints on the mass spectrum and
                        number of axionlike fields}",
      journal        = "Phys. Rev.",
      volume         = "D98",
      year           = "2018",
      number         = "8",
      pages          = "083006",
      doi            = "10.1103/PhysRevD.98.083006",
      eprint         = "1805.02016",
      archivePrefix  = "arXiv",
      primaryClass   = "hep-ph",
      reportNumber   = "KCL-PH-TH/2018-17, KCL-PH-TH-2018-17",
      SLACcitation   = "
}

@article{Superradiance,
      author         = "Brito, Richard and Cardoso, Vitor and Pani, Paolo",
      title          = "{Superradiance}",
      journal        = "Lect. Notes Phys.",
      volume         = "906",
      year           = "2015",
      pages          = "pp.1-237",
      doi            = "10.1007/978-3-319-19000-6",
      eprint         = "1501.06570",
      archivePrefix  = "arXiv",
      primaryClass   = "gr-qc",
      SLACcitation   = "
}

@article{obsmass,
      author         = "Lima, Marcos and Hu, Wayne",
      title          = "{Self-calibration of cluster dark energy studies:
                        Observable-mass distribution}",
      journal        = "Phys. Rev.",
      volume         = "D72",
      year           = "2005",
      pages          = "043006",
      doi            = "10.1103/PhysRevD.72.043006",
      eprint         = "astro-ph/0503363",
      archivePrefix  = "arXiv",
      primaryClass   = "astro-ph",
      SLACcitation   = "
}

@article{getDist,
 author         = "Lewis, Antony",
 title          = "{GetDist: a Python package for analysing Monte Carlo
                   samples}",
 year           = "2019",
 eprint         = "1910.13970",
 archivePrefix  = "arXiv",
 primaryClass   = "astro-ph.IM",
 SLACcitation   = "
 url            = "https://getdist.readthedocs.io"
}

@article{JC-paper,
    author = "Cookmeyer, Jonathan and Grin, Daniel and Smith, Tristan L.",
    title = "{How sound are our ultralight axion approximations?}",
    eprint = "1909.11094",
    archivePrefix = "arXiv",
    primaryClass = "astro-ph.CO",
    doi = "10.1103/PhysRevD.101.023501",
    journal = "Phys. Rev. D",
    volume = "101",
    number = "2",
    pages = "023501",
    year = "2020"
}

@article{baryons1,
    author = {Rees, M. J. and Ostriker, J. P.},
    title = "{Cooling, dynamics and fragmentation of massive gas clouds: clues to the masses and radii of galaxies and clusters}",
    journal = {Monthly Notices of the Royal Astronomical Society},
    volume = {179},
    number = {4},
    pages = {541-559},
    year = {1977},
    month = {08},
    issn = {0035-8711},
    doi = {10.1093/mnras/179.4.541},
    url = {https://doi.org/10.1093/mnras/179.4.541},
    eprint = {https://academic.oup.com/mnras/article-pdf/179/4/541/3312267/mnras179-0541.pdf},
}

@article{baryons2,
  title={Suppressing the formation of dwarf galaxies via photoionization},
  author={Efstathiou, George},
  journal={Monthly Notices of the Royal Astronomical Society},
  volume={256},
  number={1},
  pages={43P--47P},
  year={1992},
  publisher={The Royal Astronomical Society}
}

@article{baryons3,
  title={The cores of dwarf galaxy haloes},
  author={Navarro, Julio F and Eke, Vincent R and Frenk, Carlos S},
  journal={Monthly Notices of the Royal Astronomical Society},
  volume={283},
  number={3},
  pages={L72--L78},
  year={1996},
  publisher={Blackwell Science Ltd Oxford, UK}
}

@article{baryons4,
  title={Reionization and the abundance of galactic satellites},
  author={Bullock, James S and Kravtsov, Andrey V and Weinberg, David H},
  journal={The Astrophysical Journal},
  volume={539},
  number={2},
  pages={517},
  year={2000},
  publisher={IOP Publishing}
}

@article{baryons5,
  title={Maximum feedback and dark matter profiles of dwarf galaxies},
  author={Gnedin, Oleg Y and Zhao, HongSheng},
  journal={Monthly Notices of the Royal Astronomical Society},
  volume={333},
  number={2},
  pages={299--306},
  year={2002},
  publisher={The Royal Astronomical Society}
}

@article{baryons6,
  title={Dark halos: the flattening of the density cusp by dynamical friction},
  author={El-Zant, Amr and Shlosman, Isaac and Hoffman, Yehuda},
  journal={The Astrophysical Journal},
  volume={560},
  number={2},
  pages={636},
  year={2001},
  publisher={IOP Publishing}
}

@article{baryons7,
  title={Are halos of collisionless cold dark matter collisionless?},
  author={Ma, Chung-Pei and Boylan-Kolchin, Michael},
  journal={Physical review letters},
  volume={93},
  number={2},
  pages={021301},
  year={2004},
  publisher={APS}
}

@article{baryons8,
  title={Weakening dark matter cusps by clumpy baryonic infall},
  author={Cole, David R and Dehnen, Walter and Wilkinson, Mark I},
  journal={Monthly Notices of the Royal Astronomical Society},
  volume={416},
  number={2},
  pages={1118--1134},
  year={2011},
  publisher={The Royal Astronomical Society}
}

@article{baryons9,
  title={Cores and revived cusps of dark matter haloes in disc galaxy formation through clump clusters},
  author={Inoue, Shigeki and Saitoh, Takayuki R},
  journal={Monthly Notices of the Royal Astronomical Society},
  volume={418},
  number={4},
  pages={2527--2531},
  year={2011},
  publisher={The Royal Astronomical Society}
}

@article{baryons10,
  title={Cuspy no more: how outflows affect the central dark matter and baryon distribution in $\Lambda$ cold dark matter galaxies},
  author={Governato, F and Zolotov, A and Pontzen, A and Christensen, C and Oh, Se-Heon and Brooks, AM and Quinn, T and Shen, S and Wadsley, J},
  journal={Monthly Notices of the Royal Astronomical Society},
  volume={422},
  number={2},
  pages={1231--1240},
  year={2012},
  publisher={Blackwell Publishing Ltd Oxford, UK}
}

@article{baryons11,
  title={Baryons matter: Why luminous satellite galaxies have reduced central masses},
  author={Zolotov, Adi and Brooks, Alyson M and Willman, Beth and Governato, Fabio and Pontzen, Andrew and Christensen, Charlotte and Dekel, Avishai and Quinn, Tom and Shen, Sijing and Wadsley, James},
  journal={The Astrophysical Journal},
  volume={761},
  number={1},
  pages={71},
  year={2012},
  publisher={IOP Publishing}
}

@article{baryons12,
  title={The dependence of dark matter profiles on the stellar-to-halo mass ratio: a prediction for cusps versus cores},
  author={Di Cintio, Arianna and Brook, Chris B and Macci{\`o}, Andrea V and Stinson, Greg S and Knebe, Alexander and Dutton, Aaron A and Wadsley, James},
  journal={Monthly Notices of the Royal Astronomical Society},
  volume={437},
  number={1},
  pages={415--423},
  year={2014},
  publisher={The Royal Astronomical Society}
}

@article{baryons13,
  title={Why baryons matter: the kinematics of dwarf spheroidal satellites},
  author={Brooks, Alyson M and Zolotov, Adi},
  journal={The Astrophysical Journal},
  volume={786},
  number={2},
  pages={87},
  year={2014},
  publisher={IOP Publishing}
}

@article{baryons14,
  title={A unified solution to the small scale problems of the $\Lambda$CDM model},
  author={Del Popolo, A and Lima, Jos{\'e} Ademir Sales de and Fabris, J{\'u}lio C and Rodrigues, Davi C},
  journal={Journal of Cosmology and Astroparticle Physics},
  volume={2014},
  number={04},
  pages={021},
  year={2014},
  publisher={IOP Publishing}
}

@article{baryons15,
  title={A unified solution to the small scale problems of the $\Lambda$CDM model II: introducing parent-satellite interaction},
  author={Del Popolo, Antonino and Le Delliou, Morgan},
  journal={Journal of Cosmology and Astroparticle Physics},
  volume={2014},
  number={12},
  pages={051},
  year={2014},
  publisher={IOP Publishing}
}

@article{baryons16,
  title={Early flattening of dark matter cusps in dwarf spheroidal galaxies},
  author={Nipoti, Carlo and Binney, James},
  journal={Monthly Notices of the Royal Astronomical Society},
  volume={446},
  number={2},
  pages={1820--1828},
  year={2015},
  publisher={Oxford University Press}
}

@article{baryons17,
  title={The APOSTLE simulations: solutions to the Local Group's cosmic puzzles},
  author={Sawala, Till and Frenk, Carlos S and Fattahi, Azadeh and Navarro, Julio F and Bower, Richard G and Crain, Robert A and Vecchia, Claudio Dalla and Furlong, Michelle and Helly, John C and Jenkins, Adrian and others},
  journal={Monthly Notices of the Royal Astronomical Society},
  volume={457},
  number={2},
  pages={1931--1943},
  year={2016},
  publisher={Oxford University Press}
}

@article{baryons18,
  title={Energy transfer from baryons to dark matter as a unified solution to small-scale structure issues of the $\Lambda$ CDM model},
  author={Del Popolo, Antonino and Pace, Francesco and Le Delliou, Morgan and Lee, Xiguo},
  journal={Physical Review D},
  volume={98},
  number={6},
  pages={063517},
  year={2018},
  publisher={APS}
}

@article{FDM,
  title={Fuzzy cold dark matter: the wave properties of ultralight particles},
  author={Hu, Wayne and Barkana, Rennan and Gruzinov, Andrei},
  journal={Physical Review Letters},
  volume={85},
  number={6},
  pages={1158},
  year={2000},
  publisher={APS}
}

@article{axiverse,
  title={String axiverse},
  author={Arvanitaki, Asimina and Dimopoulos, Savas and Dubovsky, Sergei and Kaloper, Nemanja and March-Russell, John},
  journal={Physical Review D},
  volume={81},
  number={12},
  pages={123530},
  year={2010},
  publisher={APS}
}

@article{string1,
  title={Some properties of O (32) superstrings},
  author={Witten, Edward},
  journal={Physics Letters B},
  volume={149},
  number={4-5},
  pages={351--356},
  year={1984},
  publisher={Elsevier}
}

@article{string2,
  title={The QCD axion and moduli stabilisation},
  author={Conlon, Joseph P},
  journal={Journal of High Energy Physics},
  volume={2006},
  number={05},
  pages={078},
  year={2006},
  publisher={IOP Publishing}
}

@article{string3,
  title={Axions in string theory},
  author={Svrcek, Peter and Witten, Edward},
  journal={Journal of High Energy Physics},
  volume={2006},
  number={06},
  pages={051},
  year={2006},
  publisher={IOP Publishing}
}

@article{lymana2,
  title={Constraining the mass of light bosonic dark matter using SDSS Lyman-$\alpha$ forest},
  author={Armengaud, Eric and Palanque-Delabrouille, Nathalie and Y{\`e}che, Christophe and Marsh, David JE and Baur, Julien},
  journal={Monthly Notices of the Royal Astronomical Society},
  volume={471},
  number={4},
  pages={4606--4614},
  year={2017},
  publisher={Oxford University Press}
}

@article{lymana3,
  title={First constraints on fuzzy dark matter from Lyman-$\alpha$ forest data and hydrodynamical simulations},
  author={Ir{\v{s}}i{\v{c}}, Vid and Viel, Matteo and Haehnelt, Martin G and Bolton, James S and Becker, George D},
  journal={Physical review letters},
  volume={119},
  number={3},
  pages={031302},
  year={2017},
  publisher={APS}
}

@article{lymana4,
  title={Lyman $\alpha$ forest and non-linear structure characterization in Fuzzy Dark Matter cosmologies},
  author={Nori, Matteo and Murgia, Riccardo and Ir{\v{s}}i{\v{c}}, Vid and Baldi, Marco and Viel, Matteo},
  journal={Monthly Notices of the Royal Astronomical Society},
  volume={482},
  number={3},
  pages={3227--3243},
  year={2019},
  publisher={Oxford University Press}
}

@article{lymana5,
  title={Testing extreme-axion wave-like dark matter using the BOSS Lyman-alpha forest data},
  author={Leong, Ka-Hou and Schive, Hsi-Yu and Zhang, Ui-Han and Chiueh, Tzihong},
  journal={Monthly Notices of the Royal Astronomical Society},
  volume={484},
  number={3},
  pages={4273--4286},
  year={2019},
  publisher={Oxford University Press}
}

@article{klypin1999missing,
  title={Where are the missing galactic satellites?},
  author={Klypin, Anatoly and Kravtsov, Andrey V and Valenzuela, Octavio and Prada, Francisco},
  journal={The Astrophysical Journal},
  volume={522},
  number={1},
  pages={82},
  year={1999},
  publisher={IOP Publishing}
}

@article{toobigtofail2,
  title={The Milky Way’s bright satellites as an apparent failure of $\Lambda$CDM},
  author={Boylan-Kolchin, Michael and Bullock, James S and Kaplinghat, Manoj},
  journal={Monthly Notices of the Royal Astronomical Society},
  volume={422},
  number={2},
  pages={1203--1218},
  year={2012},
  publisher={Blackwell Publishing Ltd Oxford, UK}
}

@article{toobigtofail3,
  title={M31 satellite masses compared to $\Lambda$CDM subhaloes},
  author={Tollerud, Erik J and Boylan-Kolchin, Michael and Bullock, James S},
  journal={Monthly Notices of the Royal Astronomical Society},
  volume={440},
  number={4},
  pages={3511--3519},
  year={2014},
  publisher={Oxford University Press}
}

@article{toobigtofail4,
  title={Too big to fail in the local group},
  author={Garrison-Kimmel, Shea and Boylan-Kolchin, Michael and Bullock, James S and Kirby, Evan N},
  journal={Monthly Notices of the Royal Astronomical Society},
  volume={444},
  number={1},
  pages={222--236},
  year={2014},
  publisher={The Royal Astronomical Society}
}

@article{cuspcore3,
  title={The dark matter distribution in disc galaxies},
  author={Borriello, Annamaria and Salucci, Paolo},
  journal={Monthly Notices of the Royal Astronomical Society},
  volume={323},
  number={2},
  pages={285--292},
  year={2001},
  publisher={Blackwell Science Ltd Oxford, UK}
}

@article{cuspcore4,
  title={The observed properties of dark matter on small spatial scales},
  author={Gilmore, Gerard and Wilkinson, Mark I and Wyse, Rosemary FG and Kleyna, Jan T and Koch, Andreas and Evans, N Wyn and Grebel, Eva K},
  journal={The Astrophysical Journal},
  volume={663},
  number={2},
  pages={948},
  year={2007},
  publisher={IOP Publishing}
}

@article{cuspcore5,
  title={High-resolution dark matter density profiles of THINGS dwarf galaxies: correcting for noncircular motions},
  author={Oh, Se-Heon and De Blok, WJG and Walter, Fabian and Brinks, Elias and Kennicutt Jr, Robert C},
  journal={The Astronomical Journal},
  volume={136},
  number={6},
  pages={2761},
  year={2008},
  publisher={IOP Publishing}
}

@article{hwang2009axion,
  title={Axion as a cold dark matter candidate},
  author={Hwang, Jai-chan and Noh, Hyerim},
  journal={Physics Letters B},
  volume={680},
  number={1},
  pages={1--3},
  year={2009},
  publisher={Elsevier}
}

@article{poulin2018cosmological,
  title={Cosmological implications of ultralight axionlike fields},
  author={Poulin, Vivian and Smith, Tristan L and Grin, Daniel and Karwal, Tanvi and Kamionkowski, Marc},
  journal={Physical Review D},
  volume={98},
  number={8},
  pages={083525},
  year={2018},
  publisher={APS}
}

@article{UVlum1,
  title={Galaxy UV-luminosity function and reionization constraints on axion dark matter},
  author={Bozek, Brandon and Marsh, David JE and Silk, Joseph and Wyse, Rosemary FG},
  journal={Monthly Notices of the Royal Astronomical Society},
  volume={450},
  number={1},
  pages={209--222},
  year={2015},
  publisher={The Royal Astronomical Society}
}

@article{UVlum2,
  title={Contrasting galaxy formation from quantum wave dark matter, $\psi$DM, with $\Lambda$CDM, using Planck and Hubble data},
  author={Schive, Hsi-Yu and Chiueh, Tzihong and Broadhurst, Tom and Huang, Kuan-Wei},
  journal={The Astrophysical Journal},
  volume={818},
  number={1},
  pages={89},
  year={2016},
  publisher={IOP Publishing}
}

@article{UVlum3,
  title={Constraints on dark matter scenarios from measurements of the galaxy luminosity function at high redshifts},
  author={Corasaniti, PS and Agarwal, S and Marsh, DJE and Das, Subinoy},
  journal={Physical Review D},
  volume={95},
  number={8},
  pages={083512},
  year={2017},
  publisher={APS}
}

@article{garzilli2017cutoff,
  title={Cutoff in the Lyman-$\alpha$ forest power spectrum: Warm IGM or warm dark matter?},
  author={Garzilli, Antonella and Boyarsky, Alexey and Ruchayskiy, Oleg},
  journal={Physics Letters B},
  volume={773},
  pages={258--264},
  year={2017},
  publisher={Elsevier}
}

@article{BHSR,
  title={Exploring the string axiverse with precision black hole physics},
  author={Arvanitaki, Asimina and Dubovsky, Sergei},
  journal={Physical Review D},
  volume={83},
  number={4},
  pages={044026},
  year={2011},
  publisher={APS}
}

@article{marshpop,
  title={Axion dark matter, solitons and the cusp--core problem},
  author={Marsh, David JE and Pop, Ana-Roxana},
  journal={Monthly Notices of the Royal Astronomical Society},
  volume={451},
  number={3},
  pages={2479--2492},
  year={2015},
  publisher={Oxford University Press}
}

@article{marshsilk,
  title={A model for halo formation with axion mixed dark matter},
  author={Marsh, David JE and Silk, Joseph},
  journal={Monthly Notices of the Royal Astronomical Society},
  volume={437},
  number={3},
  pages={2652--2663},
  year={2014},
  publisher={The Royal Astronomical Society}
}

@article{schive2014,
  title={Cosmic structure as the quantum interference of a coherent dark wave},
  author={Schive, Hsi-Yu and Chiueh, Tzihong and Broadhurst, Tom},
  journal={Nature Physics},
  volume={10},
  number={7},
  pages={496--499},
  year={2014},
  publisher={Nature Publishing Group}
}

@article{bernal2018,
  title={Rotation curves of high-resolution LSB and SPARC galaxies with fuzzy and multistate (ultralight boson) scalar field dark matter},
  author={Bernal, Tula and Fern{\'a}ndez-Hern{\'a}ndez, Lizbeth M and Matos, Tonatiuh and Rodr{\'\i}guez-Meza, Mario A},
  journal={Monthly Notices of the Royal Astronomical Society},
  volume={475},
  number={2},
  pages={1447--1468},
  year={2018},
  publisher={Oxford University Press}
}

@article{deng2018,
  title={Can light dark matter solve the core-cusp problem?},
  author={Deng, Heling and Hertzberg, Mark P and Namjoo, Mohammad Hossein and Masoumi, Ali},
  journal={Physical Review D},
  volume={98},
  number={2},
  pages={023513},
  year={2018},
  publisher={APS}
}

@article{broadhurst2020,
  title={Ghostly galaxies as solitons of Bose-Einstein dark matter},
  author={Broadhurst, Tom and De Martino, Ivan and Luu, Hoang Nhan and Smoot, George F and Tye, S-H Henry},
  journal={Physical Review D},
  volume={101},
  number={8},
  pages={083012},
  year={2020},
  publisher={APS}
}

@ARTICLE{mana13,
       author = {{Mana}, Annalisa and {Giannantonio}, Tommaso and {Weller}, Jochen and
         {Hoyle}, Ben and {H{\"u}tsi}, Gert and {Sartoris}, Barbara},
        title = "{Combining clustering and abundances of galaxy clusters to test cosmology and primordial non-Gaussianity}",
      journal = {Monthly Notices of the Royal Astronomical Society},
     keywords = {methods: statistical, galaxies: clusters: general, cosmological parameters, Astrophysics - Cosmology and Extragalactic Astrophysics},
         year = 2013,
        month = sep,
       volume = {434},
       number = {1},
        pages = {684-695},
          doi = {10.1093/mnras/stt1062},
archivePrefix = {arXiv},
       eprint = {1303.0287},
 primaryClass = {astro-ph.CO},
       adsurl = {https://ui.adsabs.harvard.edu/abs/2013MNRAS.434..684M},
      adsnote = {Provided by the SAO/NASA Astrophysics Data System}
}

@ARTICLE{rozo14,
       author = {{Rozo}, E. and {Bartlett}, J.~G. and {Evrard}, A.~E. and {Rykoff}, E.~S.},
        title = "{Closing the loop: a self-consistent model of optical, X-ray and Sunyaev-Zel'dovich scaling relations for clusters of Galaxies}",
      journal = {Monthly Notices of the Royal Astronomical Society},
     keywords = {galaxies: clusters: general, Astrophysics - Cosmology and Nongalactic Astrophysics},
         year = 2014,
        month = feb,
       volume = {438},
       number = {1},
        pages = {78-96},
          doi = {10.1093/mnras/stt2161},
archivePrefix = {arXiv},
       eprint = {1204.6305},
 primaryClass = {astro-ph.CO},
       adsurl = {https://ui.adsabs.harvard.edu/abs/2014MNRAS.438...78R},
      adsnote = {Provided by the SAO/NASA Astrophysics Data System}
}

@article{scalrel2,
    author = "Johnston, David E. and Sheldon, Erin S. and Wechsler, Risa H. and Rozo, Eduardo and Koester, Benjamin P. and Frieman, Joshua A. and McKay, Timothy A. and Evrard, August E. and Becker, Matthew R. and Annis, James",
    collaboration = "SDSS",
    title = "{Cross-correlation Weak Lensing of SDSS galaxy Clusters II: Cluster Density Profiles and the Mass--Richness Relation}",
    eprint = "0709.1159",
    archivePrefix = "arXiv",
    primaryClass = "astro-ph",
    reportNumber = "FERMILAB-PUB-07-720-A, SLAC-PUB-12813",
    month = "9",
    year = "2007"
}

@article{hlozek2018,
  title={Using the full power of the cosmic microwave background to probe axion dark matter},
  author={Hlo{\v{z}}ek, Ren{\'e}e and Marsh, David JE and Grin, Daniel},
  journal={Monthly Notices of the Royal Astronomical Society},
  volume={476},
  number={3},
  pages={3063--3085},
  year={2018},
  publisher={Oxford University Press}
}

@article{du2016,
    author = {Du, Xiaolong and Behrens, Christoph and Niemeyer, Jens C.},
    title = "{Substructure of fuzzy dark matter haloes}",
    journal = {Monthly Notices of the Royal Astronomical Society},
    volume = {465},
    number = {1},
    pages = {941-951},
    year = {2016},
    month = {10},
    issn = {0035-8711},
    doi = {10.1093/mnras/stw2724},
    url = {https://doi.org/10.1093/mnras/stw2724},
    eprint = {https://academic.oup.com/mnras/article-pdf/465/1/941/8593407/stw2724.pdf},
}

@article{KGeq,
  title={Early time perturbations behavior in scalar field cosmologies},
  author={Perrotta, Francesca and Baccigalupi, Carlo},
  journal={Physical Review D},
  volume={59},
  number={12},
  pages={123508},
  year={1999},
  publisher={APS}
}

@article{DES,
      author         = "Abbott, T. M. C. and others",
      title          = "{The Dark Energy Survey Data Release 1}",
      collaboration  = "DES, NOAO Data Lab",
      journal        = "Astrophys. J. Suppl.",
      volume         = "239",
      year           = "2018",
      number         = "2",
      pages          = "18",
      doi            = "10.3847/1538-4365/aae9f0",
      eprint         = "1801.03181",
      archivePrefix  = "arXiv",
      primaryClass   = "astro-ph.IM",
      reportNumber   = "FERMILAB-PUB-17-603-AE-E",
      SLACcitation   = "
}

@article{ShethTormen,
    author = "Sheth, Ravi K. and Tormen, Giuseppe",
    title = "{An Excursion Set Model of Hierarchical Clustering : Ellipsoidal Collapse and the Moving Barrier}",
    eprint = "astro-ph/0105113",
    archivePrefix = "arXiv",
    reportNumber = "FERMILAB-PUB-01-061-A",
    doi = "10.1046/j.1365-8711.2002.04950.x",
    journal = "Mon. Not. Roy. Astron. Soc.",
    volume = "329",
    pages = "61",
    year = "2002"
}

@article{Springel,
	Author = {Volker Springel and Naoki Yoshida and Simon D.M. White},
	Doi = {https://doi.org/10.1016/S1384-1076(01)00042-2},
	Issn = {1384-1076},
	Journal = {New Astronomy},
	Keywords = {Methods: numerical, Galaxies: interactions, Dark matter},
	Number = {2},
	Pages = {79 - 117},
	Title = {GADGET: a code for collisionless and gasdynamical cosmological simulations},
	Url = {http://www.sciencedirect.com/science/article/pii/S1384107601000422},
	Volume = {6},
	Year = {2001},
	Bdsk-Url-1 = {http://www.sciencedirect.com/science/article/pii/S1384107601000422},
	Bdsk-Url-2 = {https://doi.org/10.1016/S1384-1076(01)00042-2}}

@ARTICLE{Jenkins,
       author = {{Jenkins}, A. and {Frenk}, C.~S. and {Pearce}, F.~R. and
         {Thomas}, P.~A. and {Colberg}, J.~M. and {White}, S.~D.~M. and
         {Couchman}, H.~M.~P. and {Peacock}, J.~A. and {Efstathiou}, G. and
         {Nelson}, A.~H.},
        title = "{Evolution of Structure in Cold Dark Matter Universes}",
      journal = {Astrophysical Journal},
     keywords = {COSMOLOGY: THEORY, COSMOLOGY: DARK MATTER, METHODS: NUMERICAL, Cosmology: Theory, Cosmology: Dark Matter, Methods: Numerical, Astrophysics},
         year = 1998,
        month = may,
       volume = {499},
       number = {1},
        pages = {20-40},
          doi = {10.1086/305615},
archivePrefix = {arXiv},
       eprint = {astro-ph/9709010},
 primaryClass = {astro-ph},
       adsurl = {https://ui.adsabs.harvard.edu/abs/1998ApJ...499...20J},
      adsnote = {Provided by the SAO/NASA Astrophysics Data System}
}

@article{MaBertschinger,
    author = "Ma, Chung-Pei and Bertschinger, Edmund",
    title = "{Cosmological perturbation theory in the synchronous versus conformal Newtonian gauge}",
    eprint = "astro-ph/9401007",
    archivePrefix = "arXiv",
    reportNumber = "GRP-375, MIT-AT-94-01, IASSNS-AST-94-1",
    month = "1",
    year = "1994"
}

@ARTICLE{1982PhRvL..48..223P,
       author = {{Pagels}, H. and {Primack}, J.~R.},
        title = "{Supersymmetry, cosmology, and new physics at teraelectronvolt energies}",
      journal = {\prl},
     keywords = {Astrophysics, Big Bang Cosmology, Broken Symmetry, Gravitation Theory, Gravitinos, High Energy Electrons, Supersymmetry, Matter (Physics), Particle Interactions, Quantum Theory, Scattering Cross Sections, Space Density, Universe, Physics (General), ASTROPHYSICS, BIG BANG COSMOLOGY, BROKEN SYMMETRY, GRAVITATION THEORY, GRAVITINOS, HIGH ENERGY ELECTRONS, SUPERSYMMETRY, MATTER (PHYSICS), PARTICLE INTERACTIONS, QUANTUM THEORY, SCATTERING CROSS SECTIONS, SPACE DENSITY, UNIVERSE, 11.30.Pb, 11.30.Qc, 98.80.Dr, Supersymmetry, Spontaneous and radiative symmetry breaking},
         year = 1982,
        month = jan,
       volume = {48},
        pages = {223-226},
          doi = {10.1103/PhysRevLett.48.223},
       adsurl = {https://ui.adsabs.harvard.edu/abs/1982PhRvL..48..223P},
      adsnote = {Provided by the SAO/NASA Astrophysics Data System}
}

@ARTICLE{1982Natur.299...37B,
       author = {{Blumenthal}, G.~R. and {Pagels}, H. and {Primack}, J.~R.},
        title = "{Galaxy formation by dissipationless particles heavier than neutrinos}",
      journal = {\nat},
     keywords = {Cosmology, Galactic Evolution, Neutral Particles, Particle Mass, Astronomical Models, Baryons, Particle Interactions, Spiral Galaxies, Astrophysics},
         year = 1982,
        month = sep,
       volume = {299},
       number = {5878},
        pages = {37-38},
          doi = {10.1038/299037a0},
       adsurl = {https://ui.adsabs.harvard.edu/abs/1982Natur.299...37B},
      adsnote = {Provided by the SAO/NASA Astrophysics Data System}
}

@ARTICLE{1983PhRvL..51..935M,
       author = {{Melott}, A.~L. and {Einasto}, J. and {Saar}, E. and {Suisalu}, I. and
         {Klypin}, A.~A. and {Shandarin}, S.~F.},
        title = "{Cluster Analysis of the Nonlinear Evolution of Large-Scale Structure in an Axion/Gravitino/Photino-Dominated Universe}",
      journal = {\prl},
     keywords = {Cluster Analysis, Cosmology, Galactic Evolution, Gravitinos, Covariance, Gravitons, Neutrinos, Nonlinear Systems, Photons, Astrophysics, 98.50.Eb, 11.30.Pb, 14.80.Pb, Supersymmetry},
         year = 1983,
        month = sep,
       volume = {51},
       number = {10},
        pages = {935-938},
          doi = {10.1103/PhysRevLett.51.935},
       adsurl = {https://ui.adsabs.harvard.edu/abs/1983PhRvL..51..935M},
      adsnote = {Provided by the SAO/NASA Astrophysics Data System}
}

@ARTICLE{1977PhLB...69...85H,
       author = {{Hut}, P.},
        title = "{Limits on masses and number of neutral weakly interacting particles}",
      journal = {Physics Letters B},
         year = 1977,
        month = jul,
       volume = {69},
       number = {1},
        pages = {85-88},
          doi = {10.1016/0370-2693(77)90139-3},
       adsurl = {https://ui.adsabs.harvard.edu/abs/1977PhLB...69...85H},
      adsnote = {Provided by the SAO/NASA Astrophysics Data System}
}

@ARTICLE{1984AdSpR...3j.379E,
       author = {{Efstathiou}, G. and {Jedrzejewski}, R.~I.},
        title = "{Observational constraints on dark matter in the universe}",
      journal = {Advances in Space Research},
     keywords = {Cosmology, Dark Matter, Galactic Clusters, Galactic Structure, Mass To Light Ratios, Red Shift, Weak Energy Interactions, Correlation, Elliptical Galaxies, Mass Distribution, Radial Velocity, Universe, Virial Theorem, Astrophysics, Missing-mass, density of the universe, galaxy redshift surveys, weakly interacting particles, galaxy dynamics},
         year = 1984,
        month = jan,
       volume = {3},
       number = {10-12},
        pages = {379-386},
          doi = {10.1016/0273-1177(84)90119-4},
       adsurl = {https://ui.adsabs.harvard.edu/abs/1984AdSpR...3j.379E},
      adsnote = {Provided by the SAO/NASA Astrophysics Data System}
}

@ARTICLE{1992ApJ...398L..81B,
       author = {{Bahcall}, Neta A. and {Cen}, Renyue},
        title = "{Galaxy Clusters and Cold Dark Matter: A Low-Density Unbiased Universe?}",
      journal = {\apjl},
     keywords = {Astronomical Catalogs, Astronomical Models, Dark Matter, Galactic Clusters, Background Radiation, Cosmic Background Explorer Satellite, Microwave Spectra, Astrophysics, COSMOLOGY: DARK MATTER, COSMOLOGY: THEORY, GALAXIES: CLUSTERING, COSMOLOGY: LARGE-SCALE STRUCTURE OF UNIVERSE, METHODS: NUMERICAL},
         year = 1992,
        month = oct,
       volume = {398},
        pages = {L81},
          doi = {10.1086/186582},
       adsurl = {https://ui.adsabs.harvard.edu/abs/1992ApJ...398L..81B},
      adsnote = {Provided by the SAO/NASA Astrophysics Data System}
}

@ARTICLE{1993ApJ...407L..49B,
       author = {{Bahcall}, Neta A. and {Cen}, Renyue},
        title = "{The Mass Function of Clusters of Galaxies}",
      journal = {\apjl},
     keywords = {Cosmology, Dark Matter, Galactic Clusters, Galactic Mass, X Ray Astronomy, Hubble Constant, Red Shift, Astrophysics, COSMOLOGY: OBSERVATIONS, COSMOLOGY: DARK MATTER, COSMOLOGY: THEORY, GALAXIES: CLUSTERING, COSMOLOGY: LARGE-SCALE STRUCTURE OF UNIVERSE},
         year = 1993,
        month = apr,
       volume = {407},
        pages = {L49},
          doi = {10.1086/186803},
       adsurl = {https://ui.adsabs.harvard.edu/abs/1993ApJ...407L..49B},
      adsnote = {Provided by the SAO/NASA Astrophysics Data System}
}

@ARTICLE{2003A&A...398..867S,
       author = {{Schuecker}, P. and {B{\"o}hringer}, H. and {Collins}, C.~A. and
         {Guzzo}, L.},
        title = "{The REFLEX galaxy cluster survey.  VII. Omega$_{m}$ and sigma$_{8}$  from cluster abundance and large-scale clustering}",
      journal = {\aap},
     keywords = {cosmology: cosmological parameters, X-rays: galaxies: clusters, Astrophysics},
         year = 2003,
        month = feb,
       volume = {398},
        pages = {867-877},
          doi = {10.1051/0004-6361:20021715},
archivePrefix = {arXiv},
       eprint = {astro-ph/0208251},
 primaryClass = {astro-ph},
       adsurl = {https://ui.adsabs.harvard.edu/abs/2003A&A...398..867S},
      adsnote = {Provided by the SAO/NASA Astrophysics Data System}
}

@ARTICLE{2014A&A...571A..20P,
       author = {{Planck Collaboration} and {Ade}, P.~A.~R. and {Aghanim}, N. and
         {Armitage-Caplan}, C. and {Arnaud}, M. and {Ashdown}, M. and {Atrio-Barand
        ela}, F. and {Aumont}, J. and {Baccigalupi}, C. and {Banday}, A.~J. and
         {Barreiro}, R.~B. and {Barrena}, R. and {Bartlett}, J.~G. and
         {Battaner}, E. and {Battye}, R. and {Benabed}, K. and
         {Beno{\^\i}t}, A. and {Benoit-L{\'e}vy}, A. and {Bernard}, J. -P. and
         {Bersanelli}, M. and {Bielewicz}, P. and {Bikmaev}, I. and
         {Blanchard}, A. and {Bobin}, J. and {Bock}, J.~J. and
         {B{\"o}hringer}, H. and {Bonaldi}, A. and {Bond}, J.~R. and
         {Borrill}, J. and {Bouchet}, F.~R. and {Bourdin}, H. and {Bridges}, M. and
         {Brown}, M.~L. and {Bucher}, M. and {Burenin}, R. and {Burigana}, C. and
         {Butler}, R.~C. and {Cardoso}, J. -F. and {Carvalho}, P. and
         {Catalano}, A. and {Challinor}, A. and {Chamballu}, A. and
         {Chary}, R. -R. and {Chiang}, L. -Y. and {Chiang}, H.~C. and
         {Chon}, G. and {Christensen}, P.~R. and {Church}, S. and
         {Clements}, D.~L. and {Colombi}, S. and {Colombo}, L.~P.~L. and
         {Couchot}, F. and {Coulais}, A. and {Crill}, B.~P. and {Curto}, A. and
         {Cuttaia}, F. and {Da Silva}, A. and {Dahle}, H. and {Danese}, L. and
         {Davies}, R.~D. and {Davis}, R.~J. and {de Bernardis}, P. and
         {de Rosa}, A. and {de Zotti}, G. and {Delabrouille}, J. and
         {Delouis}, J. -M. and {D{\'e}mocl{\`e}s}, J. and {D{\'e}sert}, F. -X. and
         {Dickinson}, C. and {Diego}, J.~M. and {Dolag}, K. and {Dole}, H. and
         {Donzelli}, S. and {Dor{\'e}}, O. and {Douspis}, M. and {Dupac}, X. and
         {Efstathiou}, G. and {En{\ss}lin}, T.~A. and {Eriksen}, H.~K. and
         {Finelli}, F. and {Flores-Cacho}, I. and {Forni}, O. and {Frailis}, M. and
         {Franceschi}, E. and {Fromenteau}, S. and {Galeotta}, S. and
         {Ganga}, K. and {G{\'e}nova-Santos}, R.~T. and {Giard}, M. and
         {Giardino}, G. and {Giraud-H{\'e}raud}, Y. and
         {Gonz{\'a}lez-Nuevo}, J. and {G{\'o}rski}, K.~M. and {Gratton}, S. and
         {Gregorio}, A. and {Gruppuso}, A. and {Hansen}, F.~K. and {Hanson}, D. and
         {Harrison}, D. and {Henrot-Versill{\'e}}, S. and
         {Hern{\'a}ndez-Monteagudo}, C. and {Herranz}, D. and {Hildebrand
        t}, S.~R. and {Hivon}, E. and {Hobson}, M. and {Holmes}, W.~A. and
         {Hornstrup}, A. and {Hovest}, W. and {Huffenberger}, K.~M. and
         {Hurier}, G. and {Jaffe}, T.~R. and {Jaffe}, A.~H. and {Jones}, W.~C. and
         {Juvela}, M. and {Keih{\"a}nen}, E. and {Keskitalo}, R. and
         {Khamitov}, I. and {Kisner}, T.~S. and {Kneissl}, R. and {Knoche}, J. and
         {Knox}, L. and {Kunz}, M. and {Kurki-Suonio}, H. and {Lagache}, G. and
         {L{\"a}hteenm{\"a}ki}, A. and {Lamarre}, J. -M. and {Lasenby}, A. and
         {Laureijs}, R.~J. and {Lawrence}, C.~R. and {Leahy}, J.~P. and
         {Leonardi}, R. and {Le{\'o}n-Tavares}, J. and {Lesgourgues}, J. and
         {Liddle}, A. and {Liguori}, M. and {Lilje}, P.~B. and
         {Linden-V{\o}rnle}, M. and {L{\'o}pez-Caniego}, M. and {Lubin}, P.~M. and
         {Mac{\'\i}as-P{\'e}rez}, J.~F. and {Maffei}, B. and {Maino}, D. and {Mand
        olesi}, N. and {Marcos-Caballero}, A. and {Maris}, M. and
         {Marshall}, D.~J. and {Martin}, P.~G. and
         {Mart{\'\i}nez-Gonz{\'a}lez}, E. and {Masi}, S. and {Matarrese}, S. and
         {Matthai}, F. and {Mazzotta}, P. and {Meinhold}, P.~R. and
         {Melchiorri}, A. and {Melin}, J. -B. and {Mendes}, L. and
         {Mennella}, A. and {Migliaccio}, M. and {Mitra}, S. and
         {Miville-Desch{\^e}nes}, M. -A. and {Moneti}, A. and {Montier}, L. and
         {Morgante}, G. and {Mortlock}, D. and {Moss}, A. and {Munshi}, D. and
         {Naselsky}, P. and {Nati}, F. and {Natoli}, P. and
         {Netterfield}, C.~B. and {N{\o}rgaard-Nielsen}, H.~U. and
         {Noviello}, F. and {Novikov}, D. and {Novikov}, I. and {Osborne}, S. and
         {Oxborrow}, C.~A. and {Paci}, F. and {Pagano}, L. and {Pajot}, F. and
         {Paoletti}, D. and {Partridge}, B. and {Pasian}, F. and
         {Patanchon}, G. and {Perdereau}, O. and {Perotto}, L. and
         {Perrotta}, F. and {Piacentini}, F. and {Piat}, M. and {Pierpaoli}, E. and
         {Pietrobon}, D. and {Plaszczynski}, S. and {Pointecouteau}, E. and
         {Polenta}, G. and {Ponthieu}, N. and {Popa}, L. and {Poutanen}, T. and
         {Pratt}, G.~W. and {Pr{\'e}zeau}, G. and {Prunet}, S. and
         {Puget}, J. -L. and {Rachen}, J.~P. and {Rebolo}, R. and
         {Reinecke}, M. and {Remazeilles}, M. and {Renault}, C. and
         {Ricciardi}, S. and {Riller}, T. and {Ristorcelli}, I. and {Rocha}, G. and
         {Roman}, M. and {Rosset}, C. and {Roudier}, G. and
         {Rowan-Robinson}, M. and {Rubi{\~n}o-Mart{\'\i}n}, J.~A. and
         {Rusholme}, B. and {Sandri}, M. and {Santos}, D. and {Savini}, G. and
         {Scott}, D. and {Seiffert}, M.~D. and {Shellard}, E.~P.~S. and
         {Spencer}, L.~D. and {Starck}, J. -L. and {Stolyarov}, V. and
         {Stompor}, R. and {Sudiwala}, R. and {Sunyaev}, R. and {Sureau}, F. and
         {Sutton}, D. and {Suur-Uski}, A. -S. and {Sygnet}, J. -F. and
         {Tauber}, J.~A. and {Tavagnacco}, D. and {Terenzi}, L. and
         {Toffolatti}, L. and {Tomasi}, M. and {Tristram}, M. and {Tucci}, M. and
         {Tuovinen}, J. and {T{\"u}rler}, M. and {Umana}, G. and
         {Valenziano}, L. and {Valiviita}, J. and {Van Tent}, B. and
         {Vielva}, P. and {Villa}, F. and {Vittorio}, N. and {Wade}, L.~A. and
         {Wandelt}, B.~D. and {Weller}, J. and {White}, M. and
         {White}, S.~D.~M. and {Yvon}, D. and {Zacchei}, A. and {Zonca}, A.},
        title = "{Planck 2013 results. XX. Cosmology from Sunyaev-Zeldovich cluster counts}",
      journal = {\aap},
     keywords = {cosmological parameters, large-scale structure of Universe, galaxies: clusters: general, Astrophysics - Cosmology and Nongalactic Astrophysics},
         year = 2014,
        month = nov,
       volume = {571},
          eid = {A20},
        pages = {A20},
          doi = {10.1051/0004-6361/201321521},
archivePrefix = {arXiv},
       eprint = {1303.5080},
 primaryClass = {astro-ph.CO},
       adsurl = {https://ui.adsabs.harvard.edu/abs/2014A&A...571A..20P},
      adsnote = {Provided by the SAO/NASA Astrophysics Data System}
}

@ARTICLE{Planck2015,
       author = {{Planck Collaboration} and {Ade}, P.~A.~R. and {Aghanim}, N. and
         {Arnaud}, M. and {Ashdown}, M. and {Aumont}, J. and {Baccigalupi}, C. and
         {Banday}, A.~J. and {Barreiro}, R.~B. and {Bartlett}, J.~G. and
         {Bartolo}, N. and {Battaner}, E. and {Battye}, R. and {Benabed}, K. and
         {Benoit}, A. and {Benoit-L{\'e}vy}, A. and {Bernard}, J. -P. and
         {Bersanelli}, M. and {Bielewicz}, P. and {Bock}, J.~J. and
         {Bonaldi}, A. and {Bonavera}, L. and {Bond}, J.~R. and {Borrill}, J. and
         {Bouchet}, F.~R. and {Bucher}, M. and {Burigana}, C. and
         {Butler}, R.~C. and {Calabrese}, E. and {Cardoso}, J. -F. and
         {Catalano}, A. and {Challinor}, A. and {Chamballu}, A. and
         {Chary}, R. -R. and {Chiang}, H.~C. and {Christensen}, P.~R. and
         {Church}, S. and {Clements}, D.~L. and {Colombi}, S. and
         {Colombo}, L.~P.~L. and {Combet}, C. and {Comis}, B. and {Couchot}, F. and
         {Coulais}, A. and {Crill}, B.~P. and {Curto}, A. and {Cuttaia}, F. and
         {Danese}, L. and {Davies}, R.~D. and {Davis}, R.~J. and
         {de Bernardis}, P. and {de Rosa}, A. and {de Zotti}, G. and
         {Delabrouille}, J. and {D{\'e}sert}, F. -X. and {Diego}, J.~M. and
         {Dolag}, K. and {Dole}, H. and {Donzelli}, S. and {Dor{\'e}}, O. and
         {Douspis}, M. and {Ducout}, A. and {Dupac}, X. and {Efstathiou}, G. and
         {Elsner}, F. and {En{\ss}lin}, T.~A. and {Eriksen}, H.~K. and
         {Falgarone}, E. and {Fergusson}, J. and {Finelli}, F. and {Forni}, O. and
         {Frailis}, M. and {Fraisse}, A.~A. and {Franceschi}, E. and
         {Frejsel}, A. and {Galeotta}, S. and {Galli}, S. and {Ganga}, K. and
         {Giard}, M. and {Giraud-H{\'e}raud}, Y. and {Gjerl{\o}w}, E. and
         {Gonz{\'a}lez-Nuevo}, J. and {G{\'o}rski}, K.~M. and {Gratton}, S. and
         {Gregorio}, A. and {Gruppuso}, A. and {Gudmundsson}, J.~E. and
         {Hansen}, F.~K. and {Hanson}, D. and {Harrison}, D.~L. and
         {Henrot-Versill{\'e}}, S. and {Hern{\'a}ndez-Monteagudo}, C. and
         {Herranz}, D. and {Hildebrandt}, S.~R. and {Hivon}, E. and
         {Hobson}, M. and {Holmes}, W.~A. and {Hornstrup}, A. and {Hovest}, W. and
         {Huffenberger}, K.~M. and {Hurier}, G. and {Jaffe}, A.~H. and
         {Jaffe}, T.~R. and {Jones}, W.~C. and {Juvela}, M. and
         {Keih{\"a}nen}, E. and {Keskitalo}, R. and {Kisner}, T.~S. and
         {Kneissl}, R. and {Knoche}, J. and {Kunz}, M. and {Kurki-Suonio}, H. and
         {Lagache}, G. and {L{\"a}hteenm{\"a}ki}, A. and {Lamarre}, J. -M. and
         {Lasenby}, A. and {Lattanzi}, M. and {Lawrence}, C.~R. and
         {Leonardi}, R. and {Lesgourgues}, J. and {Levrier}, F. and
         {Liguori}, M. and {Lilje}, P.~B. and {Linden-V{\o}rnle}, M. and
         {L{\'o}pez-Caniego}, M. and {Lubin}, P.~M. and
         {Mac{\'\i}as-P{\'e}rez}, J.~F. and {Maggio}, G. and {Maino}, D. and {Mand
        olesi}, N. and {Mangilli}, A. and {Maris}, M. and {Martin}, P.~G. and
         {Mart{\'\i}nez-Gonz{\'a}lez}, E. and {Masi}, S. and {Matarrese}, S. and
         {McGehee}, P. and {Meinhold}, P.~R. and {Melchiorri}, A. and
         {Melin}, J. -B. and {Mendes}, L. and {Mennella}, A. and
         {Migliaccio}, M. and {Mitra}, S. and {Miville-Desch{\^e}nes}, M. -A. and
         {Moneti}, A. and {Montier}, L. and {Morgante}, G. and {Mortlock}, D. and
         {Moss}, A. and {Munshi}, D. and {Murphy}, J.~A. and {Naselsky}, P. and
         {Nati}, F. and {Natoli}, P. and {Netterfield}, C.~B. and
         {N{\o}rgaard-Nielsen}, H.~U. and {Noviello}, F. and {Novikov}, D. and
         {Novikov}, I. and {Oxborrow}, C.~A. and {Paci}, F. and {Pagano}, L. and
         {Pajot}, F. and {Paoletti}, D. and {Partridge}, B. and {Pasian}, F. and
         {Patanchon}, G. and {Pearson}, T.~J. and {Perdereau}, O. and
         {Perotto}, L. and {Perrotta}, F. and {Pettorino}, V. and
         {Piacentini}, F. and {Piat}, M. and {Pierpaoli}, E. and
         {Pietrobon}, D. and {Plaszczynski}, S. and {Pointecouteau}, E. and
         {Polenta}, G. and {Popa}, L. and {Pratt}, G.~W. and {Pr{\'e}zeau}, G. and
         {Prunet}, S. and {Puget}, J. -L. and {Rachen}, J.~P. and {Rebolo}, R. and
         {Reinecke}, M. and {Remazeilles}, M. and {Renault}, C. and {Renzi}, A. and
         {Ristorcelli}, I. and {Rocha}, G. and {Roman}, M. and {Rosset}, C. and
         {Rossetti}, M. and {Roudier}, G. and {Rubi{\~n}o-Mart{\'\i}n}, J.~A. and
         {Rusholme}, B. and {Sandri}, M. and {Santos}, D. and {Savelainen}, M. and
         {Savini}, G. and {Scott}, D. and {Seiffert}, M.~D. and
         {Shellard}, E.~P.~S. and {Spencer}, L.~D. and {Stolyarov}, V. and
         {Stompor}, R. and {Sudiwala}, R. and {Sunyaev}, R. and {Sutton}, D. and
         {Suur-Uski}, A. -S. and {Sygnet}, J. -F. and {Tauber}, J.~A. and
         {Terenzi}, L. and {Toffolatti}, L. and {Tomasi}, M. and {Tristram}, M. and
         {Tucci}, M. and {Tuovinen}, J. and {T{\"u}rler}, M. and {Umana}, G. and
         {Valenziano}, L. and {Valiviita}, J. and {Van Tent}, B. and
         {Vielva}, P. and {Villa}, F. and {Wade}, L.~A. and {Wandelt}, B.~D. and
         {Wehus}, I.~K. and {Weller}, J. and {White}, S.~D.~M. and {Yvon}, D. and
         {Zacchei}, A. and {Zonca}, A.},
        title = "{Planck 2015 results. XXIV. Cosmology from Sunyaev-Zeldovich cluster counts}",
      journal = {\aap},
     keywords = {cosmological parameters, large-scale structure of Universe, Astrophysics - Cosmology and Nongalactic Astrophysics},
         year = 2016,
        month = sep,
       volume = {594},
          eid = {A24},
        pages = {A24},
          doi = {10.1051/0004-6361/201525833},
archivePrefix = {arXiv},
       eprint = {1502.01597},
 primaryClass = {astro-ph.CO},
       adsurl = {https://ui.adsabs.harvard.edu/abs/2016A&A...594A..24P},
      adsnote = {Provided by the SAO/NASA Astrophysics Data System}
}

@ARTICLE{2019ApJ...878...55B,
       author = {{Bocquet}, S. and {Dietrich}, J.~P. and {Schrabback}, T. and
         {Bleem}, L.~E. and {Klein}, M. and {Allen}, S.~W. and
         {Applegate}, D.~E. and {Ashby}, M.~L.~N. and {Bautz}, M. and
         {Bayliss}, M. and {Benson}, B.~A. and {Brodwin}, M. and {Bulbul}, E. and
         {Canning}, R.~E.~A. and {Capasso}, R. and {Carlstrom}, J.~E. and
         {Chang}, C.~L. and {Chiu}, I. and {Cho}, H. -M. and {Clocchiatti}, A. and
         {Crawford}, T.~M. and {Crites}, A.~T. and {de Haan}, T. and
         {Desai}, S. and {Dobbs}, M.~A. and {Foley}, R.~J. and {Forman}, W.~R. and
         {Garmire}, G.~P. and {George}, E.~M. and {Gladders}, M.~D. and
         {Gonzalez}, A.~H. and {Grandis}, S. and {Gupta}, N. and
         {Halverson}, N.~W. and {Hlavacek-Larrondo}, J. and {Hoekstra}, H. and
         {Holder}, G.~P. and {Holzapfel}, W.~L. and {Hou}, Z. and
         {Hrubes}, J.~D. and {Huang}, N. and {Jones}, C. and {Khullar}, G. and
         {Knox}, L. and {Kraft}, R. and {Lee}, A.~T. and {von der Linden}, A. and
         {Luong-Van}, D. and {Mantz}, A. and {Marrone}, D.~P. and
         {McDonald}, M. and {McMahon}, J.~J. and {Meyer}, S.~S. and
         {Mocanu}, L.~M. and {Mohr}, J.~J. and {Morris}, R.~G. and {Padin}, S. and
         {Patil}, S. and {Pryke}, C. and {Rapetti}, D. and {Reichardt}, C.~L. and
         {Rest}, A. and {Ruhl}, J.~E. and {Saliwanchik}, B.~R. and {Saro}, A. and
         {Sayre}, J.~T. and {Schaffer}, K.~K. and {Shirokoff}, E. and
         {Stalder}, B. and {Stanford}, S.~A. and {Staniszewski}, Z. and
         {Stark}, A.~A. and {Story}, K.~T. and {Strazzullo}, V. and
         {Stubbs}, C.~W. and {Vanderlinde}, K. and {Vieira}, J.~D. and
         {Vikhlinin}, A. and {Williamson}, R. and {Zenteno}, A.},
        title = "{Cluster Cosmology Constraints from the 2500 deg$^{2}$ SPT-SZ Survey: Inclusion of Weak Gravitational Lensing Data from Magellan and the Hubble Space Telescope}",
      journal = {\apj},
     keywords = {cosmological parameters, cosmology: observations, galaxies: clusters: general, large-scale structure of universe, Astrophysics - Cosmology and Nongalactic Astrophysics},
         year = 2019,
        month = jun,
       volume = {878},
       number = {1},
          eid = {55},
        pages = {55},
          doi = {10.3847/1538-4357/ab1f10},
archivePrefix = {arXiv},
       eprint = {1812.01679},
 primaryClass = {astro-ph.CO},
       adsurl = {https://ui.adsabs.harvard.edu/abs/2019ApJ...878...55B},
      adsnote = {Provided by the SAO/NASA Astrophysics Data System}
}

@ARTICLE{scalrel3,
       author = {{Rozo}, Eduardo and {Wechsler}, Risa H. and {Rykoff}, Eli S. and
         {Annis}, James T. and {Becker}, Matthew R. and {Evrard}, August E. and
         {Frieman}, Joshua A. and {Hansen}, Sarah M. and {Hao}, Jiangang and
         {Johnston}, David E. and {Koester}, Benjamin P. and
         {McKay}, Timothy A. and {Sheldon}, Erin S. and {Weinberg}, David H.},
        title = "{Cosmological Constraints from the Sloan Digital Sky Survey maxBCG Cluster Catalog}",
      journal = {\apj},
     keywords = {cosmological parameters, cosmology: observations, large-scale structure of universe, Astrophysics - Cosmology and Extragalactic Astrophysics},
         year = 2010,
        month = jan,
       volume = {708},
       number = {1},
        pages = {645-660},
          doi = {10.1088/0004-637X/708/1/645},
archivePrefix = {arXiv},
       eprint = {0902.3702},
 primaryClass = {astro-ph.CO},
       adsurl = {https://ui.adsabs.harvard.edu/abs/2010ApJ...708..645R},
      adsnote = {Provided by the SAO/NASA Astrophysics Data System}
}

@ARTICLE{2020PhRvD.102b3509A,
       author = {{Abbott}, T.~M.~C. and {Aguena}, M. and {Alarcon}, A. and {Allam}, S. and
         {Allen}, S. and {Annis}, J. and {Avila}, S. and {Bacon}, D. and
         {Bechtol}, K. and {Bermeo}, A. and {Bernstein}, G.~M. and {Bertin}, E. and
         {Bhargava}, S. and {Bocquet}, S. and {Brooks}, D. and {Brout}, D. and
         {Buckley-Geer}, E. and {Burke}, D.~L. and {Carnero Rosell}, A. and
         {Carrasco Kind}, M. and {Carretero}, J. and {Castander}, F.~J. and
         {Cawthon}, R. and {Chang}, C. and {Chen}, X. and {Choi}, A. and
         {Costanzi}, M. and {Crocce}, M. and {da Costa}, L.~N. and
         {Davis}, T.~M. and {De Vicente}, J. and {DeRose}, J. and {Desai}, S. and
         {Diehl}, H.~T. and {Dietrich}, J.~P. and {Dodelson}, S. and {Doel}, P. and
         {Drlica-Wagner}, A. and {Eckert}, K. and {Eifler}, T.~F. and
         {Elvin-Poole}, J. and {Estrada}, J. and {Everett}, S. and
         {Evrard}, A.~E. and {Farahi}, A. and {Ferrero}, I. and {Flaugher}, B. and
         {Fosalba}, P. and {Frieman}, J. and {Garc{\'\i}a-Bellido}, J. and
         {Gatti}, M. and {Gaztanaga}, E. and {Gerdes}, D.~W. and
         {Giannantonio}, T. and {Giles}, P. and {Grandis}, S. and {Gruen}, D. and
         {Gruendl}, R.~A. and {Gschwend}, J. and {Gutierrez}, G. and
         {Hartley}, W.~G. and {Hinton}, S.~R. and {Hollowood}, D.~L. and
         {Honscheid}, K. and {Hoyle}, B. and {Huterer}, D. and {James}, D.~J. and
         {Jarvis}, M. and {Jeltema}, T. and {Johnson}, M.~W.~G. and
         {Johnson}, M.~D. and {Kent}, S. and {Krause}, E. and {Kron}, R. and
         {Kuehn}, K. and {Kuropatkin}, N. and {Lahav}, O. and {Li}, T.~S. and
         {Lidman}, C. and {Lima}, M. and {Lin}, H. and {MacCrann}, N. and
         {Maia}, M.~A.~G. and {Mantz}, A. and {Marshall}, J.~L. and
         {Martini}, P. and {Mayers}, J. and {Melchior}, P. and
         {Mena-Fern{\'a}ndez}, J. and {Menanteau}, F. and {Miquel}, R. and
         {Mohr}, J.~J. and {Nichol}, R.~C. and {Nord}, B. and {Ogand
        o}, R.~L.~C. and {Palmese}, A. and {Paz-Chinch{\'o}n}, F. and
         {Plazas}, A.~A. and {Prat}, J. and {Rau}, M.~M. and {Romer}, A.~K. and
         {Roodman}, A. and {Rooney}, P. and {Rozo}, E. and {Rykoff}, E.~S. and
         {Sako}, M. and {Samuroff}, S. and {S{\'a}nchez}, C. and {Sanchez}, E. and
         {Saro}, A. and {Scarpine}, V. and {Schubnell}, M. and {Scolnic}, D. and
         {Serrano}, S. and {Sevilla-Noarbe}, I. and {Sheldon}, E. and
         {Smith}, J. Allyn. and {Smith}, M. and {Suchyta}, E. and
         {Swanson}, M.~E.~C. and {Tarle}, G. and {Thomas}, D. and {To}, C. and
         {Troxel}, M.~A. and {Tucker}, D.~L. and {Varga}, T.~N. and
         {von der Linden}, A. and {Walker}, A.~R. and {Wechsler}, R.~H. and
         {Weller}, J. and {Wilkinson}, R.~D. and {Wu}, H. and {Yanny}, B. and
         {Zhang}, Y. and {Zhang}, Z. and {Zuntz}, J. and {DES Collaboration}},
        title = "{Dark Energy Survey Year 1 Results: Cosmological constraints from cluster abundances and weak lensing}",
      journal = {\prd},
     keywords = {Astrophysics - Cosmology and Nongalactic Astrophysics},
         year = 2020,
        month = jul,
       volume = {102},
       number = {2},
          eid = {023509},
        pages = {023509},
          doi = {10.1103/PhysRevD.102.023509},
archivePrefix = {arXiv},
       eprint = {2002.11124},
 primaryClass = {astro-ph.CO},
       adsurl = {https://ui.adsabs.harvard.edu/abs/2020PhRvD.102b3509A},
      adsnote = {Provided by the SAO/NASA Astrophysics Data System}
}

@INPROCEEDINGS{2010AIPC.1248..543P,
       author = {{Predehl}, Peter and {B{\"o}hringer}, Hans and {Brunner}, Hermann and
         {Brusa}, Marcella and {Burwitz}, Vadim and {Cappelluti}, Nico and
         {Churazov}, Evgeniy and {Dennerl}, Konrad and {Freyberg}, Michael and
         {Friedrich}, Peter and {Hasinger}, G{\"u}nther and
         {Kendziorra}, Eckhard and {Kreykenbohm}, Ingo and {Schmid}, Christian and
         {Wilms}, J{\"o}rn and {Lamer}, Georg and {Meidinger}, Norbert and
         {M{\"u}hlegger}, Martin and {Pavlinsky}, Mikhail and {Robrade}, Jan and
         {Santangelo}, Andrea and {Schmitt}, J{\"u}rgen and {Schwope}, Axel and
         {Steinmetz}, Matthias and {Str{\"u}der}, Lothar and {Sunyaev}, Rashid and
         {Tenzer}, Chris},
        title = "{eROSITA on SRG}",
     keywords = {X-ray astronomy, sky surveys, satellites, artificial, dark matter, astronomical telescopes, 95.85.Nv, 95.80.+p, 95.40.+s, 95.35.+d, 95.55.Ka, X-ray, Astronomical catalogs atlases sky surveys databases retrieval systems archives etc., Artificial Earth satellites, Dark matter, X- and gamma-ray telescopes and instrumentation},
    booktitle = {X-ray Astronomy 2009; Present Status, Multi-Wavelength Approach and Future Perspectives},
         year = 2010,
       editor = {{Comastri}, A. and {Angelini}, L. and {Cappi}, M.},
       series = {American Institute of Physics Conference Series},
       volume = {1248},
        month = jul,
        pages = {543-548},
          doi = {10.1063/1.3475336},
       adsurl = {https://ui.adsabs.harvard.edu/abs/2010AIPC.1248..543P},
      adsnote = {Provided by the SAO/NASA Astrophysics Data System}
}

@ARTICLE{2011arXiv1110.3193L,
       author = {{Laureijs}, R. and {Amiaux}, J. and {Arduini}, S. and
         {Augu{\`e}res}, J. -L. and {Brinchmann}, J. and {Cole}, R. and
         {Cropper}, M. and {Dabin}, C. and {Duvet}, L. and {Ealet}, A. and
         {Garilli}, B. and {Gondoin}, P. and {Guzzo}, L. and {Hoar}, J. and
         {Hoekstra}, H. and {Holmes}, R. and {Kitching}, T. and {Maciaszek}, T. and
         {Mellier}, Y. and {Pasian}, F. and {Percival}, W. and {Rhodes}, J. and
         {Saavedra Criado}, G. and {Sauvage}, M. and {Scaramella}, R. and
         {Valenziano}, L. and {Warren}, S. and {Bender}, R. and {Castander}, F. and
         {Cimatti}, A. and {Le F{\`e}vre}, O. and {Kurki-Suonio}, H. and
         {Levi}, M. and {Lilje}, P. and {Meylan}, G. and {Nichol}, R. and
         {Pedersen}, K. and {Popa}, V. and {Rebolo Lopez}, R. and {Rix}, H. -W. and
         {Rottgering}, H. and {Zeilinger}, W. and {Grupp}, F. and {Hudelot}, P. and
         {Massey}, R. and {Meneghetti}, M. and {Miller}, L. and {Paltani}, S. and
         {Paulin-Henriksson}, S. and {Pires}, S. and {Saxton}, C. and
         {Schrabback}, T. and {Seidel}, G. and {Walsh}, J. and {Aghanim}, N. and
         {Amendola}, L. and {Bartlett}, J. and {Baccigalupi}, C. and
         {Beaulieu}, J. -P. and {Benabed}, K. and {Cuby}, J. -G. and
         {Elbaz}, D. and {Fosalba}, P. and {Gavazzi}, G. and {Helmi}, A. and
         {Hook}, I. and {Irwin}, M. and {Kneib}, J. -P. and {Kunz}, M. and
         {Mannucci}, F. and {Moscardini}, L. and {Tao}, C. and {Teyssier}, R. and
         {Weller}, J. and {Zamorani}, G. and {Zapatero Osorio}, M.~R. and
         {Boulade}, O. and {Foumond}, J.~J. and {Di Giorgio}, A. and
         {Guttridge}, P. and {James}, A. and {Kemp}, M. and {Martignac}, J. and
         {Spencer}, A. and {Walton}, D. and {Bl{\"u}mchen}, T. and {Bonoli}, C. and
         {Bortoletto}, F. and {Cerna}, C. and {Corcione}, L. and {Fabron}, C. and
         {Jahnke}, K. and {Ligori}, S. and {Madrid}, F. and {Martin}, L. and
         {Morgante}, G. and {Pamplona}, T. and {Prieto}, E. and {Riva}, M. and
         {Toledo}, R. and {Trifoglio}, M. and {Zerbi}, F. and {Abdalla}, F. and
         {Douspis}, M. and {Grenet}, C. and {Borgani}, S. and {Bouwens}, R. and
         {Courbin}, F. and {Delouis}, J. -M. and {Dubath}, P. and {Fontana}, A. and
         {Frailis}, M. and {Grazian}, A. and {Koppenh{\"o}fer}, J. and
         {Mansutti}, O. and {Melchior}, M. and {Mignoli}, M. and {Mohr}, J. and
         {Neissner}, C. and {Noddle}, K. and {Poncet}, M. and {Scodeggio}, M. and
         {Serrano}, S. and {Shane}, N. and {Starck}, J. -L. and {Surace}, C. and
         {Taylor}, A. and {Verdoes-Kleijn}, G. and {Vuerli}, C. and
         {Williams}, O.~R. and {Zacchei}, A. and {Altieri}, B. and
         {Escudero Sanz}, I. and {Kohley}, R. and {Oosterbroek}, T. and
         {Astier}, P. and {Bacon}, D. and {Bardelli}, S. and {Baugh}, C. and
         {Bellagamba}, F. and {Benoist}, C. and {Bianchi}, D. and {Biviano}, A. and
         {Branchini}, E. and {Carbone}, C. and {Cardone}, V. and {Clements}, D. and
         {Colombi}, S. and {Conselice}, C. and {Cresci}, G. and {Deacon}, N. and
         {Dunlop}, J. and {Fedeli}, C. and {Fontanot}, F. and {Franzetti}, P. and
         {Giocoli}, C. and {Garcia-Bellido}, J. and {Gow}, J. and {Heavens}, A. and
         {Hewett}, P. and {Heymans}, C. and {Holland}, A. and {Huang}, Z. and
         {Ilbert}, O. and {Joachimi}, B. and {Jennins}, E. and {Kerins}, E. and
         {Kiessling}, A. and {Kirk}, D. and {Kotak}, R. and {Krause}, O. and
         {Lahav}, O. and {van Leeuwen}, F. and {Lesgourgues}, J. and
         {Lombardi}, M. and {Magliocchetti}, M. and {Maguire}, K. and
         {Majerotto}, E. and {Maoli}, R. and {Marulli}, F. and
         {Maurogordato}, S. and {McCracken}, H. and {McLure}, R. and
         {Melchiorri}, A. and {Merson}, A. and {Moresco}, M. and {Nonino}, M. and
         {Norberg}, P. and {Peacock}, J. and {Pello}, R. and {Penny}, M. and
         {Pettorino}, V. and {Di Porto}, C. and {Pozzetti}, L. and
         {Quercellini}, C. and {Radovich}, M. and {Rassat}, A. and {Roche}, N. and
         {Ronayette}, S. and {Rossetti}, E. and {Sartoris}, B. and
         {Schneider}, P. and {Semboloni}, E. and {Serjeant}, S. and
         {Simpson}, F. and {Skordis}, C. and {Smadja}, G. and {Smartt}, S. and
         {Spano}, P. and {Spiro}, S. and {Sullivan}, M. and {Tilquin}, A. and
         {Trotta}, R. and {Verde}, L. and {Wang}, Y. and {Williger}, G. and
         {Zhao}, G. and {Zoubian}, J. and {Zucca}, E.},
        title = "{Euclid Definition Study Report}",
      journal = {arXiv e-prints},
     keywords = {Astrophysics - Cosmology and Extragalactic Astrophysics, Astrophysics - Galaxy Astrophysics},
         year = 2011,
        month = oct,
          eid = {arXiv:1110.3193},
        pages = {arXiv:1110.3193},
archivePrefix = {arXiv},
       eprint = {1110.3193},
 primaryClass = {astro-ph.CO},
       adsurl = {https://ui.adsabs.harvard.edu/abs/2011arXiv1110.3193L},
      adsnote = {Provided by the SAO/NASA Astrophysics Data System}
}

@ARTICLE{2010arXiv1001.0061R,
       author = {{Refregier}, A. and {Amara}, A. and {Kitching}, T.~D. and {Rassat}, A. and
         {Scaramella}, R. and {Weller}, J.},
        title = "{Euclid Imaging Consortium Science Book}",
      journal = {arXiv e-prints},
     keywords = {Astrophysics - Instrumentation and Methods for Astrophysics, Astrophysics - Cosmology and Nongalactic Astrophysics},
         year = 2010,
        month = jan,
          eid = {arXiv:1001.0061},
        pages = {arXiv:1001.0061},
archivePrefix = {arXiv},
       eprint = {1001.0061},
 primaryClass = {astro-ph.IM},
       adsurl = {https://ui.adsabs.harvard.edu/abs/2010arXiv1001.0061R},
      adsnote = {Provided by the SAO/NASA Astrophysics Data System}
}

@article{axionsim,
    author = "Schwabe, Bodo and Gosenca, Mateja and Behrens, Christoph and Niemeyer, Jens C. and Easther, Richard",
    title = "{Simulating mixed fuzzy and cold dark matter}",
    eprint = "2007.08256",
    archivePrefix = "arXiv",
    primaryClass = "astro-ph.CO",
    doi = "10.1103/PhysRevD.102.083518",
    journal = "Phys. Rev. D",
    volume = "102",
    number = "8",
    pages = "083518",
    year = "2020"
}

@article{LSSprobesDM1,
    author = "Viel, Matteo and Becker, George D. and Bolton, James S. and Haehnelt, Martin G.",
    title = "{Warm dark matter as a solution to the small scale crisis: New constraints from high redshift Lyman-\ensuremath{\alpha} forest data}",
    eprint = "1306.2314",
    archivePrefix = "arXiv",
    primaryClass = "astro-ph.CO",
    doi = "10.1103/PhysRevD.88.043502",
    journal = "Phys. Rev. D",
    volume = "88",
    pages = "043502",
    year = "2013"
}

@ARTICLE{LSSprobesDM2,
       author = {{Vald{\'e}s}, M. and {Evoli}, C. and {Mesinger}, A. and {Ferrara}, A. and
         {Yoshida}, N.},
        title = "{The nature of dark matter from the global high-redshift H I 21 cm signal}",
      journal = {Mon. Not. of the Royal Academic Society},
     keywords = {intergalactic medium, cosmology: theory, dark matter, diffuse radiation, Astrophysics - Cosmology and Extragalactic Astrophysics, Astrophysics - High Energy Astrophysical Phenomena},
         year = 2013,
        month = feb,
       volume = {429},
       number = {2},
        pages = {1705-1716},
          doi = {10.1093/mnras/sts458},
archivePrefix = {arXiv},
       eprint = {1209.2120},
 primaryClass = {astro-ph.CO},
       adsurl = {https://ui.adsabs.harvard.edu/abs/2013MNRAS.429.1705V},
      adsnote = {Provided by the SAO/NASA Astrophysics Data System}
}

@article{LSSprobesDM3,
    author = "Scoville, Nick and others",
    title = "{The Cosmic Evolution Survey (COSMOS): Overview}",
    eprint = "astro-ph/0612305",
    archivePrefix = "arXiv",
    doi = "10.1086/516585",
    journal = "Astrophys. J. Suppl.",
    volume = "172",
    pages = "1--8",
    year = "2007"
}

@article{LSSprobesDM4,
    author = "Novosyadlyj, B. and Durrer, R. and Apunevych, S.",
    title = "{Cosmological parameters from observational data on the large scale structure of the universe}",
    eprint = "astro-ph/0009485",
    archivePrefix = "arXiv",
    month = "9",
    year = "2000"
}

@article{LSSprobesDM5,
    author = "Buen-Abad, Manuel A. and Schmaltz, Martin and Lesgourgues, Julien and Brinckmann, Thejs",
    title = "{Interacting Dark Sector and Precision Cosmology}",
    eprint = "1708.09406",
    archivePrefix = "arXiv",
    primaryClass = "astro-ph.CO",
    doi = "10.1088/1475-7516/2018/01/008",
    journal = "JCAP",
    volume = "01",
    pages = "008",
    year = "2018"
}

@article{LSSprobesDM6,
    author = "Escudero, Miguel and Mena, Olga and Vincent, Aaron C. and Wilkinson, Ryan J. and B\oe{}hm, C\'eline",
    title = "{Exploring dark matter microphysics with galaxy surveys}",
    eprint = "1505.06735",
    archivePrefix = "arXiv",
    primaryClass = "astro-ph.CO",
    reportNumber = "IFIC-15-32, IPPP-15-29, DCPT-15-58",
    doi = "10.1088/1475-7516/2015/9/034",
    journal = "JCAP",
    volume = "09",
    pages = "034",
    year = "2015"
}

@article{LSSprobesDM7,
    author = "Hamaus, Nico and Pisani, Alice and Sutter, Paul M. and Lavaux, Guilhem and Escoffier, St\'ephanie and Wandelt, Benjamin D. and Weller, Jochen",
    title = "{Constraints on Cosmology and Gravity from the Dynamics of Voids}",
    eprint = "1602.01784",
    archivePrefix = "arXiv",
    primaryClass = "astro-ph.CO",
    doi = "10.1103/PhysRevLett.117.091302",
    journal = "Phys. Rev. Lett.",
    volume = "117",
    number = "9",
    pages = "091302",
    year = "2016"
}

@ARTICLE{2020arXiv201102116K,
       author = {{Kulkarni}, Mihir and {Ostriker}, Jeremiah P.},
        title = "{What is the Halo Mass Function in a Fuzzy Dark Matter Cosmology?}",
      journal = {arXiv e-prints},
     keywords = {Astrophysics - Cosmology and Nongalactic Astrophysics},
         year = 2020,
        month = nov,
          eid = {arXiv:2011.02116},
        pages = {arXiv:2011.02116},
archivePrefix = {arXiv},
       eprint = {2011.02116},
 primaryClass = {astro-ph.CO},
       adsurl = {https://ui.adsabs.harvard.edu/abs/2020arXiv201102116K},
      adsnote = {Provided by the SAO/NASA Astrophysics Data System}
}

@ARTICLE{1972CoASP...4..173S,
       author = {{Sunyaev}, R.~A. and {Zeldovich}, Ya. B.},
        title = "{The Observations of Relic Radiation as a Test of the Nature of X-Ray Radiation from the Clusters of Galaxies}",
      journal = {Comments on Astrophysics and Space Physics},
     keywords = {Cosmology, Microwave Background Radiation, Clusters of Galaxies, X-Ray Astronomy, Intergalactic Gas},
         year = 1972,
        month = nov,
       volume = {4},
        pages = {173},
       adsurl = {https://ui.adsabs.harvard.edu/abs/1972CoASP...4..173S},
      adsnote = {Provided by the SAO/NASA Astrophysics Data System}
}

@ARTICLE{1969Ap&SS...4..301Z,
       author = {{Zeldovich}, Ya. B. and {Sunyaev}, R.~A.},
        title = "{The Interaction of Matter and Radiation in a Hot-Model Universe}",
      journal = {\apss},
         year = 1969,
        month = jul,
       volume = {4},
       number = {3},
        pages = {301-316},
          doi = {10.1007/BF00661821},
       adsurl = {https://ui.adsabs.harvard.edu/abs/1969Ap&SS...4..301Z},
      adsnote = {Provided by the SAO/NASA Astrophysics Data System}
}

@ARTICLE{1970Ap&SS...7...20S,
       author = {{Sunyaev}, R.~A. and {Zeldovich}, Ya. B.},
        title = "{The interaction of matter and radiation in the hot model of the Universe, II}",
      journal = {\apss},
         year = 1970,
        month = apr,
       volume = {7},
       number = {1},
        pages = {20-30},
          doi = {10.1007/BF00653472},
       adsurl = {https://ui.adsabs.harvard.edu/abs/1970Ap&SS...7...20S},
      adsnote = {Provided by the SAO/NASA Astrophysics Data System}
}

@ARTICLE{2014A&A...571A..29P,
       author = {{Planck Collaboration} and {Ade}, P.~A.~R. and {Aghanim}, N. and {Armitage-Caplan}, C. and {Arnaud}, M. and {Ashdown}, M. and {Atrio-Barandela}, F. and {Aumont}, J. and {Aussel}, H. and {Baccigalupi}, C. and {Banday}, A.~J. and {Barreiro}, R.~B. and {Barrena}, R. and {Bartelmann}, M. and {Bartlett}, J.~G. and {Battaner}, E. and {Benabed}, K. and {Beno{\^\i}t}, A. and {Benoit-L{\'e}vy}, A. and {Bernard}, J. -P. and {Bersanelli}, M. and {Bielewicz}, P. and {Bikmaev}, I. and {Bobin}, J. and {Bock}, J.~J. and {B{\"o}hringer}, H. and {Bonaldi}, A. and {Bond}, J.~R. and {Borrill}, J. and {Bouchet}, F.~R. and {Bridges}, M. and {Bucher}, M. and {Burenin}, R. and {Burigana}, C. and {Butler}, R.~C. and {Cardoso}, J. -F. and {Carvalho}, P. and {Catalano}, A. and {Challinor}, A. and {Chamballu}, A. and {Chary}, R. -R. and {Chen}, X. and {Chiang}, H.~C. and {Chiang}, L. -Y. and {Chon}, G. and {Christensen}, P.~R. and {Churazov}, E. and {Church}, S. and {Clements}, D.~L. and {Colombi}, S. and {Colombo}, L.~P.~L. and {Comis}, B. and {Couchot}, F. and {Coulais}, A. and {Crill}, B.~P. and {Curto}, A. and {Cuttaia}, F. and {Da Silva}, A. and {Dahle}, H. and {Danese}, L. and {Davies}, R.~D. and {Davis}, R.~J. and {de Bernardis}, P. and {de Rosa}, A. and {de Zotti}, G. and {Delabrouille}, J. and {Delouis}, J. -M. and {D{\'e}mocl{\`e}s}, J. and {D{\'e}sert}, F. -X. and {Dickinson}, C. and {Diego}, J.~M. and {Dolag}, K. and {Dole}, H. and {Donzelli}, S. and {Dor{\'e}}, O. and {Douspis}, M. and {Dupac}, X. and {Efstathiou}, G. and {Eisenhardt}, P.~R.~M. and {En{\ss}lin}, T.~A. and {Eriksen}, H.~K. and {Feroz}, F. and {Finelli}, F. and {Flores-Cacho}, I. and {Forni}, O. and {Frailis}, M. and {Franceschi}, E. and {Fromenteau}, S. and {Galeotta}, S. and {Ganga}, K. and {G{\'e}nova-Santos}, R.~T. and {Giard}, M. and {Giardino}, G. and {Gilfanov}, M. and {Giraud-H{\'e}raud}, Y. and {Gonz{\'a}lez-Nuevo}, J. and {G{\'o}rski}, K.~M. and {Grainge}, K.~J.~B. and {Gratton}, S. and {Gregorio}, A. and {Groeneboom}, N., E. and {Gruppuso}, A. and {Hansen}, F.~K. and {Hanson}, D. and {Harrison}, D. and {Hempel}, A. and {Henrot-Versill{\'e}}, S. and {Hern{\'a}ndez-Monteagudo}, C. and {Herranz}, D. and {Hildebrandt}, S.~R. and {Hivon}, E. and {Hobson}, M. and {Holmes}, W.~A. and {Hornstrup}, A. and {Hovest}, W. and {Huffenberger}, K.~M. and {Hurier}, G. and {Hurley-Walker}, N. and {Jaffe}, A.~H. and {Jaffe}, T.~R. and {Jones}, W.~C. and {Juvela}, M. and {Keih{\"a}nen}, E. and {Keskitalo}, R. and {Khamitov}, I. and {Kisner}, T.~S. and {Kneissl}, R. and {Knoche}, J. and {Knox}, L. and {Kunz}, M. and {Kurki-Suonio}, H. and {Lagache}, G. and {L{\"a}hteenm{\"a}ki}, A. and {Lamarre}, J. -M. and {Lasenby}, A. and {Laureijs}, R.~J. and {Lawrence}, C.~R. and {Leahy}, J.~P. and {Leonardi}, R. and {Le{\'o}n-Tavares}, J. and {Lesgourgues}, J. and {Li}, C. and {Liddle}, A. and {Liguori}, M. and {Lilje}, P.~B. and {Linden-V{\o}rnle}, M. and {L{\'o}pez-Caniego}, M. and {Lubin}, P.~M. and {Mac{\'\i}as-P{\'e}rez}, J.~F. and {MacTavish}, C.~J. and {Maffei}, B. and {Maino}, D. and {Mandolesi}, N. and {Maris}, M. and {Marshall}, D.~J. and {Martin}, P.~G. and {Mart{\'\i}nez-Gonz{\'a}lez}, E. and {Masi}, S. and {Massardi}, M. and {Matarrese}, S. and {Matthai}, F. and {Mazzotta}, P. and {Mei}, S. and {Meinhold}, P.~R. and {Melchiorri}, A. and {Melin}, J. -B. and {Mendes}, L. and {Mennella}, A. and {Migliaccio}, M. and {Mikkelsen}, K. and {Mitra}, S. and {Miville-Desch{\^e}nes}, M. -A. and {Moneti}, A. and {Montier}, L. and {Morgante}, G. and {Mortlock}, D. and {Munshi}, D. and {Murphy}, J.~A. and {Naselsky}, P. and {Nati}, F. and {Natoli}, P. and {Nesvadba}, N.~P.~H. and {Netterfield}, C.~B. and {N{\o}rgaard-Nielsen}, H.~U. and {Noviello}, F. and {Novikov}, D. and {Novikov}, I. and {O'Dwyer}, I.~J. and {Olamaie}, M. and {Osborne}, S. and {Oxborrow}, C.~A. and {Paci}, F. and {Pagano}, L. and {Pajot}, F. and {Paoletti}, D. and {Pasian}, F. and {Patanchon}, G. and {Pearson}, T.~J. and {Perdereau}, O. and {Perotto}, L. and {Perrott}, Y.~C. and {Perrotta}, F. and {Piacentini}, F. and {Piat}, M. and {Pierpaoli}, E. and {Pietrobon}, D. and {Plaszczynski}, S. and {Pointecouteau}, E. and {Polenta}, G. and {Ponthieu}, N. and {Popa}, L. and {Poutanen}, T. and {Pratt}, G.~W. and {Pr{\'e}zeau}, G. and {Prunet}, S. and {Puget}, J. -L. and {Rachen}, J.~P. and {Reach}, W.~T. and {Rebolo}, R. and {Reinecke}, M. and {Remazeilles}, M. and {Renault}, C. and {Ricciardi}, S. and {Riller}, T. and {Ristorcelli}, I. and {Rocha}, G. and {Rosset}, C. and {Roudier}, G. and {Rowan-Robinson}, M. and {Rubi{\~n}o-Mart{\'\i}n}, J.~A. and {Rumsey}, C. and {Rusholme}, B. and {Sandri}, M. and {Santos}, D. and {Saunders}, R.~D.~E. and {Savini}, G. and {Schammel}, M.~P. and {Scott}, D. and {Seiffert}, M.~D. and {Shellard}, E.~P.~S. and {Shimwell}, T.~W. and {Spencer}, L.~D. and {Stanford}, S.~A. and {Starck}, J. -L. and {Stolyarov}, V. and {Stompor}, R. and {Sudiwala}, R. and {Sunyaev}, R. and {Sureau}, F. and {Sutton}, D. and {Suur-Uski}, A. -S. and {Sygnet}, J. -F. and {Tauber}, J.~A. and {Tavagnacco}, D. and {Terenzi}, L. and {Toffolatti}, L. and {Tomasi}, M. and {Tristram}, M. and {Tucci}, M. and {Tuovinen}, J. and {T{\"u}rler}, M. and {Umana}, G. and {Valenziano}, L. and {Valiviita}, J. and {Van Tent}, B. and {Vibert}, L. and {Vielva}, P. and {Villa}, F. and {Vittorio}, N. and {Wade}, L.~A. and {Wandelt}, B.~D. and {White}, M. and {White}, S.~D.~M. and {Yvon}, D. and {Zacchei}, A. and {Zonca}, A.},
        title = "{Planck 2013 results. XXIX. The Planck catalogue of Sunyaev-Zeldovich sources}",
      journal = {\aap},
     keywords = {large-scale structure of Universe, galaxies: clusters: general, catalogs, Astrophysics - Cosmology and Nongalactic Astrophysics},
         year = 2014,
        month = nov,
       volume = {571},
          eid = {A29},
        pages = {A29},
          doi = {10.1051/0004-6361/201321523},
archivePrefix = {arXiv},
       eprint = {1303.5089},
 primaryClass = {astro-ph.CO},
       adsurl = {https://ui.adsabs.harvard.edu/abs/2014A&A...571A..29P},
      adsnote = {Provided by the SAO/NASA Astrophysics Data System}
}

@ARTICLE{2016A&A...594A..27P,
       author = {{Planck Collaboration} and {Ade}, P.~A.~R. and {Aghanim}, N. and {Arnaud}, M. and {Ashdown}, M. and {Aumont}, J. and {Baccigalupi}, C. and {Banday}, A.~J. and {Barreiro}, R.~B. and {Barrena}, R. and {Bartlett}, J.~G. and {Bartolo}, N. and {Battaner}, E. and {Battye}, R. and {Benabed}, K. and {Beno{\^\i}t}, A. and {Benoit-L{\'e}vy}, A. and {Bernard}, J. -P. and {Bersanelli}, M. and {Bielewicz}, P. and {Bikmaev}, I. and {B{\"o}hringer}, H. and {Bonaldi}, A. and {Bonavera}, L. and {Bond}, J.~R. and {Borrill}, J. and {Bouchet}, F.~R. and {Bucher}, M. and {Burenin}, R. and {Burigana}, C. and {Butler}, R.~C. and {Calabrese}, E. and {Cardoso}, J. -F. and {Carvalho}, P. and {Catalano}, A. and {Challinor}, A. and {Chamballu}, A. and {Chary}, R. -R. and {Chiang}, H.~C. and {Chon}, G. and {Christensen}, P.~R. and {Clements}, D.~L. and {Colombi}, S. and {Colombo}, L.~P.~L. and {Combet}, C. and {Comis}, B. and {Couchot}, F. and {Coulais}, A. and {Crill}, B.~P. and {Curto}, A. and {Cuttaia}, F. and {Dahle}, H. and {Danese}, L. and {Davies}, R.~D. and {Davis}, R.~J. and {de Bernardis}, P. and {de Rosa}, A. and {de Zotti}, G. and {Delabrouille}, J. and {D{\'e}sert}, F. -X. and {Dickinson}, C. and {Diego}, J.~M. and {Dolag}, K. and {Dole}, H. and {Donzelli}, S. and {Dor{\'e}}, O. and {Douspis}, M. and {Ducout}, A. and {Dupac}, X. and {Efstathiou}, G. and {Eisenhardt}, P.~R.~M. and {Elsner}, F. and {En{\ss}lin}, T.~A. and {Eriksen}, H.~K. and {Falgarone}, E. and {Fergusson}, J. and {Feroz}, F. and {Ferragamo}, A. and {Finelli}, F. and {Forni}, O. and {Frailis}, M. and {Fraisse}, A.~A. and {Franceschi}, E. and {Frejsel}, A. and {Galeotta}, S. and {Galli}, S. and {Ganga}, K. and {G{\'e}nova-Santos}, R.~T. and {Giard}, M. and {Giraud-H{\'e}raud}, Y. and {Gjerl{\o}w}, E. and {Gonz{\'a}lez-Nuevo}, J. and {G{\'o}rski}, K.~M. and {Grainge}, K.~J.~B. and {Gratton}, S. and {Gregorio}, A. and {Gruppuso}, A. and {Gudmundsson}, J.~E. and {Hansen}, F.~K. and {Hanson}, D. and {Harrison}, D.~L. and {Hempel}, A. and {Henrot-Versill{\'e}}, S. and {Hern{\'a}ndez-Monteagudo}, C. and {Herranz}, D. and {Hildebrandt}, S.~R. and {Hivon}, E. and {Hobson}, M. and {Holmes}, W.~A. and {Hornstrup}, A. and {Hovest}, W. and {Huffenberger}, K.~M. and {Hurier}, G. and {Jaffe}, A.~H. and {Jaffe}, T.~R. and {Jin}, T. and {Jones}, W.~C. and {Juvela}, M. and {Keih{\"a}nen}, E. and {Keskitalo}, R. and {Khamitov}, I. and {Kisner}, T.~S. and {Kneissl}, R. and {Knoche}, J. and {Kunz}, M. and {Kurki-Suonio}, H. and {Lagache}, G. and {Lamarre}, J. -M. and {Lasenby}, A. and {Lattanzi}, M. and {Lawrence}, C.~R. and {Leonardi}, R. and {Lesgourgues}, J. and {Levrier}, F. and {Liguori}, M. and {Lilje}, P.~B. and {Linden-V{\o}rnle}, M. and {L{\'o}pez-Caniego}, M. and {Lubin}, P.~M. and {Mac{\'\i}as-P{\'e}rez}, J.~F. and {Maggio}, G. and {Maino}, D. and {Mak}, D.~S.~Y. and {Mandolesi}, N. and {Mangilli}, A. and {Martin}, P.~G. and {Mart{\'\i}nez-Gonz{\'a}lez}, E. and {Masi}, S. and {Matarrese}, S. and {Mazzotta}, P. and {McGehee}, P. and {Mei}, S. and {Melchiorri}, A. and {Melin}, J. -B. and {Mendes}, L. and {Mennella}, A. and {Migliaccio}, M. and {Mitra}, S. and {Miville-Desch{\^e}nes}, M. -A. and {Moneti}, A. and {Montier}, L. and {Morgante}, G. and {Mortlock}, D. and {Moss}, A. and {Munshi}, D. and {Murphy}, J.~A. and {Naselsky}, P. and {Nastasi}, A. and {Nati}, F. and {Natoli}, P. and {Netterfield}, C.~B. and {N{\o}rgaard-Nielsen}, H.~U. and {Noviello}, F. and {Novikov}, D. and {Novikov}, I. and {Olamaie}, M. and {Oxborrow}, C.~A. and {Paci}, F. and {Pagano}, L. and {Pajot}, F. and {Paoletti}, D. and {Pasian}, F. and {Patanchon}, G. and {Pearson}, T.~J. and {Perdereau}, O. and {Perotto}, L. and {Perrott}, Y.~C. and {Perrotta}, F. and {Pettorino}, V. and {Piacentini}, F. and {Piat}, M. and {Pierpaoli}, E. and {Pietrobon}, D. and {Plaszczynski}, S. and {Pointecouteau}, E. and {Polenta}, G. and {Pratt}, G.~W. and {Pr{\'e}zeau}, G. and {Prunet}, S. and {Puget}, J. -L. and {Rachen}, J.~P. and {Reach}, W.~T. and {Rebolo}, R. and {Reinecke}, M. and {Remazeilles}, M. and {Renault}, C. and {Renzi}, A. and {Ristorcelli}, I. and {Rocha}, G. and {Rosset}, C. and {Rossetti}, M. and {Roudier}, G. and {Rozo}, E. and {Rubi{\~n}o-Mart{\'\i}n}, J.~A. and {Rumsey}, C. and {Rusholme}, B. and {Rykoff}, E.~S. and {Sandri}, M. and {Santos}, D. and {Saunders}, R.~D.~E. and {Savelainen}, M. and {Savini}, G. and {Schammel}, M.~P. and {Scott}, D. and {Seiffert}, M.~D. and {Shellard}, E.~P.~S. and {Shimwell}, T.~W. and {Spencer}, L.~D. and {Stanford}, S.~A. and {Stern}, D. and {Stolyarov}, V. and {Stompor}, R. and {Streblyanska}, A. and {Sudiwala}, R. and {Sunyaev}, R. and {Sutton}, D. and {Suur-Uski}, A. -S. and {Sygnet}, J. -F. and {Tauber}, J.~A. and {Terenzi}, L. and {Toffolatti}, L. and {Tomasi}, M. and {Tramonte}, D. and {Tristram}, M. and {Tucci}, M. and {Tuovinen}, J. and {Umana}, G. and {Valenziano}, L. and {Valiviita}, J. and {Van Tent}, B. and {Vielva}, P. and {Villa}, F. and {Wade}, L.~A. and {Wandelt}, B.~D. and {Wehus}, I.~K. and {White}, S.~D.~M. and {Wright}, E.~L. and {Yvon}, D. and {Zacchei}, A. and {Zonca}, A.},
        title = "{Planck 2015 results. XXVII. The second Planck catalogue of Sunyaev-Zeldovich sources}",
      journal = {\aap},
     keywords = {cosmology: observations, galaxies: clusters: general, catalogs, Astrophysics - Cosmology and Nongalactic Astrophysics},
         year = 2016,
        month = sep,
       volume = {594},
          eid = {A27},
        pages = {A27},
          doi = {10.1051/0004-6361/201525823},
archivePrefix = {arXiv},
       eprint = {1502.01598},
 primaryClass = {astro-ph.CO},
       adsurl = {https://ui.adsabs.harvard.edu/abs/2016A&A...594A..27P},
      adsnote = {Provided by the SAO/NASA Astrophysics Data System}
}

@ARTICLE{1958ApJS....3..211A,
       author = {{Abell}, George O.},
        title = "{The Distribution of Rich Clusters of Galaxies.}",
      journal = {\apjs},
         year = 1958,
        month = may,
       volume = {3},
        pages = {211},
          doi = {10.1086/190036},
       adsurl = {https://ui.adsabs.harvard.edu/abs/1958ApJS....3..211A},
      adsnote = {Provided by the SAO/NASA Astrophysics Data System}
}

@ARTICLE{1989ApJS...70....1A,
       author = {{Abell}, George O. and {Corwin}, Harold G., Jr. and {Olowin}, Ronald P.},
        title = "{A Catalog of Rich Clusters of Galaxies}",
      journal = {\apjs},
     keywords = {Astronomical Catalogs, Galactic Clusters, Sky Surveys (Astronomy), Galactic Structure, Luminosity, Red Shift, Schmidt Telescopes, Spatial Distribution, Astronomy, GALAXIES: CLUSTERING, GALAXIES: REDSHIFTS, GALAXIES: STRUCTURE},
         year = 1989,
        month = may,
       volume = {70},
        pages = {1},
          doi = {10.1086/191333},
       adsurl = {https://ui.adsabs.harvard.edu/abs/1989ApJS...70....1A},
      adsnote = {Provided by the SAO/NASA Astrophysics Data System}
}

@ARTICLE{2007ApJ...660..239K,
       author = {{Koester}, B.~P. and {McKay}, T.~A. and {Annis}, J. and {Wechsler}, R.~H. and {Evrard}, A. and {Bleem}, L. and {Becker}, M. and {Johnston}, D. and {Sheldon}, E. and {Nichol}, R. and {Miller}, C. and {Scranton}, R. and {Bahcall}, N. and {Barentine}, J. and {Brewington}, H. and {Brinkmann}, J. and {Harvanek}, M. and {Kleinman}, S. and {Krzesinski}, J. and {Long}, D. and {Nitta}, A. and {Schneider}, D.~P. and {Sneddin}, S. and {Voges}, W. and {York}, D.},
        title = "{A MaxBCG Catalog of 13,823 Galaxy Clusters from the Sloan Digital Sky Survey}",
      journal = {\apj},
     keywords = {Galaxies: Clusters: General, Astrophysics},
         year = 2007,
        month = may,
       volume = {660},
       number = {1},
        pages = {239-255},
          doi = {10.1086/509599},
archivePrefix = {arXiv},
       eprint = {astro-ph/0701265},
 primaryClass = {astro-ph},
       adsurl = {https://ui.adsabs.harvard.edu/abs/2007ApJ...660..239K},
      adsnote = {Provided by the SAO/NASA Astrophysics Data System}
}

@ARTICLE{2016ApJS..224....1R,
       author = {{Rykoff}, E.~S. and {Rozo}, E. and {Hollowood}, D. and {Bermeo-Hernandez}, A. and {Jeltema}, T. and {Mayers}, J. and {Romer}, A.~K. and {Rooney}, P. and {Saro}, A. and {Vergara Cervantes}, C. and {Wechsler}, R.~H. and {Wilcox}, H. and {Abbott}, T.~M.~C. and {Abdalla}, F.~B. and {Allam}, S. and {Annis}, J. and {Benoit-L{\'e}vy}, A. and {Bernstein}, G.~M. and {Bertin}, E. and {Brooks}, D. and {Burke}, D.~L. and {Capozzi}, D. and {Carnero Rosell}, A. and {Carrasco Kind}, M. and {Castander}, F.~J. and {Childress}, M. and {Collins}, C.~A. and {Cunha}, C.~E. and {D'Andrea}, C.~B. and {da Costa}, L.~N. and {Davis}, T.~M. and {Desai}, S. and {Diehl}, H.~T. and {Dietrich}, J.~P. and {Doel}, P. and {Evrard}, A.~E. and {Finley}, D.~A. and {Flaugher}, B. and {Fosalba}, P. and {Frieman}, J. and {Glazebrook}, K. and {Goldstein}, D.~A. and {Gruen}, D. and {Gruendl}, R.~A. and {Gutierrez}, G. and {Hilton}, M. and {Honscheid}, K. and {Hoyle}, B. and {James}, D.~J. and {Kay}, S.~T. and {Kuehn}, K. and {Kuropatkin}, N. and {Lahav}, O. and {Lewis}, G.~F. and {Lidman}, C. and {Lima}, M. and {Maia}, M.~A.~G. and {Mann}, R.~G. and {Marshall}, J.~L. and {Martini}, P. and {Melchior}, P. and {Miller}, C.~J. and {Miquel}, R. and {Mohr}, J.~J. and {Nichol}, R.~C. and {Nord}, B. and {Ogando}, R. and {Plazas}, A.~A. and {Reil}, K. and {Sahl{\'e}n}, M. and {Sanchez}, E. and {Santiago}, B. and {Scarpine}, V. and {Schubnell}, M. and {Sevilla-Noarbe}, I. and {Smith}, R.~C. and {Soares-Santos}, M. and {Sobreira}, F. and {Stott}, J.~P. and {Suchyta}, E. and {Swanson}, M.~E.~C. and {Tarle}, G. and {Thomas}, D. and {Tucker}, D. and {Uddin}, S. and {Viana}, P.~T.~P. and {Vikram}, V. and {Walker}, A.~R. and {Zhang}, Y. and {DES Collaboration}},
        title = "{The RedMaPPer Galaxy Cluster Catalog From DES Science Verification Data}",
      journal = {\apjs},
     keywords = {galaxies: clusters: general, Astrophysics - Cosmology and Nongalactic Astrophysics},
         year = 2016,
        month = may,
       volume = {224},
       number = {1},
          eid = {1},
        pages = {1},
          doi = {10.3847/0067-0049/224/1/1},
archivePrefix = {arXiv},
       eprint = {1601.00621},
 primaryClass = {astro-ph.CO},
       adsurl = {https://ui.adsabs.harvard.edu/abs/2016ApJS..224....1R},
      adsnote = {Provided by the SAO/NASA Astrophysics Data System}
}

@ARTICLE{2000ApJS..129..435B,
       author = {{B{\"o}hringer}, H. and {Voges}, W. and {Huchra}, J.~P. and {McLean}, B. and {Giacconi}, R. and {Rosati}, P. and {Burg}, R. and {Mader}, J. and {Schuecker}, P. and {Simi{\c{c}}}, D. and {Komossa}, S. and {Reiprich}, T.~H. and {Retzlaff}, J. and {Tr{\"u}mper}, J.},
        title = "{The Northern ROSAT All-Sky (NORAS) Galaxy Cluster Survey. I. X-Ray Properties of Clusters Detected as Extended X-Ray Sources}",
      journal = {\apjs},
     keywords = {Galaxies: Clusters: General, Cosmology: Large-Scale Structure of Universe, Surveys, X-Rays: Galaxies, Astrophysics},
         year = 2000,
        month = aug,
       volume = {129},
       number = {2},
        pages = {435-474},
          doi = {10.1086/313427},
archivePrefix = {arXiv},
       eprint = {astro-ph/0003219},
 primaryClass = {astro-ph},
       adsurl = {https://ui.adsabs.harvard.edu/abs/2000ApJS..129..435B},
      adsnote = {Provided by the SAO/NASA Astrophysics Data System}
}

@ARTICLE{2004A&A...425..367B,
       author = {{B{\"o}hringer}, H. and {Schuecker}, P. and {Guzzo}, L. and {Collins}, C.~A. and {Voges}, W. and {Cruddace}, R.~G. and {Ortiz-Gil}, A. and {Chincarini}, G. and {De Grandi}, S. and {Edge}, A.~C. and {MacGillivray}, H.~T. and {Neumann}, D.~M. and {Schindler}, S. and {Shaver}, P.},
        title = "{The ROSAT-ESO Flux Limited X-ray (REFLEX) Galaxy cluster survey. V. The cluster catalogue}",
      journal = {\aap},
     keywords = {catalogs, surveys, galaxies: clusters: general, cosmology: large-scale structure of Universe, X-rays: general, X-rays: galaxies: clusters, Astrophysics},
         year = 2004,
        month = oct,
       volume = {425},
        pages = {367-383},
          doi = {10.1051/0004-6361:20034484},
archivePrefix = {arXiv},
       eprint = {astro-ph/0405546},
 primaryClass = {astro-ph},
       adsurl = {https://ui.adsabs.harvard.edu/abs/2004A&A...425..367B},
      adsnote = {Provided by the SAO/NASA Astrophysics Data System}
}

@ARTICLE{2012MNRAS.423.1024M,
       author = {{Mehrtens}, Nicola and {Romer}, A. Kathy and {Hilton}, Matt and {Lloyd-Davies}, E.~J. and {Miller}, Christopher J. and {Stanford}, S.~A. and {Hosmer}, Mark and {Hoyle}, Ben and {Collins}, Chris A. and {Liddle}, Andrew R. and {Viana}, Pedro T.~P. and {Nichol}, Robert C. and {Stott}, John P. and {Dubois}, E. Naomi and {Kay}, Scott T. and {Sahl{\'e}n}, Martin and {Young}, Owain and {Short}, C.~J. and {Christodoulou}, L. and {Watson}, William A. and {Davidson}, Michael and {Harrison}, Craig D. and {Baruah}, Leon and {Smith}, Mathew and {Burke}, Claire and {Mayers}, Julian A. and {Deadman}, Paul-James and {Rooney}, Philip J. and {Edmondson}, Edward M. and {West}, Michael and {Campbell}, Heather C. and {Edge}, Alastair C. and {Mann}, Robert G. and {Sabirli}, Kivanc and {Wake}, David and {Benoist}, Christophe and {da Costa}, Luiz and {Maia}, Marcio A.~G. and {Ogando}, Ricardo},
        title = "{The XMM Cluster Survey: optical analysis methodology and the first data release}",
      journal = {\mnras},
     keywords = {techniques: photometric, techniques: spectroscopic, surveys, galaxies: clusters: individual: XMMXCS J091821.9+211446.0, galaxies: distances and redshifts, X-rays: galaxies: clusters, Astrophysics - Cosmology and Nongalactic Astrophysics},
         year = 2012,
        month = jun,
       volume = {423},
       number = {2},
        pages = {1024-1052},
          doi = {10.1111/j.1365-2966.2012.20931.x},
archivePrefix = {arXiv},
       eprint = {1106.3056},
 primaryClass = {astro-ph.CO},
       adsurl = {https://ui.adsabs.harvard.edu/abs/2012MNRAS.423.1024M},
      adsnote = {Provided by the SAO/NASA Astrophysics Data System}
}

@ARTICLE{2013MNRAS.430..134W,
       author = {{Willis}, J.~P. and {Clerc}, N. and {Bremer}, M.~N. and {Pierre}, M. and {Adami}, C. and {Ilbert}, O. and {Maughan}, B. and {Maurogordato}, S. and {Pacaud}, F. and {Valtchanov}, I. and {Chiappetti}, L. and {Thanjavur}, K. and {Gwyn}, S. and {Stanway}, E.~R. and {Winkworth}, C.},
        title = "{Distant galaxy clusters in the XMM Large Scale Structure survey}",
      journal = {\mnras},
     keywords = {galaxies: clusters: general, galaxies: high-redshift, Astrophysics - Cosmology and Nongalactic Astrophysics},
         year = 2013,
        month = mar,
       volume = {430},
       number = {1},
        pages = {134-156},
          doi = {10.1093/mnras/sts540},
archivePrefix = {arXiv},
       eprint = {1212.4185},
 primaryClass = {astro-ph.CO},
       adsurl = {https://ui.adsabs.harvard.edu/abs/2013MNRAS.430..134W},
      adsnote = {Provided by the SAO/NASA Astrophysics Data System}
}

@ARTICLE{2003PhRvD..68h3506B,
       author = {{Battye}, Richard A. and {Weller}, Jochen},
        title = "{Constraining cosmological parameters using Sunyaev-Zel'dovich cluster surveys}",
      journal = {\prd},
     keywords = {98.80.Es, 98.65.Cw, 98.80.Cq, Observational cosmology, Galaxy clusters, Particle-theory and field-theory models of the early Universe, Astrophysics},
         year = 2003,
        month = oct,
       volume = {68},
       number = {8},
          eid = {083506},
        pages = {083506},
          doi = {10.1103/PhysRevD.68.083506},
archivePrefix = {arXiv},
       eprint = {astro-ph/0305568},
 primaryClass = {astro-ph},
       adsurl = {https://ui.adsabs.harvard.edu/abs/2003PhRvD..68h3506B},
      adsnote = {Provided by the SAO/NASA Astrophysics Data System}
}

@ARTICLE{2010MNRAS.407.2339S,
       author = {{Sartoris}, B. and {Borgani}, S. and {Fedeli}, C. and {Matarrese}, S. and {Moscardini}, L. and {Rosati}, P. and {Weller}, J.},
        title = "{The potential of X-ray cluster surveys to constrain primordial non-Gaussianity}",
      journal = {\mnras},
     keywords = {galaxies: clusters: general, cosmological parameters, X-rays: galaxies: clusters, Astrophysics - Cosmology and Nongalactic Astrophysics},
         year = 2010,
        month = oct,
       volume = {407},
       number = {4},
        pages = {2339-2354},
          doi = {10.1111/j.1365-2966.2010.17050.x},
archivePrefix = {arXiv},
       eprint = {1003.0841},
 primaryClass = {astro-ph.CO},
       adsurl = {https://ui.adsabs.harvard.edu/abs/2010MNRAS.407.2339S},
      adsnote = {Provided by the SAO/NASA Astrophysics Data System}
}

@ARTICLE{2007ApJ...668..772M,
       author = {{Maughan}, B.~J.},
        title = "{The L$_{X}$-Y$_{X}$ Relation: Using Galaxy Cluster X-Ray Luminosity as a Robust, Low-Scatter Mass Proxy}",
      journal = {\apj},
     keywords = {Cosmology: Observations, Galaxies: Clusters: General, Galaxies: High-Redshift, Galaxies: Intergalactic Medium, X-Rays: Galaxies, Astrophysics},
         year = 2007,
        month = oct,
       volume = {668},
       number = {2},
        pages = {772-780},
          doi = {10.1086/520831},
archivePrefix = {arXiv},
       eprint = {astro-ph/0703504},
 primaryClass = {astro-ph},
       adsurl = {https://ui.adsabs.harvard.edu/abs/2007ApJ...668..772M},
      adsnote = {Provided by the SAO/NASA Astrophysics Data System}
}

@article{May,
    author = "May, Simon and Springel, Volker",
    title = "{Structure formation in large-volume cosmological simulations of fuzzy dark matter: Impact of the non-linear dynamics}",
    eprint = "2101.01828",
    archivePrefix = "arXiv",
    primaryClass = "astro-ph.CO",
    month = "1",
    year = "2021"
}

@article{growthfactor,
    author = "Linares Cede\~no, Francisco X. and Gonz\'alez-Morales, Alma X. and Ure\~na-L\'opez, L. Arturo",
    title = "{Ultralight DM bosons with an Axion-like potential: scale-dependent constraints revisited}",
    eprint = "2006.05037",
    archivePrefix = "arXiv",
    primaryClass = "astro-ph.CO",
    month = "6",
    year = "2020"
}

@article{Lima:2004wn,
    author = "Lima, Marcos and Hu, Wayne",
    title = "{Self - calibration of cluster dark energy studies: Counts in cells}",
    eprint = "astro-ph/0401559",
    archivePrefix = "arXiv",
    doi = "10.1103/PhysRevD.70.043504",
    journal = "Phys. Rev. D",
    volume = "70",
    pages = "043504",
    year = "2004"
}

@article{Davoudiasl:2019nlo,
    author = "Davoudiasl, Hooman and Denton, Peter B",
    title = "{Ultralight Boson Dark Matter and Event Horizon Telescope Observations of M87*}",
    eprint = "1904.09242",
    archivePrefix = "arXiv",
    primaryClass = "astro-ph.CO",
    doi = "10.1103/PhysRevLett.123.021102",
    journal = "Phys. Rev. Lett.",
    volume = "123",
    number = "2",
    pages = "021102",
    year = "2019"
}

@ARTICLE{Zoutendijk,
       author = {{Zoutendijk}, Sebastiaan L. and {Brinchmann}, Jarle and {Bouch{\'e}}, Nicolas F. and {den Brok}, Mark and {Krajnovi{\'c}}, Davor and {Kuijken}, Konrad and {Maseda}, Michael V. and {Schaye}, Joop},
        title = "{The MUSE-Faint survey. II. The dark matter-density profile of the ultra-faint dwarf galaxy Eridanus 2}",
      journal = {arXiv e-prints},
     keywords = {Astrophysics - Astrophysics of Galaxies, Astrophysics - Cosmology and Nongalactic Astrophysics},
         year = 2021,
        month = jan,
          eid = {arXiv:2101.00253},
        pages = {arXiv:2101.00253},
archivePrefix = {arXiv},
       eprint = {2101.00253},
 primaryClass = {astro-ph.GA},
       adsurl = {https://ui.adsabs.harvard.edu/abs/2021arXiv210100253Z},
      adsnote = {Provided by the SAO/NASA Astrophysics Data System}
}

@ARTICLE{Jeans1,
       author = {{Chavanis}, Pierre-Henri},
        title = "{Mass-radius relation of Newtonian self-gravitating Bose-Einstein condensates with short-range interactions. I. Analytical results}",
      journal = {\prd},
     keywords = {95.35.+d, 95.30.Sf, Dark matter, Relativity and gravitation, Astrophysics - Cosmology and Nongalactic Astrophysics},
         year = 2011,
        month = aug,
       volume = {84},
       number = {4},
          eid = {043531},
        pages = {043531},
          doi = {10.1103/PhysRevD.84.043531},
archivePrefix = {arXiv},
       eprint = {1103.2050},
 primaryClass = {astro-ph.CO},
       adsurl = {https://ui.adsabs.harvard.edu/abs/2011PhRvD..84d3531C},
      adsnote = {Provided by the SAO/NASA Astrophysics Data System}
}

@article{Jeans2,
    author = "Suarez, Abril and Matos, Tonatiuh",
    title = "{Structure Formation with Scalar Field Dark Matter: The Fluid Approach}",
    eprint = "1101.4039",
    archivePrefix = "arXiv",
    primaryClass = "gr-qc",
    doi = "10.1111/j.1365-2966.2011.19012.x",
    journal = {\mnras},
    volume = "416",
    pages = "87",
    year = "2011"
}

@ARTICLE{Jeans3,
       author = {{Khlopov}, M. Iu. and {Malomed}, B.~A. and {Zeldovich}, Ia. B.},
        title = "{Gravitational instability of scalar fields and formation of primordial black holes}",
      journal = {\mnras},
     keywords = {Black Holes (Astronomy), Computational Astrophysics, Gravitation Theory, Scalars, Universe, Dynamic Stability, Flow Stability, Hydrodynamics, Jeans Theory, Symmetry, Unified Field Theory, Astrophysics},
         year = 1985,
        month = aug,
       volume = {215},
        pages = {575-589},
          doi = {10.1093/mnras/215.4.575},
       adsurl = {https://ui.adsabs.harvard.edu/abs/1985MNRAS.215..575K},
      adsnote = {Provided by the SAO/NASA Astrophysics Data System}
}

@article{Jeans4,
    author = "Bianchi, M. and Grasso, D. and Ruffini, R.",
    title = "{Jeans mass of a cosmological coherent scalar field}",
    journal = "Astron. Astrophys.",
    volume = "231",
    number = "2",
    pages = "301--308",
    year = "1990"
}

@article{Jeans5,
    author = "Johnson, Matthew C. and Kamionkowski, Marc",
    title = "{Dynamical and Gravitational Instability of Oscillating-Field Dark Energy and Dark Matter}",
    eprint = "0805.1748",
    archivePrefix = "arXiv",
    primaryClass = "astro-ph",
    doi = "10.1103/PhysRevD.78.063010",
    journal = "Phys. Rev. D",
    volume = "78",
    pages = "063010",
    year = "2008"
}

@article{Jeans6,
    author = "Lee, Jae-Weon and Lim, Sooil",
    title = "{Minimum mass of galaxies from BEC or scalar field dark matter}",
    eprint = "0812.1342",
    archivePrefix = "arXiv",
    primaryClass = "astro-ph",
    doi = "10.1088/1475-7516/2010/01/007",
    journal = "JCAP",
    volume = "01",
    pages = "007",
    year = "2010"
}

@article{Poddar20,
  title = {Constraints on ultralight axions from compact binary systems},
  author = {Poddar, Tanmay Kumar and Mohanty, Subhendra and Jana, Soumya},
  journal = {Phys. Rev. D},
  volume = {101},
  issue = {8},
  pages = {083007},
  numpages = {9},
  year = {2020},
  month = {Apr},
  publisher = {American Physical Society},
  doi = {10.1103/PhysRevD.101.083007},
  url = {https://link.aps.org/doi/10.1103/PhysRevD.101.083007}
}

@article{Rogers:2020ltq,
    author = "Rogers, Keir K. and Peiris, Hiranya V.",
    title = "{Strong Bound on Canonical Ultralight Axion Dark Matter from the Lyman-Alpha Forest}",
    eprint = "2007.12705",
    archivePrefix = "arXiv",
    primaryClass = "astro-ph.CO",
    doi = "10.1103/PhysRevLett.126.071302",
    journal = "Phys. Rev. Lett.",
    volume = "126",
    number = "7",
    pages = "071302",
    year = "2021"
}

@ARTICLE{2012MNRAS.422...44P,
       author = {{Pillepich}, Annalisa and {Porciani}, Cristiano and {Reiprich}, Thomas H.},
        title = "{The X-ray cluster survey with eRosita: forecasts for cosmology, cluster physics and primordial non-Gaussianity}",
      journal = {\mnras},
     keywords = {galaxies: clusters: general, cosmological parameters, early Universe, large-scale structure of Universe, X-rays: galaxies: clusters, Astrophysics - Cosmology and Nongalactic Astrophysics},
         year = 2012,
        month = may,
       volume = {422},
       number = {1},
        pages = {44-69},
          doi = {10.1111/j.1365-2966.2012.20443.x},
archivePrefix = {arXiv},
       eprint = {1111.6587},
 primaryClass = {astro-ph.CO},
       adsurl = {https://ui.adsabs.harvard.edu/abs/2012MNRAS.422...44P},
      adsnote = {Provided by the SAO/NASA Astrophysics Data System}
}

@article{Staniszewski:2008ma,
    author = "Staniszewski, Z. and others",
    title = "{Galaxy clusters discovered with a Sunyaev-Zel'dovich effect survey}",
    eprint = "0810.1578",
    archivePrefix = "arXiv",
    primaryClass = "astro-ph",
    doi = "10.1088/0004-637X/701/1/32",
    journal = "Astrophys. J.",
    volume = "701",
    pages = "32--41",
    year = "2009"
}

@ARTICLE{Krippendorf1,
       author = {{Conlon}, Joseph P. and {Day}, Francesca and {Jennings}, Nicholas and {Krippendorf}, Sven and {Muia}, Francesco},
        title = "{Projected bounds on ALPs from Athena}",
      journal = {\mnras},
     keywords = {astroparticle physics, elementary particles, galaxies: clusters: individual: Perseus, Astrophysics - High Energy Astrophysical Phenomena, High Energy Physics - Phenomenology},
         year = 2018,
        month = feb,
       volume = {473},
       number = {4},
        pages = {4932-4936},
          doi = {10.1093/mnras/stx2652},
archivePrefix = {arXiv},
       eprint = {1707.00176},
 primaryClass = {astro-ph.HE},
       adsurl = {https://ui.adsabs.harvard.edu/abs/2018MNRAS.473.4932C},
      adsnote = {Provided by the SAO/NASA Astrophysics Data System}
}

@ARTICLE{Krippendorf2,
       author = {{Conlon}, Joseph P. and {Day}, Francesca and {Jennings}, Nicholas and {Krippendorf}, Sven and {Rummel}, Markus},
        title = "{Constraints on axion-like particles from non-observation of spectral modulations for X-ray point sources}",
      journal = {\jcap},
     keywords = {Astrophysics - High Energy Astrophysical Phenomena, High Energy Physics - Phenomenology},
         year = 2017,
        month = jul,
       volume = {2017},
       number = {7},
          eid = {005},
        pages = {005},
          doi = {10.1088/1475-7516/2017/07/005},
archivePrefix = {arXiv},
       eprint = {1704.05256},
 primaryClass = {astro-ph.HE},
       adsurl = {https://ui.adsabs.harvard.edu/abs/2017JCAP...07..005C},
      adsnote = {Provided by the SAO/NASA Astrophysics Data System}
}

@ARTICLE{Krippendorf3,
       author = {{Berg}, Marcus and {Conlon}, Joseph P. and {Day}, Francesca and {Jennings}, Nicholas and {Krippendorf}, Sven and {Powell}, Andrew J. and {Rummel}, Markus},
        title = "{Constraints on Axion-like Particles from X-Ray Observations of NGC1275}",
      journal = {\apj},
     keywords = {elementary particles, X-rays: galaxies: clusters, X-rays: individual: NGC1275, Astrophysics - High Energy Astrophysical Phenomena, High Energy Physics - Phenomenology},
         year = 2017,
        month = oct,
       volume = {847},
       number = {2},
          eid = {101},
        pages = {101},
          doi = {10.3847/1538-4357/aa8b16},
archivePrefix = {arXiv},
       eprint = {1605.01043},
 primaryClass = {astro-ph.HE},
       adsurl = {https://ui.adsabs.harvard.edu/abs/2017ApJ...847..101B},
      adsnote = {Provided by the SAO/NASA Astrophysics Data System}
}